\documentclass[traditabstract]{aa}
\usepackage{longtable}
\usepackage{natbib}
\usepackage{graphicx}

\bibpunct[, ]{(}{)}{;}{a}{}{,}

\usepackage{color}

\begin{document}

\title{Environment, morphology and stellar populations of bulgeless low surface brightness galaxies}

\author{
 X. Shao\inst{1,2,3}, K. Disseau\inst{2}, Y. B. Yang\inst{2}, F. Hammer\inst{2}, M. Puech\inst{2}, M. Rodrigues\inst{2},
 Y. C. Liang\inst{1}, L. C. Deng\inst{1}}
\institute{ Key Laboratory of Optical Astronomy, National Astronomical
  Observatories, Chinese Academy of Sciences, 20A Datun Road, Chaoyang District, Beijing, 100012,
  China
\and Laboratoire GEPI,Observatoire de Paris,  CNRS-UMR8111, Univ. Paris Diderot,
  Sorbonne Paris Cit\'{e}, 5 place Jules Janssen, F-92195 Meudon, France
 \and University of the Chinese Academy of Sciences, Beijing, 100049, China}

\date{Received  / Accepted }

\abstract{Based on the Sloan Digital Sky Survey DR 7, we investigate the environment, morphology and
stellar population of bulgeless low surface brightness (LSB) galaxies in a volume-limited sample with
redshift ranging from 0.024 to 0.04 and $M_r$ $\leq$ $-18.8$. The
local density parameter $\Sigma_5$
is used to trace their environments. We find that, for bulgeless galaxies, the surface brightness does not depend on
the environment. The stellar populations are compared for bulgeless LSB galaxies in
different environments and for bulgeless LSB galaxies with different morphologies. The stellar populations
of LSB galaxies in low density regions are similar to those
of LSB galaxies in high density regions. Irregular LSB galaxies
have more young stars and are more metal-poor than regular LSB galaxies.
These results suggest that
the evolution of LSB galaxies may be driven by their dynamics including
mergers rather than by their large scale environment.}


\titlerunning{stellar populations of bulgeless LSB galaxies}
\authorrunning{Shao et al.}

\maketitle

\section{Introduction}

Low surface brightness (LSB) galaxies usually refer to galaxies possessing an
exponential disk with a face-on central surface brightness
fainter than the ambient night sky. This type of galaxy has been widely discussed
since Freeman (1970) found that 28 of 36 disk galaxies have
central surface brightness in the range $\mu_0(B)$=21.65$\pm$0.30 mag arcsec$^{-2}$.
\citet{1976Natur.263..573D} showed that this result is biased by a strong selection effect, which acts
against the discovery of LSB galaxies. Thanks to improvements in telescopes and instruments,
many works have been done to understand this kind of galaxy \citep{1988AJ.....95.1389S,
1987AJ.....94.1126C, 1992AJ....103.1107S}. \citet{1995AJ....109.2019M} found that
the morphologies of LSB galaxies span a wide range, from dwarfs and irregulars to disk
galaxies. Regardless of their morphology and size, all LSB galaxies have low star
formation rates, low metallicities and extended HI gas disks \citep{1997ARA&A..35..267I}.
There is no conventional definition for LSB galaxies yet (Impey \& Bothun 1997), although they have
been intensively studied. The most common thresholds of $\mu_0(B)$ for LSB galaxies found in the literature are
between 22 and 23 mag arcsec$^{-2}$ \citep{1995AJ....109.2019M, 1995MNRAS.274..235D, 
2004A&A...422L...5R, 2011ApJ...728...74G}. Since the definition of LSB galaxies only depends on the
central surface brightness, more work needs to be done to understand whether LSB
galaxies trace the formation/evolution of galaxies in a more representative manner
than high surface brightness (HSB) galaxies, or if they just follow the trends of HSB galaxies.
This is crucial for understanding the formation and evolution of galaxies.

The nature of the local and large-scale environment in which LSB galaxies are
embedded has been analyzed by \citet{1993AJ....106..530B} and \citet{1994MNRAS.267..129M}. They found
that LSB galaxies are located in more isolate environments compared to HSB galaxies.
The deficiency of nearby companions for LSB galaxies was discovered by \citet{1993egte.conf...82Z}.
Recently, based on the large amount of data released by the SDSS, the
influence of environment on the evolution of LSB galaxies was further revealed.
Using the Early Data Release of the SDSS, \citet{2004A&A...422L...5R} show significant
differences in local density between LSB and HSB galaxies on scales from 2 to 5 Mpc.
In their following analysis, \citet{2009A&A...504..807R} compared LSB
galaxies at 0.01$<$$z$$\leq$0.055 and at 0.055$<$$z$$\leq$0.1. The results suggested that
LSB galaxies formed in a low-density region of the initial universe and have drifted
to the outer parts of the filaments and walls of large-scale structures.

In addition to this issue, the origin of the LSB phenomenon is still unknown. Could LSB galaxies
result from a similar formation scenario as HSB galaxies?
Considerable progress has been done in understanding the formation
of HSB disk galaxies. In examining the progenitors of the latter, \citet{2005A&A...430..115H} suggested
that their disks have been rebuilt after a past collision of gas-rich mergers. This
has been supported by the strong evolution of HSB disk morphologies
revealed by \citet{2010A&A...509A..78D}, showing that half of the disk
progenitors were morphologically peculiar 6 Gyr ago. In fact, the combination of
morphology and kinematics \citep{2008A&A...484..159N, 2009A&A...507.1313H} convincingly
demonstrated that a significant fraction of present day disks were formed after being
re-processed through a gas-rich major merger, in agreement
with expectations from $\Lambda$CDM \citep{2012ApJ...753..128P}. This has been
discussed by \citet{2010ApJ...723...54K} who claimed that most local disks do not have a
classical bulge, implying that they could not be related to mergers. However, the question needs further
clarifications because pseudo bulges or even bulgeless galaxies can be formed
after very gas-rich mergers (e.g., \citet{2012arXiv1211.1978H, 2012MNRAS.424.1232K}).

As part of our series of works (\citet{2008MNRAS.391..986Z} and \citet{2010MNRAS.409..213L},
that analyzed metallicities in LSB galaxies based on a large sample), we would like
to analyze the environment, morphology and stellar populations of bulgeless LSB galaxies.
Bulgeless galaxies have lower average surface brightness than galaxies with bulges \citep{2013MNRAS.435.1186S}.
The simulations of \citet{2010Natur.463..203G} show that strong outflows from star formation can drag dark and luminous matter
with them to remove matter with low angular momentum and prevent the formation of a high-density core or bulge. This
is supported by \citet{2005ApJ...618..237W} who found massive star clusters at the cores of bulgeless galaxies.
The analysis of central star formation in relation to the environment and dynamics may be helpful for understanding
the evolution of bulgeless LSB galaxies. When analyzing
stellar populations using SDSS spectra, the selection of bulgeless LSB galaxies has the advantage
that most of the light covered by the SDSS fiber will
sample the properties of central disks without contaminations from the bulge light.
Several studies have found that LSB galaxies are
dominated by bulgeless galaxies. \citet{1995AJ....109.2019M} analyzed
the images of 22 LSB galaxies and found that most of them show B/D $<$ 0.1.
\citet{1997AJ....113.1212O} analyzed the morphology of 127 LSB galaxies and found that
the majority of them (80\%) were well fitted by a single exponential profile.
This implies that studying a large sample of bulgeless LSB galaxies may significantly
help in general to understand the properties of LSB galaxies.

Our bulgeless LSB galaxies are selected from the nearby universe.
The local densities are calculated to see in what kinds of environment bulgeless LSB galaxies
are located. Their stellar populations will be analyzed to see which factor,
environment or dynamics, drives the evolution of bulgeless LSB galaxies.
This paper is organized as follows. In Section 2, we describe the selection of
our samples. The relations between surface brightness of bulgeless galaxies and 
their environment are investigated in section 3. 
In Section 4, we analyze the stellar populations of bulgeless LSB galaxies and see whether
their evolution is driven by their environment or by their morphological behavior. Finally,
we discuss our results and summarize our conclusions in Sections 5 \& 6 respectively. Throughout the paper, a cosmological
model with $H_0$ = 70 kms$^{-1}$ Mpc$^{-1}$, $\Omega_M$ = 0.3 and $\Omega_{\Lambda}$ = 0.7 has
been adopted.

\section{The sample}

\subsection{Volume-limited bulgeless sample}

\begin{figure}
\centering
\includegraphics [width=7.5cm, height=7.5cm] {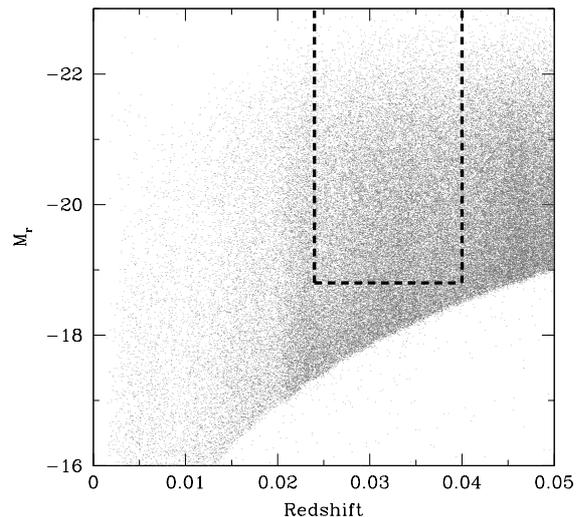}
\caption {The location of our volume-limited sample in
the SDSS main sample (surrounded by black dashed line).} \label{fig.volume}
\end{figure}

The bulgeless LSB galaxies analyzed in this work are obtained from the main-galaxy sample of
SDSS Data Release 7 \citep{2000AJ....120.1579Y, 2002AJ....123..485S,
2002AJ....124.1810S, 2009ApJS..182..543A}.
The SDSS is the most ambitious astronomical survey ever undertaken in
imaging and spectroscopy and targets hundreds of thousands of galaxies. The imaging data, which are
obtained in drift scan mode and have covered more than 10$^4$ deg$^2$ until now, are
95\% complete for point sources at magnitudes of 22.0 22.2 22.2 21.3 and 20.5 in five bands ($ugriz$),
respectively. Our samples are selected following the criteria below:

\begin{table}
\begin{center}
\caption{Number of galaxies in the different samples.}
\label{sample}
\begin{tabular}{c c c}
  \hline \hline
  Sample   &      Number    &          Description                \\ \hline
    $S_0$                     &  31674    &  Volume-limited sample  \\
   $S_1$         &  2606      &  Bulgeless volume-limited sample   \\
\ \    \ \ \ \ $S_{LSB}$         &  1235     &  bulgeless LSB galaxies   \\
\ \    \ \ \ \ $S_{HSB}$    &  1371     &  bulgeless HSB galaxies   \\ \hline
\end{tabular}
\end{center}
\end{table}

\begin{figure*}
\centering
\includegraphics [width=7.5cm, height=6.5cm] {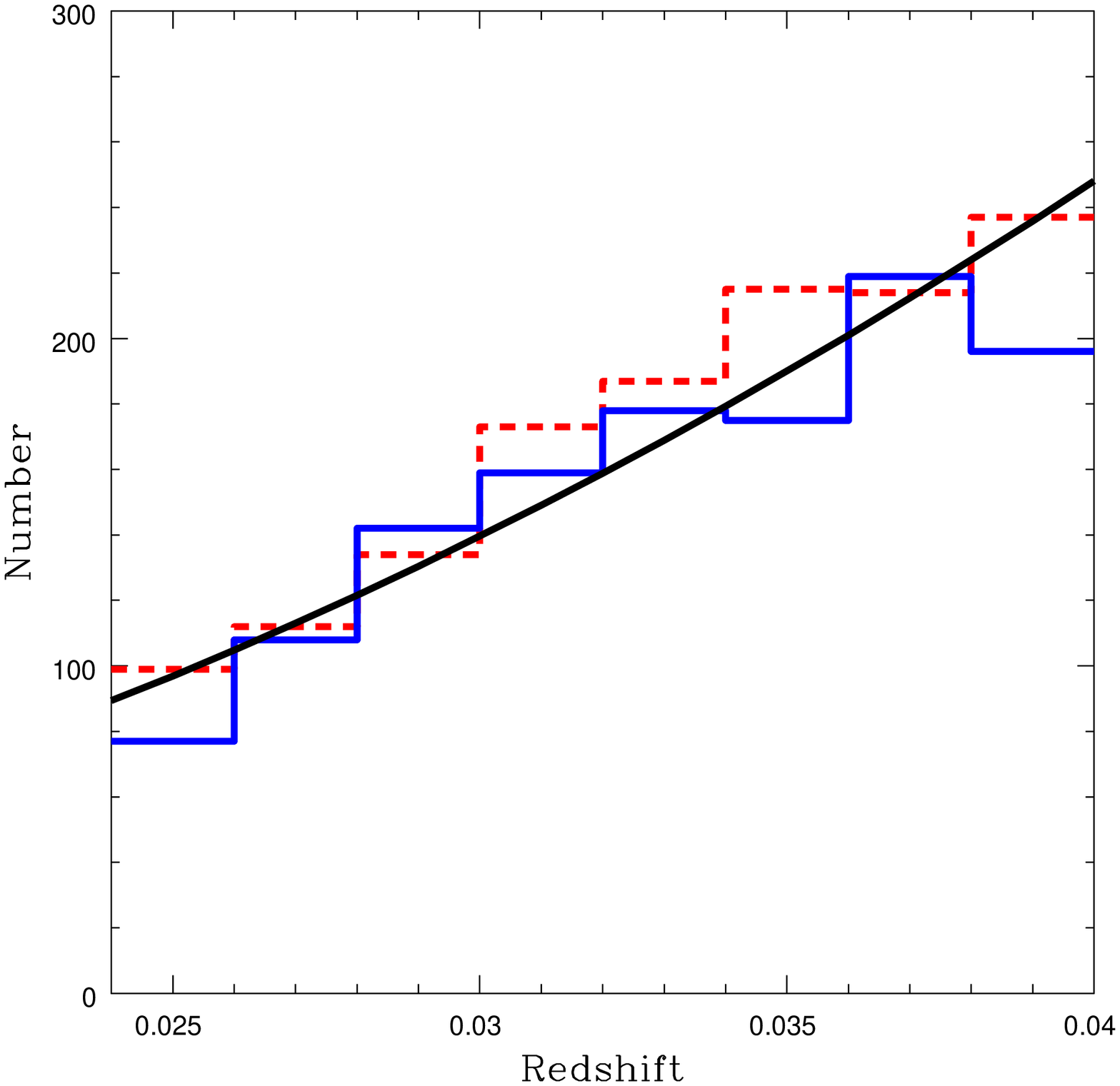}
\includegraphics [width=7.5cm, height=6.5cm] {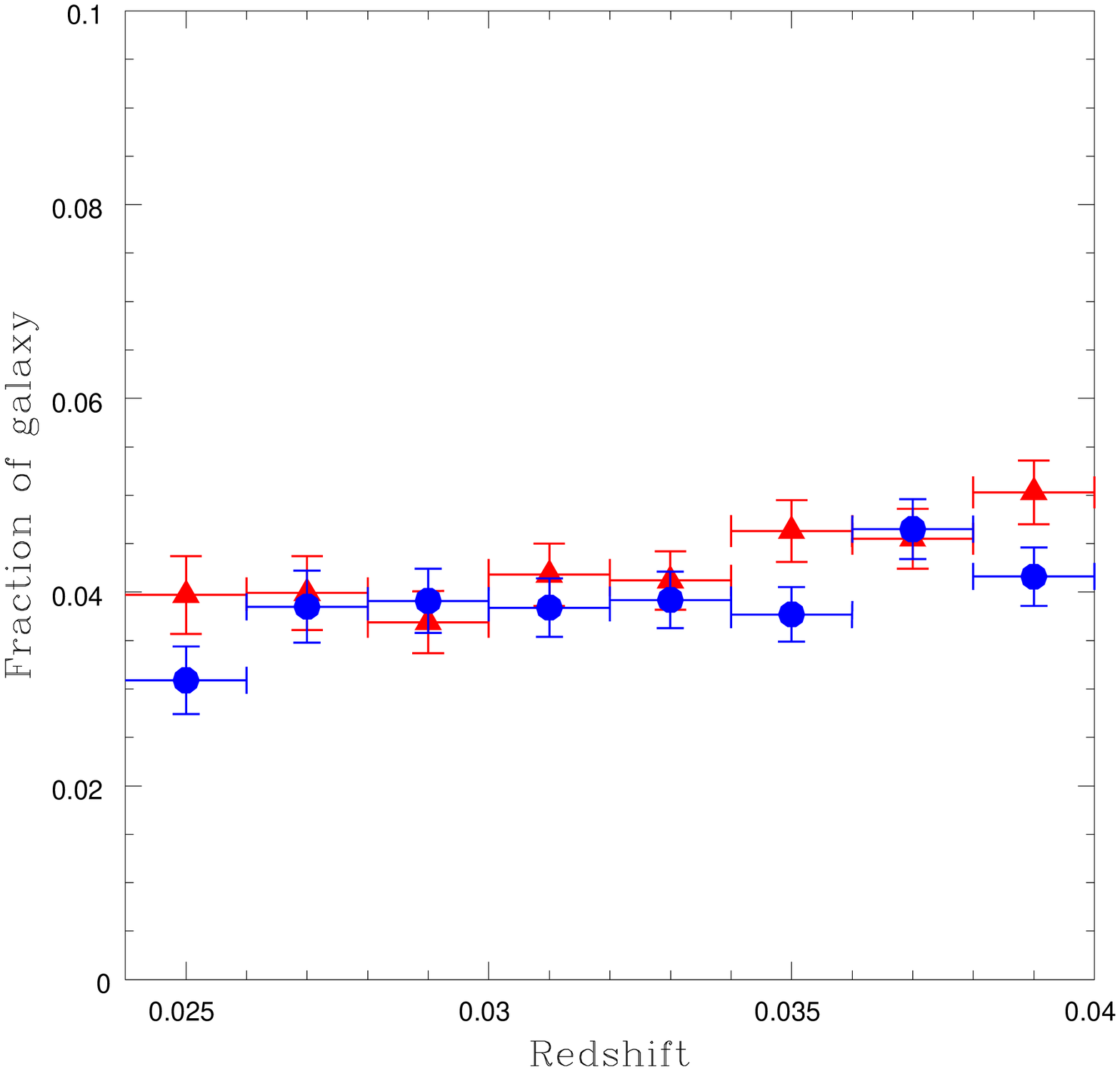}
\caption {The histogram distribution of redshifts for LSB (blue, sample $S_{LSB}$) and
HSB (red, sample $S_{HSB}$) galaxies and their fraction
of the whole sample as a function of redshift. The black solid line denotes the 
expected number histogram of a complete volume-limited sample.} \label{fig.histz}
\end{figure*}

\begin{enumerate}

\item We construct a volume-limited sample to investigate the environments and
analyze the stellar populations. A volume-limited sample eliminates the luminosity
bias in the SDSS database. The redshift range of this sample is 0.024$<$$z$$<$0.04 and the
corresponding absolute magnitude limit is $M_r$$\leq$$-18.8$. The lower redshift limit
avoids effects from proper motions, which are not negligible
in the nearby universe. The upper redshift limit is set to
include as many faint galaxies as possible when
studying the environment. In Figure~\ref{fig.volume}, we show the limits of our 
volume-limited sample (sample $S_0$),
which is surrounded by the black dashed line.

\item Bulgeless galaxies are selected from this volume-limited sample. Strictly, decomposition
is a good method to estimate the fraction of bulge light and disk light in one galaxy. Here
we use the parameter $fracDeV_r$, which is provided by the SDSS database, 
to select bulgeless galaxies. However, $fracDeV$ is an qualitative indicator of the
existence of a bulge and it cannot be used to represent the fraction of the
bulge light quantitatively. Considering the definition of $fracDeV$
\footnote{$F_{composite}=fracDeV F_{deV}+(1-fracDeV) F_{exp}$},
only for the case of $fracDeV$$=$0, it corresponds to
a pure disk galaxy without a bulge. So, we choose galaxies with $fracDeV_r$$=$0 in our
volume-limited sample and get the bulgeless galaxies.

\item Low inclination galaxies are required to avoid a possible bias caused by
internal extinction. As pointed out by
\citet{2008ApJ...687..976U}, extinction related to the inclination becomes
important when the axis ratio $b/a$ decreases ($a$ and $b$ are the semimajor and semiminor axis, respectively).
But for galaxies with $b/a$$>$0.5, extinction related to the inclination cannot
be distinguished from intrinsic extinction. Hence we limit our sample to $b/a$$>$0.5
to avoid any bias.

\end{enumerate}

Therefore, we selected a bulgeless ($fracDeV_r$$=$0) and low inclination ($b/a$$>$0.5)
volume-limited (0.024$<$$z$$<$0.04, $M_r$$\leq$$-18.8$) sample from SDSS
DR7 to investigate the environment and stellar populations of LSB galaxies.
The number of galaxies selected from the different criteria are list in Table 1.

\begin{figure}
\centering
\includegraphics [width=7.5cm, height=6.5cm] {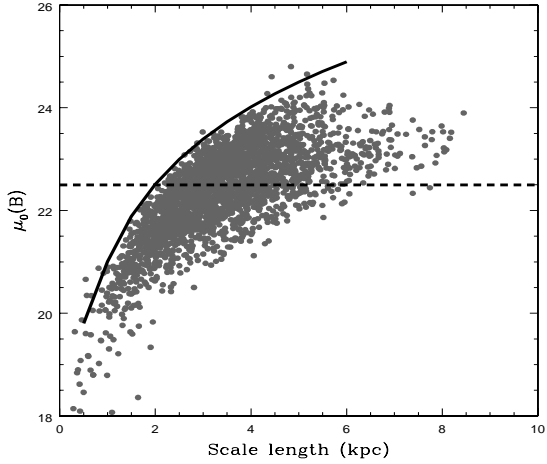}
\caption {The relation between scale length and surface brightness for sample $S_1$.
The dashed line classifies the sample into LSB galaxies and HSB galaxies, and the solid line 
corresponds to the luminosity cut.} \label{fig.kpc}
\end{figure}

\subsection{Low surface brightness sample}

The next step, after the selection of a volume-limited sample, is to calculate the surface
brightness and classify the whole sample into LSB galaxies and HSB galaxies.
For pure disk galaxies, the profile shape is assumed to be exponential:

\begin{equation}
\Sigma(r)=\Sigma_0exp(-r/{\alpha})
\end{equation}
where $\Sigma_0$ is the central surface brightness of
the disk in units of M$_{\odot}$pc$^{-2}$ and $\alpha$ is the scale
length of the disk. In logarithmic units,
the exponential profile is

\begin{equation}
\mu(r)=\mu_0+1.086(r/{\alpha})
\end{equation}
where $\mu_0$ is the central surface brightness in
mag~arcsec$^{-2}$ \citep{1995AJ....110..573M}. The total flux
is

\begin{equation}
F_{tot}=2{\pi}{\alpha^2}\Sigma_0
\end{equation}
if the disk is assumed to be infinitely thin.
By including the correction of extinction and cosmological dimming effect \citep{1990MNRAS.242..235P}, we
get the general formula to calculate central surface brightness:

\begin{equation}
\mu_0(m)=m_{disk}+2.5log(2{\pi}{\alpha}^2)+\mathcal{E}(b/a)-10log(1+z)
\end{equation}
where $m_{disk}$ refers to the apparent magnitude of the disk,
$\mathcal{E}(b/a)$ denotes extinction related to the inclination and $z$ is the redshift.
Our sample consists of bulgeless galaxies, so the total apparent magnitude of galaxies
is denoted as $m_{disk}$. For galaxies with $b/a$$>$0.5, extinction related to the inclination  cannot
be distinguished from the intrinsic extinction \citep{2008ApJ...687..976U}.
So, the extinction correction $\mathcal{E}(b/a)$ is neglected in our sample
when calculating the central surface brightness.
The B-band surface brightness is calculated from magnitudes in $g$ and $r$ bands
of the SDSS system by using the conversion provided by
\citet{2002AJ....123.2121S}.

The widely used definition of LSB galaxies is to select them as 1$\sigma$ fainter than the
median value of the Freeman law (21.65$\pm$0.3 mag arcsec$^{-2}$). However, this is not
an ideal definition, because Freeman's sample was not
complete. \citet{1995AJ....110..573M} (see their Fig. 3) and
\citet{2000ApJ...529..811O} (see their Fig. 2) estimated the distribution
of surface brightness in more complete samples. For galaxies with
$\mu_0(B)$$<$21.65 mag arcsec$^{-2}$, their space densities become lower when $\mu_0(B)$
decreases. For galaxies with $\mu_0(B)$$>$21.65 mag arcsec$^{-2}$, their space densities
are almost constant up to $\mu_0(B)$=25 mag arcsec$^{-2}$.
In this work, using the same definition as \citet{1995MNRAS.274..235D}, 
\citet{1999MNRAS.304..297M} and \citet{2004A&A...422L...5R}, 
we define LSB galaxies as
$\mu_0(B)$$>$22.5 mag~arcsec$^{-2}$ and get 1235 LSB galaxies (sample $S_{LSB}$, see Table~\ref{sample}).
This definition is 3$\sigma$ fainter than the median
value of the Freeman law (21.65 mag arcsec$^{-2}$). For the rest, i.e. 1371 galaxies with $\mu_0(B)$$<$22.5 mag arcsec$^{-2}$,
we define them as HSB galaxies (sample $S_{HSB}$). But we should note that the HSB galaxies defined that way are not all
high surface brightness galaxies, since some LSB galaxies could be part of them if one follows
the classical definition of 1$\sigma$ below 21.65 mag arcsec$^{-2}$. However, this
does not affect our results when we analyze the dependence between surface brightness and environment
by dividing these bulgeless galaxies into two parts.

\subsection{Selection effects}

Since we select bulgeless LSB galaxies in a volume-limited sample, the possible selection
effects of our sample need to be checked. We use $fracDeV_r$=0 to define bulgeless LSB galaxies, so some
bulgeless LSB galaxies with $fracDeV_r$ $>$ 0 might still be excluded from our sample.
Unfortunately, we have not found a complete sample of LSB galaxies with
the same selection criteria as our sample with which we can compare. \citet{Disseau et al.2015} selected a representative
sample of 150 galaxies from a volume-limited sample with 0.022$<$z$<$0.033 and
$M_r$ $\leq$ -18. Although this sample was not selected in exactly the same way,
we use it to roughly estimate the selection bias.
In order to remove the extinction bias, they selected 68 low inclination galaxies with
$b/a$ $>$ 0.5. Using bulge-disk decomposition, they calculate the surface brightness and finally
obtained 11 low inclination LSB galaxies by defining $\mu_0(r)$ $>$ 21.7 mag arcsec$^{-2}$.
The fraction of LSB galaxies in their volume-limited sample is
$\sim$16.2$\pm$4.9\%. In our work, we obtain 1235 bulgeless LSB galaxies in the
volume-limited sample of 31674 galaxies (which contains 13731 low inclination galaxies).
The fraction of bulgeless LSB galaxies is $\sim$9.0$\pm$0.2\%.
Compared with these two samples, the fraction of LSB galaxies only differs by $\sim$1.5$\sigma$, 
indicating that perhaps we may have lost $\sim$40\% of bulgeless LSB galaxies by using 
the $fracDeV_r$=0 criterion. 

Figure~\ref{fig.histz} shows the histogram distribution of redshifts for
LSB and HSB galaxies in sample $S_1$ and their fraction
of the whole sample as a function of redshift. In the left panel, one can see that the number in our
sample grows as z$^2$. This is in line with volume geometry at extremely low redshift. In the right panel,
one can see that the fraction of LSB and HSB galaxies in each bin
is almost identical, although only bulgeless galaxies are selected here. These two figures suggest that the
selection of bulgeless galaxies does not introduce any significant bias into the distribution of LSB and HSB galaxies.
Figure~\ref{fig.kpc} shows the relation between scale length and surface brightness for sample $S_1$.
The dashed line classifies the sample into LSB
galaxies and HSB galaxies. Most galaxies with surface brightness larger than 24 mag arcsec$^{-2}$ are not
included in this sample. This is probably due to the effective isophotal limit of
the SDSS. Very diffuse galaxies below the detection limit cannot be identified. 
In Figure~\ref{fig.kpc}, we also show a line corresponding to the luminosity cut for a mean profile. 

\section{The environments of LSB and HSB galaxies}

The local density parameter $\Sigma_5$ \citep{2004A&A...422L...5R, 2008ApJ...675.1025S,
2010MNRAS.409..936P, 2011ApJ...728...74G} is used to estimate the density of the local environment.
To obtain this parameter, we need to calculate the projected
distance from an individual sample galaxy to the 5th nearest neighbor galaxy. During this step,
the galaxies in volume-limited main-galaxy sample of SDSS DR7 (sample $S_0$), of which the K-corrected absolute
magnitude are brighter than $-18.8$ (i.e. $M_r$ $\leq$ $-18.8$), are used as the tracers
to count the Nth nearest neighbor. The selection of tracers is consistent with that of
LSB and HSB sample galaxies in terms of limiting the absolute magnitude
and will help to eliminate the luminosity bias.
Here we would like to note that, for the tracers,
we do not constrain the $fracDeV$ and $b/a$. This means that
not only the disk galaxies but also the galaxies with a bulge or elliptical
galaxies are also included in this tracer-sample. The information about the environment
of one certain galaxy should include all types of neighboring galaxies, as all of them will
influence the formation and evolution of the target-galaxies.

After the tracers are constructed, we are able to calculate the projected distance
to the Nth nearest neighbor galaxies and then compute the local density parameter $\Sigma_5$. We define
the Nth nearest neighbor within a velocity shell of $\pm$ 500km s$^{-1}$ and count the nearest
neighbor from the first one to the fifth one. Then we get the projected distance of
the fifth nearest neighbor $d_5$ in the unit of Mpc, and define the local density by:

\begin{equation}
\Sigma_5 = 5/({\pi}{d_5}^2 )
\end{equation}
in the unit of Mpc$^{-2}$.

\begin{figure*}
\centering
\includegraphics [width=7.5cm, height=6.5cm] {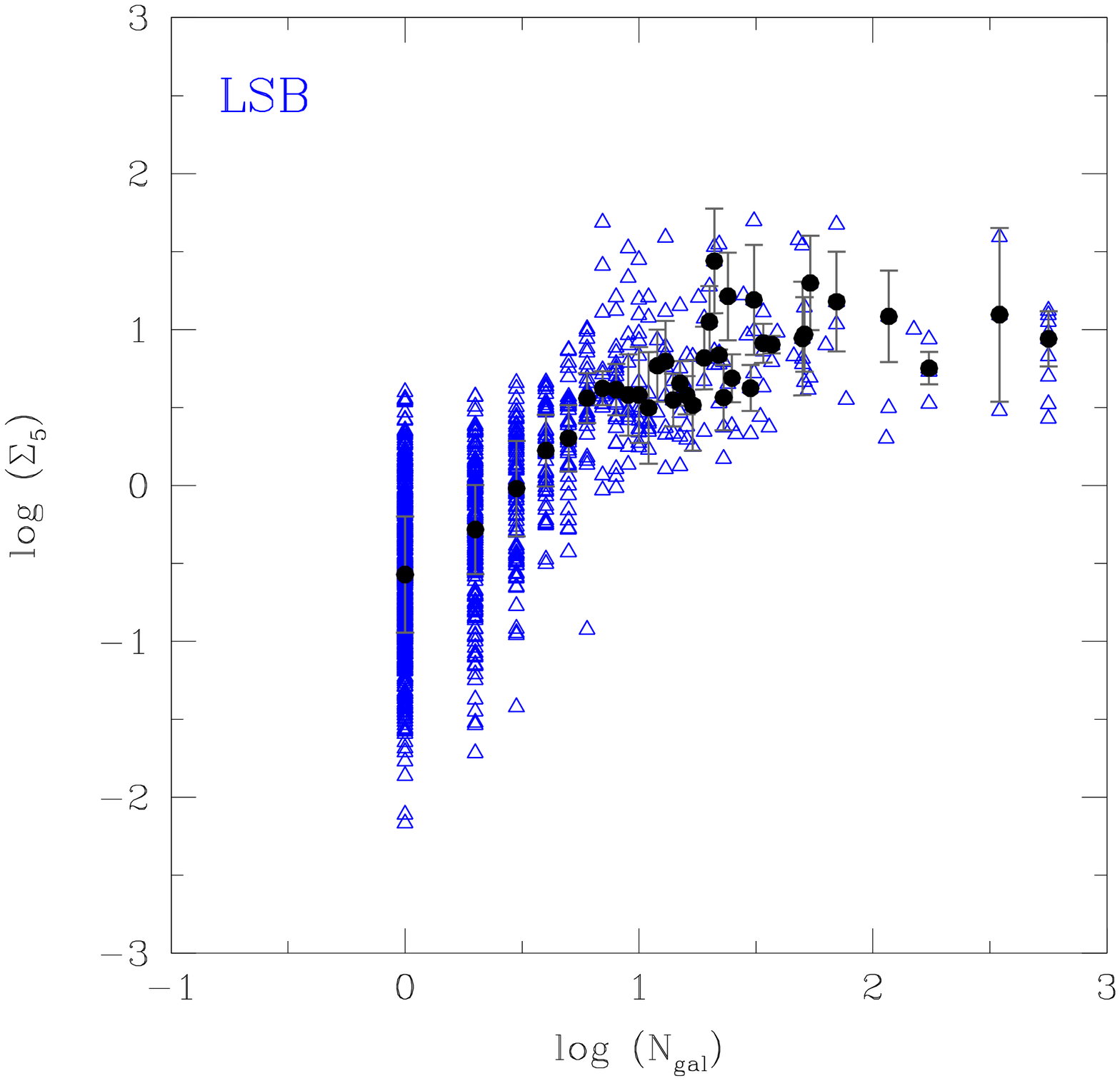}
\includegraphics [width=7.5cm, height=6.5cm] {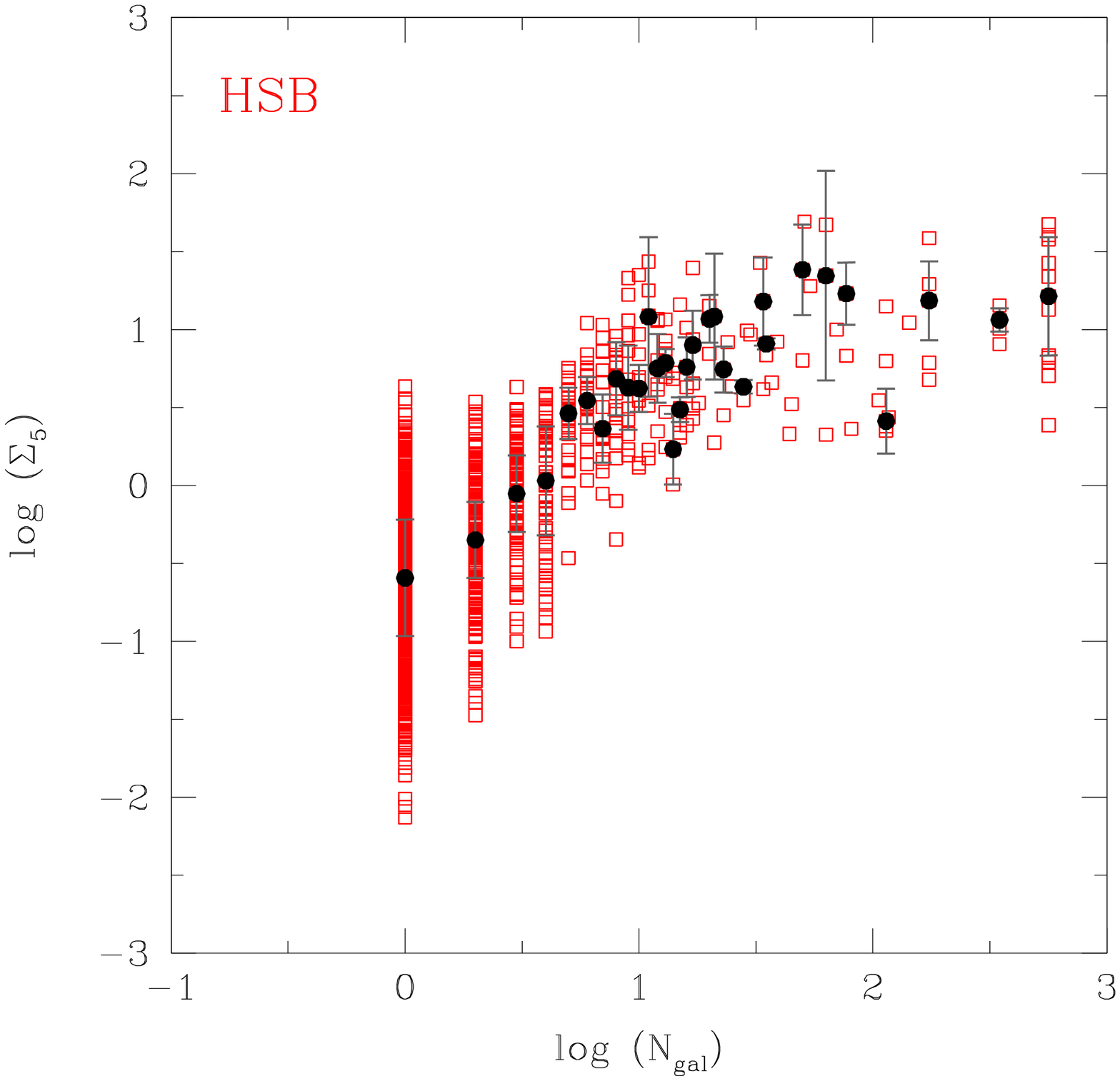}
\caption {The relation between log($N_{gal}$) and log($\Sigma_5$) for LSB galaxies
(left panel, blue triangles)and HSB galaxies (right panel, red squares). The black dots give the median of
log($\Sigma_5$) for each log($N_{gal}$) and the error bars show the dispersions.} \label{fig.ngal}
\end{figure*}

Whether the local density parameter $\Sigma_5$ can represent realistic environments of
LSB and HSB galaxies needs to be investigated. We cross-correlate
our volume-limited sample with the volume-limited galaxy group and clusters catalogues
provided by \citet{2014A&A...566A...1T}. They used a modified friends-of-friends
method with a variable linking length (LL) to identify the realistic groups/clusters with SDSS data release 10.
For each galaxy, all neighbors within the LL radius are regarded as belonging to the same system.
Their Vol-lim-18.0 catalogue, of which the $r$-band absolute magnitude limit is $-$18.0 and the
LL radius is 0.38 Mpc, is consistent with
the selection of our samples.
We use $objID$ to make the cross-correlation
between their Vol-lim-18.0 catalogue and our samples, and the parameter $N_{gal}$ is recovered.
$N_{gal}$ denotes the richness (number of members) of
groups/clusters to which the individual galaxy belongs and
is a useful parameter to reflect the realistic environment where LSB galaxies are located.
Figure~\ref{fig.ngal} shows the relation between log($N_{gal}$) and log($\Sigma_5$) for LSB galaxies
(blue triangles) and HSB galaxies (red squares). The black dots give the median of
log($\Sigma_5$) for each log($N_{gal}$) and the error bars show the dispersions.
We can see that log($\Sigma_5$) increases when log($N_{gal}$) becomes large. This
relation is very clear when log($\Sigma_5$)$<$1, suggesting the local
density parameter $\Sigma_5$ is consistent with realistic environment parameter $N_{gal}$
and can reflect the density of the environment in which LSB and HSB galaxies are located.
For the region with log($\Sigma_5$)$>$1, $d_5$ becomes less than 0.4 Mpc, which is
comparable to or less than the LL radius (0.38 Mpc), and the relation between log($\Sigma_5$) and
log($N_{gal}$) becomes less obvious.

\begin{figure}
\centering
\includegraphics [width=6.5cm, height=6.5cm] {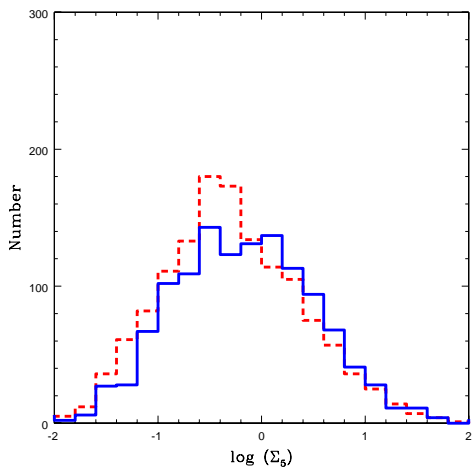}
\caption {The histogram distributions of log($\Sigma_5$) for LSB
galaxies (blue solid line) and HSB galaxies (red dashed line).} \label{fig.sigma5}
\end{figure}

In Figure~\ref{fig.sigma5}, we show the histogram distribution of log($\Sigma_5$) for LSB
galaxies (blue solid line) and HSB galaxies (red dashed line). We can see that the
distributions of log($\Sigma_5$) are nearly the same, suggesting the
surface brightness of bulgeless galaxies does not depend on the environment. 

\section{The Stellar Populations}

In this section, we study the stellar populations of LSB galaxies and see
which factors, environment or dynamics, drive their evolution.

\subsection{Spectral synthesis analysis}

\begin{figure*}
\centering
\includegraphics [width=5.5cm, height=5.5cm] {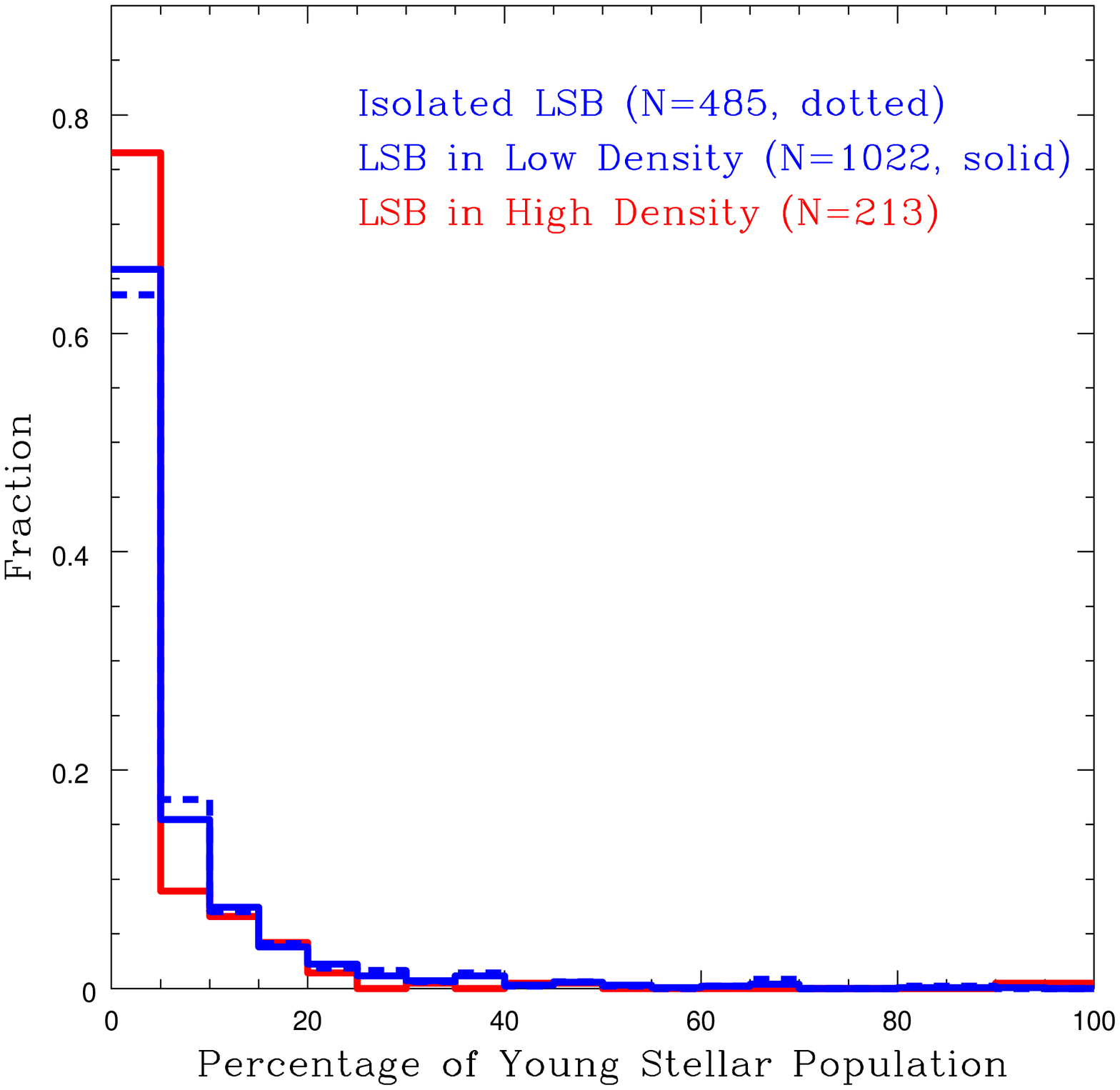}
\includegraphics [width=5.5cm, height=5.5cm] {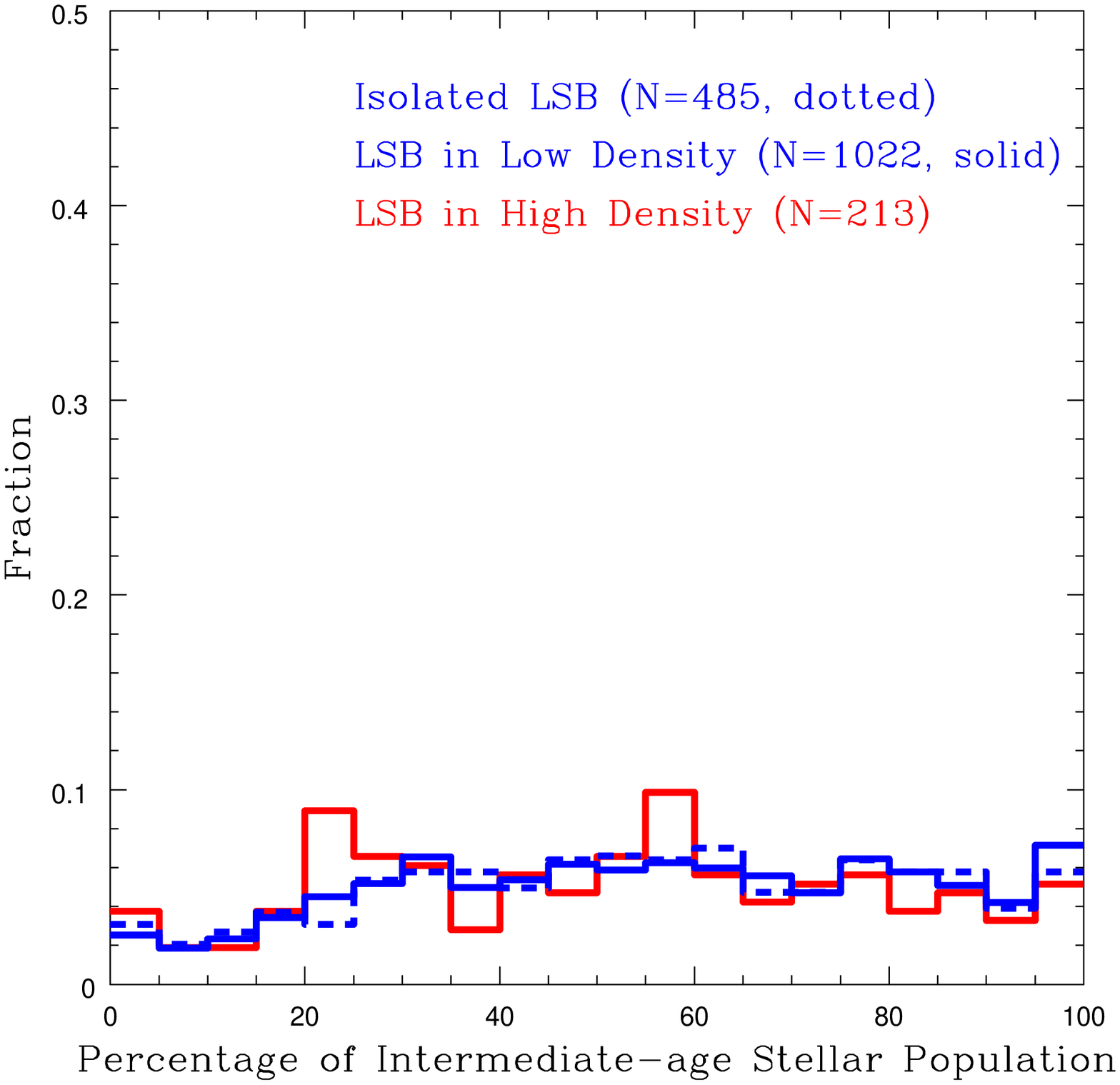}
\includegraphics [width=5.5cm, height=5.5cm] {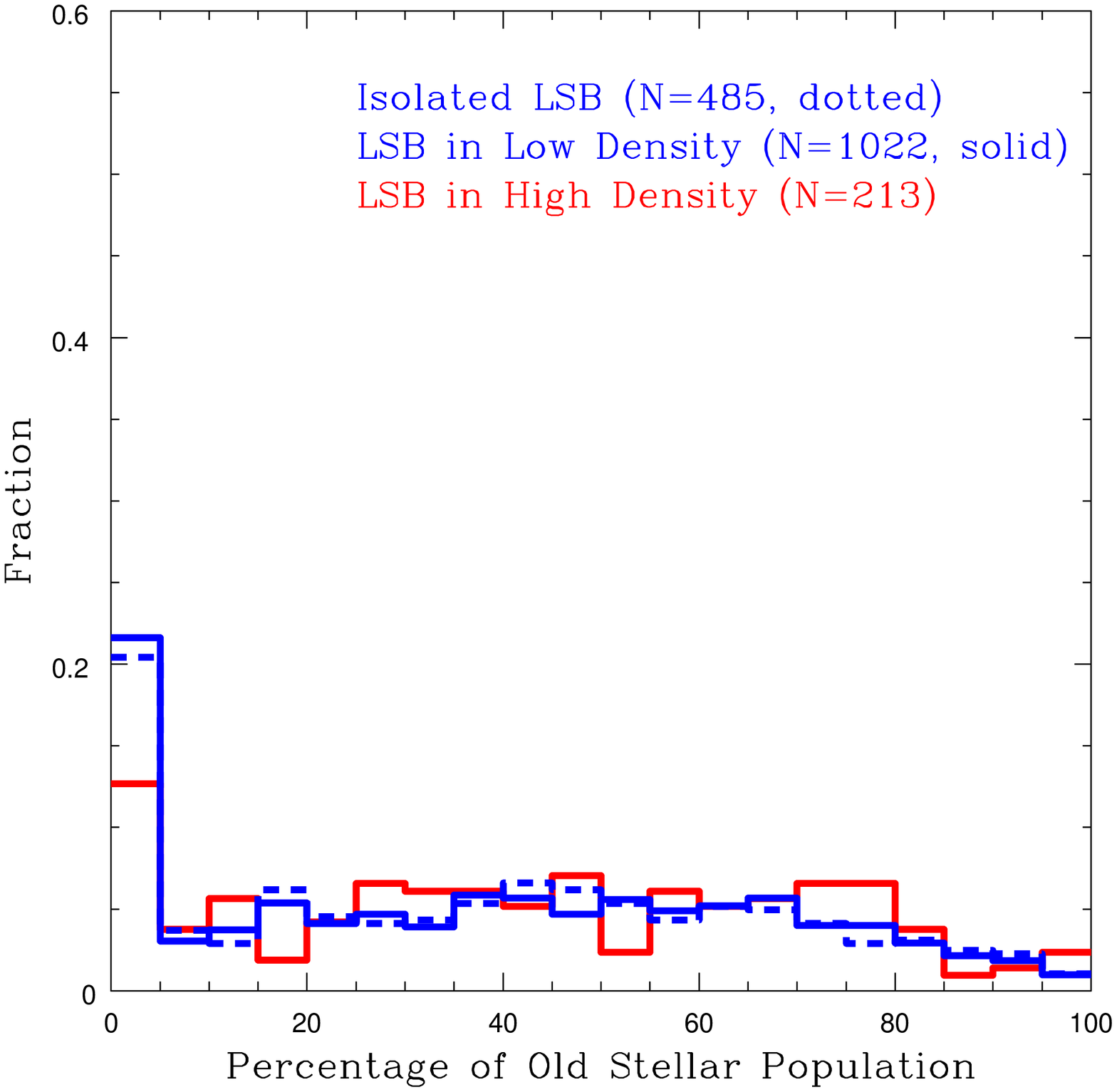}
\caption {The histogram distributions of young, intermediate-age and old stellar populations for
LSB galaxies located in a low density region (log($\Sigma_5$)$<$0.5) and
a high density region (log($\Sigma_5$)$>$0.5). Isolated LSB galaxies are defined
with log($\Sigma_5$)$<$-0.4.} \label{fig.age.environment1}
\end{figure*}

In order to investigate the evolution of LSB galaxies, we
would like to analyze their stellar populations by fitting the full optical spectra
using spectral synthesis. Spectral synthesis provides an efficient way to retrieve information
about stellar populations of galaxies from observed spectra, which is very important for
understanding the formation and evolution of galaxies. The information on both age and metallicity
distributions of stars contained in galaxy spectra can reflect the star formation
and chemical histories of LSB galaxies. Here we analyze the stellar populations of LSB galaxies to
see which factor, environment or dynamics, affects their evolution.

We fit the absorption lines and continua of the spectrum of every individual galaxy using the software
STARLIGHT \citep{2005MNRAS.356..270C, 2007MNRAS.375L..16C}.
This software fits an observed spectrum $O_{\lambda}$ with a model $M_{\lambda}$ that adds
up $N_*$ Simple Stellar Populations (SSPs) with different ages and metallicities taken
from \citet{2003MNRAS.344.1000B} (BC03). In our analysis, we take 24 SSPs, including 12 different ages from
40 Myr to 13 Gyr (40, 280, 900 Myr and 1.27, 1.43, 2.5, 4.25, 5, 6.25, 7.5, 10, 13 Gyr) and
two metallicities (0.2 and 1.0 $Z_{\odot}$), the stellar evolutionary tracks of Padova 1994
\citep{1993A&AS...97..851A, 1996A&AS..117..113G}, the initial mass function (IMF) from \citet{2003PASP..115..763C}
and the extinction law of \citet{1989ApJ...345..245C} with $R_V$ $=$ 3.1. The Galactic extinctions are
corrected by the reddening map of \citet{1998ApJ...500..525S} and the spectra are shifted to the rest frame. The
range of the spectra is from 3700 to 8000\AA \ with a step of 1\AA \ and normalized to the
median flux in the 4010 to 4060\AA \ region. During spectral synthesis fitting, we exclude the
emission lines and four windows (5870-5905\AA, to avoid the $Na D \lambda\lambda$ 5890, 5896
doublet, which is from the interstellar medium; 6845-6945\AA \ and 7550-7725\AA, which are the strong 
absorption bands from the Earth's atmosphere and were flagged by BC03
as issues in the STELIB library \citep{2003A&A...402..433L};
7165-7210\AA, which shows a systematic broad residual in emission as mentioned
by \citet{2006MNRAS.370..721M}), see also \citet{2009A&A...495..457C, 2010A&A...515A.101C}.

As discussed in \citet{2005MNRAS.356..270C}, the individual components of stellar populations
are fluctuant, but the combined stellar populations are robust enough. We need to combine
the individual stellar population into young population ($<$ 1Gyr),
intermediate-age population (1 $\sim$ 5Gyr) and old population ($\geq$ 5Gyr). When doing
the spectral synthesis, we fit every spectrum 30 times with different $random$ $seed$ to get a robust result and
estimate the associated uncertainty. For every individual galaxy, we take the median value of these
30 fitting results as a robust result and the standard deviation as the uncertainty.
The typical value of the relative uncertainty of mass fraction in our sample is 13\%, 21\% and 30\% for young,
intermediate-age and old stellar populations, respectively.

\begin{figure}
\centering
\includegraphics [width=6.5cm, height=6.5cm] {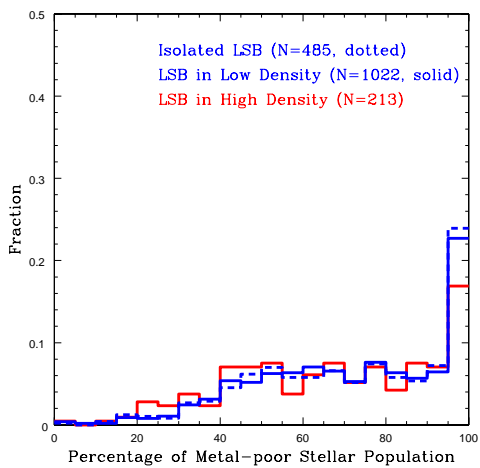}
\caption {The histogram distributions of metal-poor stellar populations for
LSB galaxies located in a low density region (log($\Sigma_5$)$<$0.5) and
a high density region (log($\Sigma_5$)$>$0.5). Isolated LSB galaxies are defined
with log($\Sigma_5$)$<$-0.4.} \label{fig.metal.environment1}
\end{figure}

\subsection{Stellar population of LSB galaxies in different environments}

Figure~\ref{fig.age.environment1} shows the histogram distributions of young, intermediate-age and old
stellar populations for LSB galaxies located in low (log($\Sigma_5$)$<$0.5 and $<$-0.4, i.e., $d_5$ $>$ 0.7 and 2 Mpc, respectively)
and high (log($\Sigma_5$)$>$0.5) density region. The red solid line denotes
LSB galaxies in a high density region and the blue solid line denotes LSB galaxies in
a low density region while the blue dotted line delineates isolated LSB galaxies.
Table~\ref{sp} shows the median values of different stellar populations.
We can see that, for young, intermediate-age and old stellar populations,
the distributions and the median value are nearly the same
for LSB galaxies in low and high density region. On the other hand, Figure~\ref{fig.metal.environment1}
shows the histogram distributions of metal-poor stellar populations for
LSB galaxies in low and high density regions. We cannot see obvious differences of metallicities
for LSB galaxies in these three different environments. These results suggest that the stellar populations
of LSB galaxies are similar when they are located in different environments.
Environment is not found to play an important role in the evolution of LSB galaxies.

\begin{figure*}
\centering
\includegraphics [width=5.5cm, height=5.5cm] {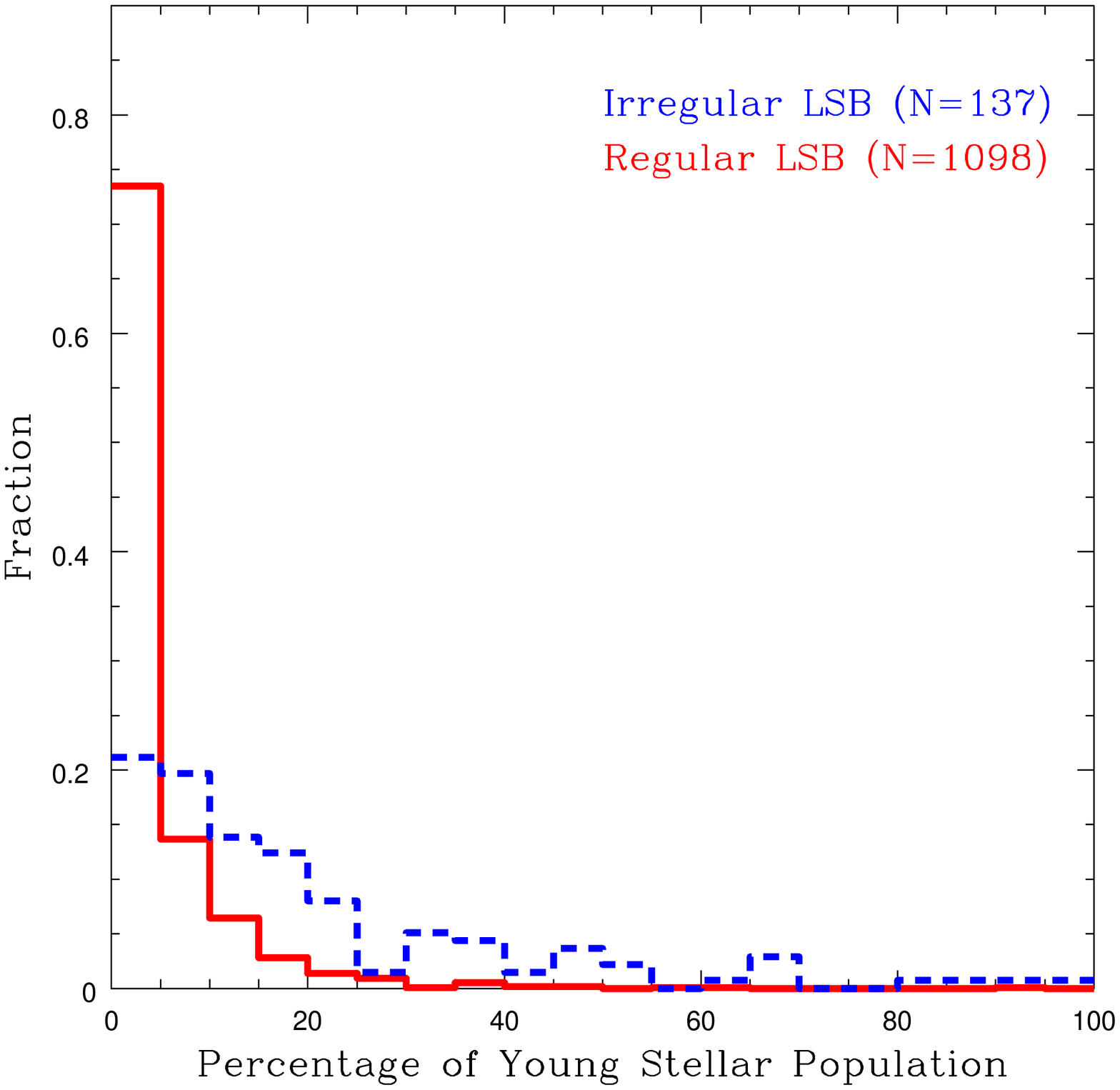}
\includegraphics [width=5.5cm, height=5.5cm] {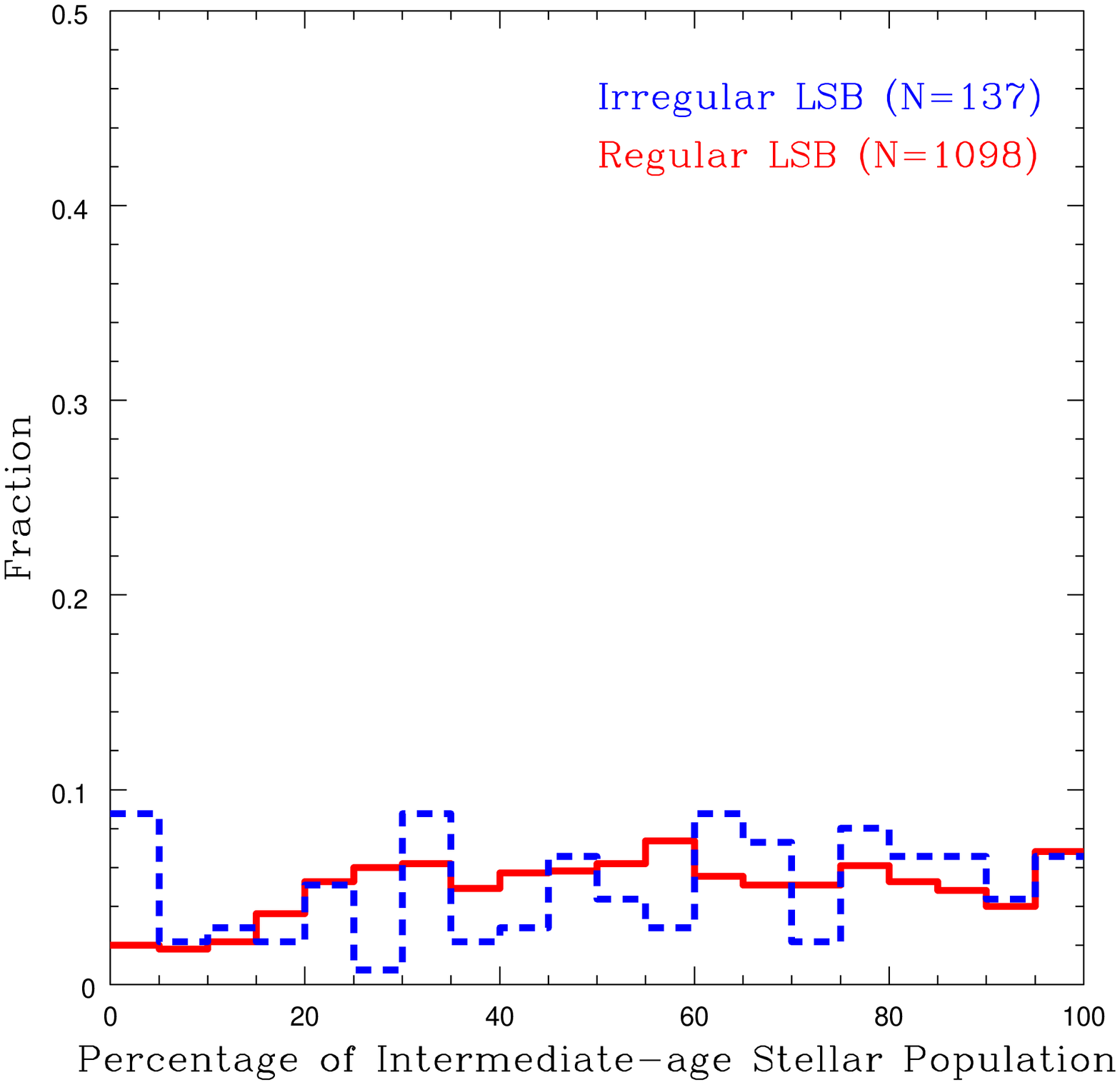}
\includegraphics [width=5.5cm, height=5.5cm] {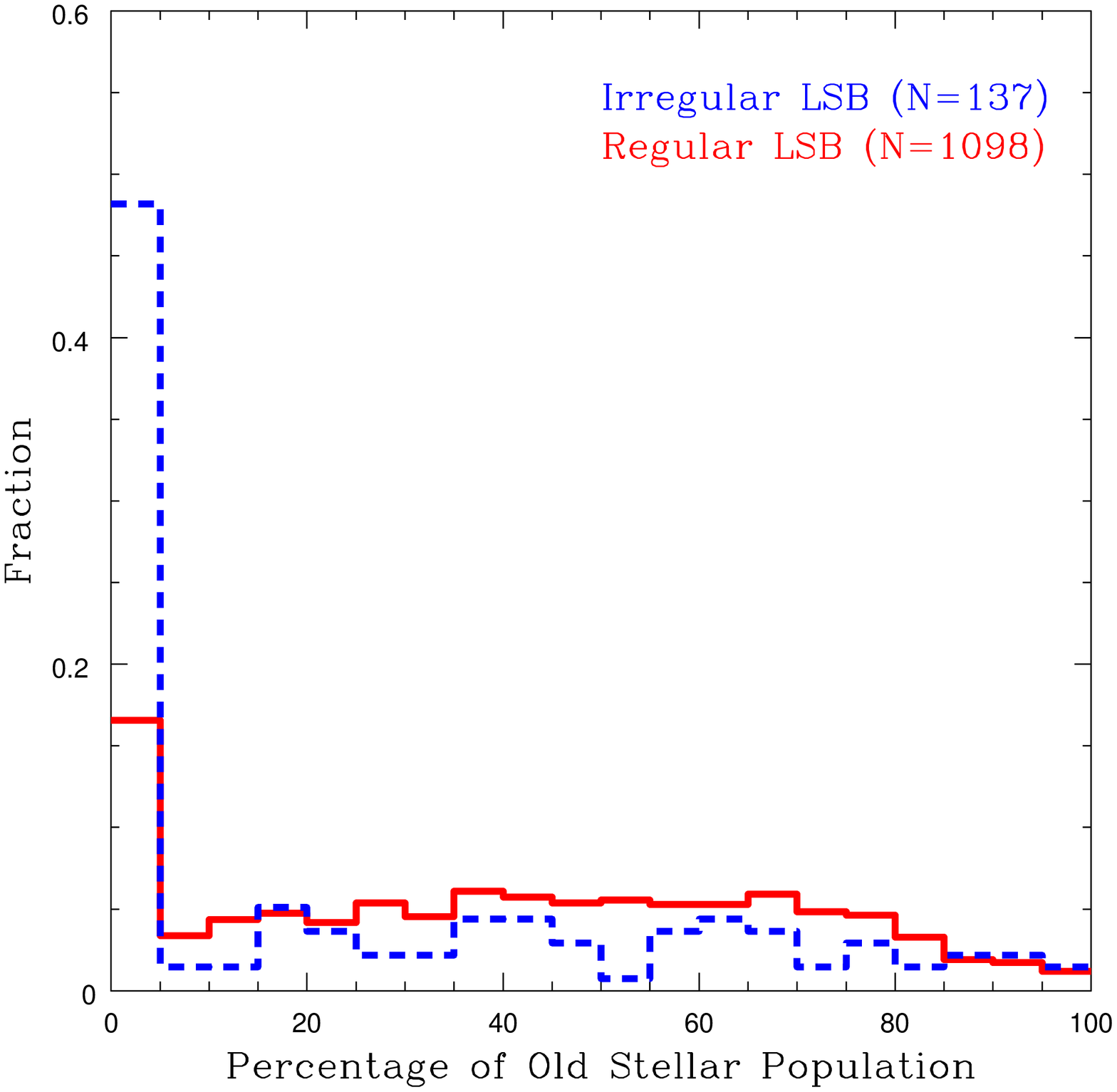}
\caption {The histogram distributions of young, intermediate-age and old stellar populations for
regular LSB galaxies and irregular LSB galaxies. Notice that regular LSB galaxies
have almost no young stellar component compared to morphologically
peculiar LSB galaxies that spread over significant fractions.} \label{fig.age.morphology}
\end{figure*}

Whether these results are biased by the definition of low and high density needs to be
checked. With two different definitions, we check the histogram distribution of
stellar populations for LSB galaxies located in low (log($\Sigma_5$)$<$0, log($\Sigma_5$)$<$0.8)
and high (log($\Sigma_5$)$>$0, log($\Sigma_5$)$>$0.8) density regions.
We find that there is still no obvious difference between the
stellar populations of LSB galaxies in different environments, even with different
definition of low/high density.
This suggests that the above results are not biased by the definition of low/high density and
the similarities of stellar populations are real for LSB galaxies
in different environments.

\subsection{Stellar population of LSB galaxies with different morphologies}

LSB galaxies exhibit a wide range of morphologies from dwarf
irregular to luminous disk \citep{1995AJ....109.2019M}. Like LSB galaxies in
different environments, whether the stellar populations of LSB galaxies with different
morphologies are similar or not is an important issue to investigate. To make
this clear, we would like to compare the stellar population of LSB galaxies which have
different morphologies.

Morphological classification of LSB galaxies has been done by
visual inspection of galaxy images from the SDSS. The whole LSB sample is
classified into two groups: regular LSB galaxies and irregular LSB galaxies. Most
LSB galaxies can be classified well, except some very small ones because of the
 spatial resolution. The classification has been done by XS and FH and some
 examples of irregular galaxies are shown at the end of the paper (Figure~\ref{fig.highmass}
 and Figure~\ref{fig.lowmass}). The major difficulty is to distinguish Irr from SBm (very
late type galaxies) and this is certainly the main uncertainty in this
visual classification.

\begin{table}
\begin{center}
\caption{Median value of different stellar populations} \tiny
\label{sp}
\begin{tabular}{c c c c|c}
  \hline \hline
             &     &   Age  &     &  Metallicity   \\
	     &          Young                &      Intermediate   &    Old    &    Poor    \\ \hline
   Isolated LSB                  &  3.3\%  &  55.5\%   &   38.3\%   &   74.9\%    \\
   LSB in Low Density        &  3.2\%  &  55.9\%   &   37.5\%   &   73.9\%    \\ 
   LSB in High Density        &  2.2\%   &  53.6\%   &   42.2\%   &    68.2\%    \\ \hline
   Irregular LSB         &  12.9\%   &  60.1\%    &    11.0\%   &    88.6\%     \\ 
   Regular LSB    &  2.7\%   &  55.1\%    &   40.5\%    &   71.0\%    \\ \hline
\end{tabular}
\end{center}
\end{table}

\begin{figure}
\centering
\includegraphics [width=6.5cm, height=6.5cm] {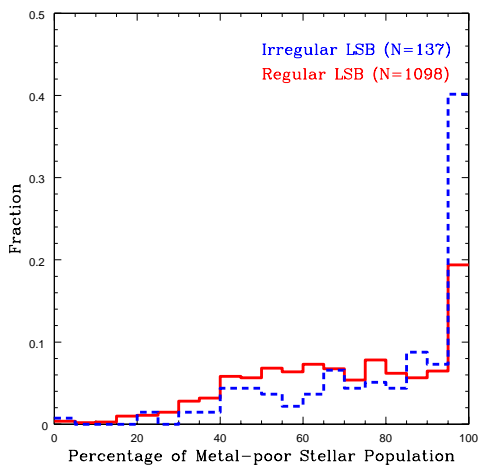}
\caption {The histogram distributions of metal-poor stellar populations for
regular LSB galaxies and irregular LSB galaxies.} \label{fig.metal.morphology}
\end{figure}

Figure~\ref{fig.age.morphology} shows the histogram distributions of young, intermediate-age and old
stellar populations for regular and irregular LSB galaxies. Table~\ref{sp} shows
median values of these distributions. The differences are significant.
Irregular LSB galaxies have more young stellar populations and less old stellar
populations than regular LSB galaxies, suggesting they are younger and have more
young stars than regular LSB galaxies. For the intermediate-age stellar populations,
the differences are not obvious. Figure~\ref{fig.metal.morphology} shows the
histogram distributions of metal-poor stellar populations for regular
and irregular LSB galaxies. Regular LSB
galaxies have less metal-poor stellar populations than irregular LSB
galaxies. These results suggest that irregular LSB galaxies
formed more stars than regular LSB galaxies at recent epochs. The increase of metallicity in
regular LSB galaxies may go faster than in irregular LSB galaxies.

\begin{figure*}
\centering
\includegraphics [width=7.5cm, height=6.5cm] {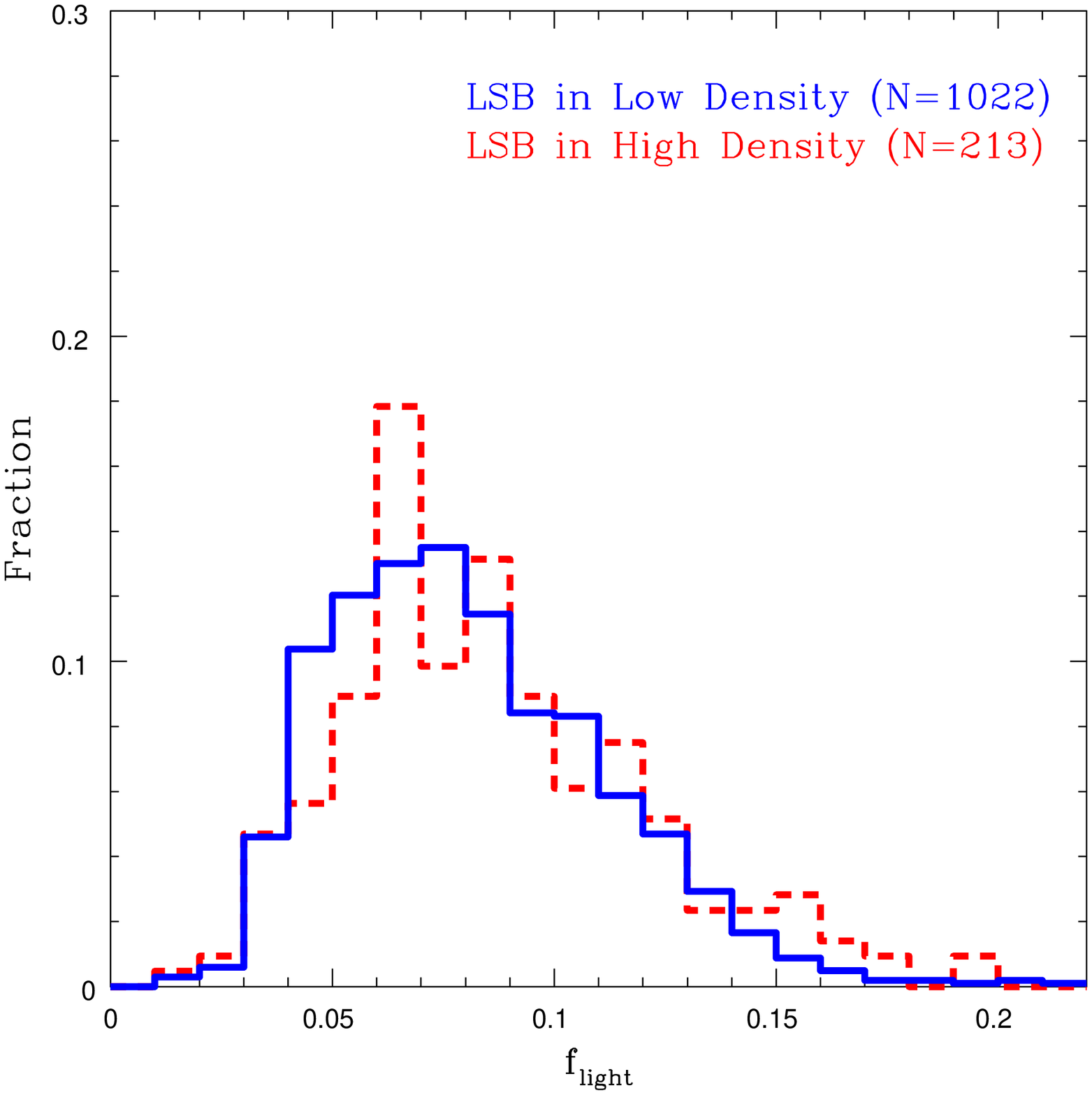}
\includegraphics [width=7.5cm, height=6.5cm] {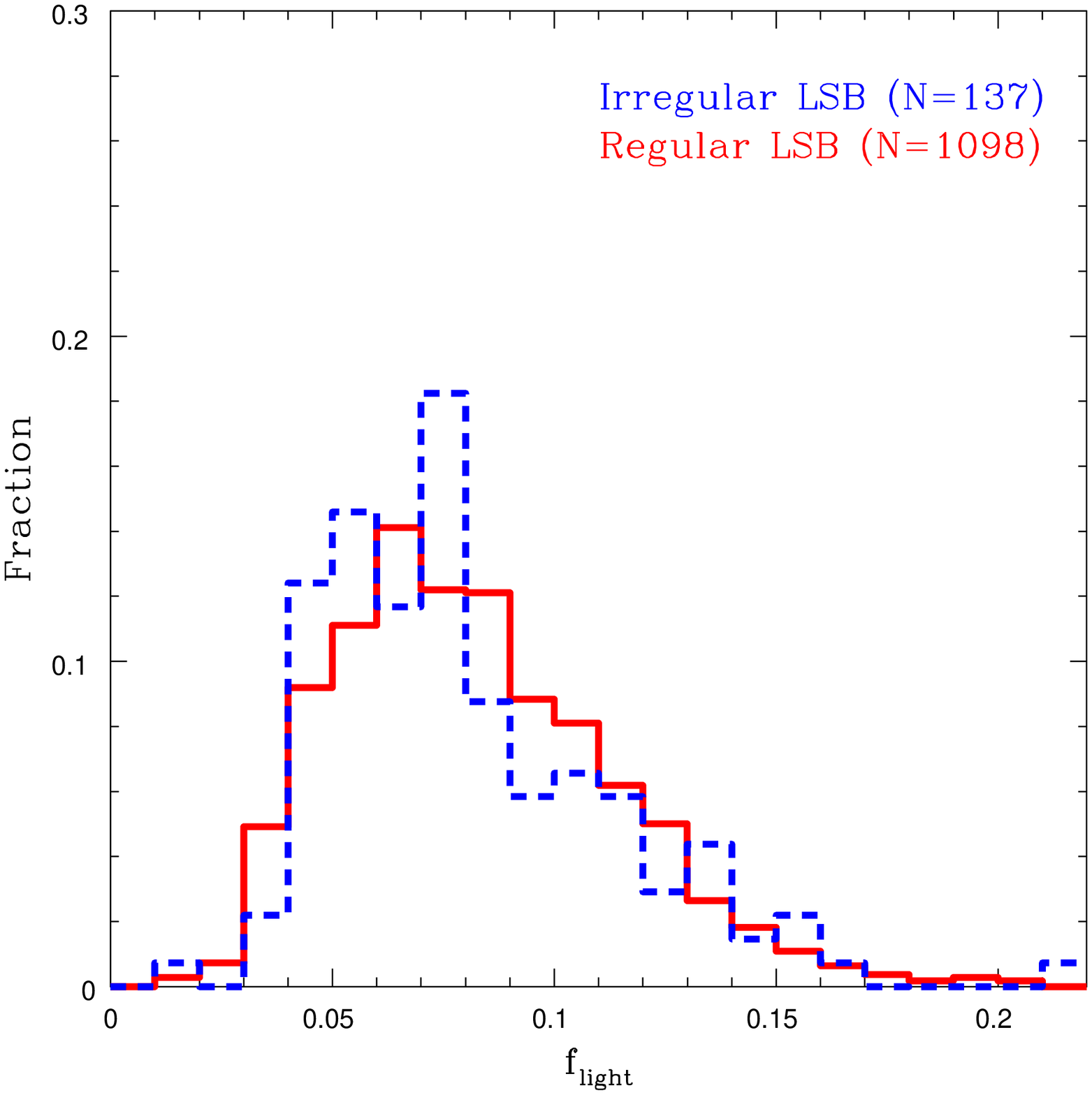}
\caption {The histogram distributions of light fraction for
different LSB galaxies.} \label{fig.lf}
\end{figure*}

\section{Discussion}

\subsection{Differences of environment between LSB and HSB galaxies}

In the above analysis, we found that the surface brightness of bulgeless galaxies does not depend on
the environment. We notice that this appears to be different from previous studies.
\citet{1993AJ....106..530B} and \citet{1994MNRAS.267..129M} found that LSB galaxies are located in
more isolated environments than HSB galaxies. The differences are probably due to the different
selection criteria. They selected not only bulgeless galaxies but also galaxies with bulges, but we only select
bulgeless galaxies. However, only selecting 
bulgeless LSB and HSB galaxies, which are late type galaxies, should not bring significant bias when comparing
their environments. Indeed, massive galaxies are located
in higher density regions and have higher B/T than low mass galaxies \citep{2005A&A...430..115H}. This suggests that
galaxies with high B/T may be located in high density regions. Considering the large fraction 
of HSB galaxies with bulges, which may make them have higher average density than LSB galaxies, 
previous works may have bias when comparing their environment by selecting galaxies with bulges. 

\subsection{Aperture effect in spectral synthesis}

As is well known, SDSS is a fiber-based survey. The spectra of objects are taken by fibers with diameters of 3 arcsec.
When extended sources are analyzed, such as nearby galaxies, there are aperture effects.
\citet{2004ApJ...613..898T} and \citet{2005PASP..117..227K} have discussed the aperture effects of SDSS
spectroscopy in detail. In order to check how much light is covered by
SDSS observation, one simple and accurate way is to compare the fiber magnitude and Petrosian
magnitude of our samples. The fiber magnitude is a measurement of the light falling into the
fiber and the Petrosian magnitude is a good estimate of the total light. Thus, we adopt the
formula below \citep{2003ApJ...599..971H} to estimate how much light is covered by fiber observations:

\begin{equation}
f_{light}=10^{(-0.4(m_{fiber}-m_{petro})_r)}
\end{equation}

The typical value of $f_{light}$ is $\sim$0.08, indicating that the light of the spectra comes
from the inner regions of galaxies. Generally there are age and/or metallicity gradients in
LSB galaxies \citep{1996A&A...313..377D, 2000MNRAS.312..470B}. Inner regions have old and metal-rich
stellar populations, while young and metal-poor stellar populations dominate outer regions.
Therefore, our estimations of the stellar populations represent the properties of the central regions.
The whole disk of LSB galaxies could be younger and
more metal-poor than central regions. On the other hand,
our samples consist of bulgeless galaxies and the spectra sample only the central part of the disks.
So, our results represent the properties of central disks well
without contaminations from bulge light.

In order to check whether the above results are biased by different aperture effects, we show the
histogram distributions of $f_{light}$ for different LSB galaxies in Figure~\ref{fig.lf}. We
can see that, for LSB galaxies in different environments or with different morphologies,
the distributions are nearly the same, suggesting our results are not biased by different
aperture effects and are robust enough.

\begin{figure}
\centering
\includegraphics [width=6.5cm, height=6.5cm] {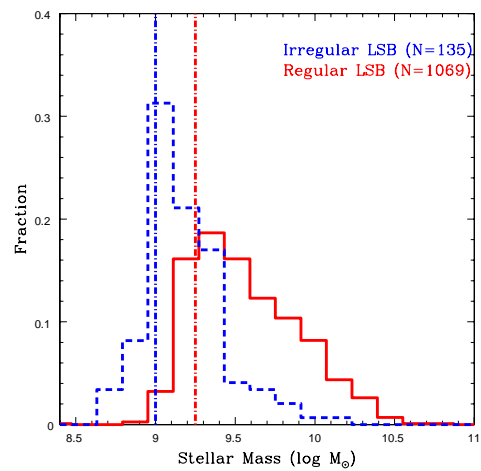}
\caption {The histogram distributions of stellar mass for
regular LSB galaxies and irregular LSB galaxies. The red and blue dot-dashed lines denote the 
mass-completeness limits of regular and irregular LSB galaxies, respectively.} \label{fig.mass.morphology}
\end{figure}

\subsection{The origin of star formation in irregular LSB galaxies}

Table~\ref{sp} shows the median values of different stellar populations for irregular LSB galaxies and
regular LSB galaxies. Irregular LSB galaxies have more young and metal-poor
stellar populations in the center than regular LSB galaxies.
Because of the degeneracy between age and metallicity, it is difficult to
analyze the star formation histories in detail using these results. But for a young stellar
population with poor metallicity, there is no degeneracy.
Generally the central region of disk galaxies has old and metal-rich
stellar populations, and the outer region has young and metal-poor
stellar populations \citep{1996A&A...313..377D, 2000MNRAS.312..470B}. It is interesting that the irregular
LSB galaxies have such a high fraction of young stars ($<$1Gyr) in their central regions. What
kind of mechanisms can make young stars appear in the center of irregular LSB galaxies?

The detection of young stars suggests that the central
regions of irregular LSB galaxies may experience recent star
formation episodes, which result in the so-called clumps. What drives the formation of clumps has
been widely studied. By analyzing the kinematics of clumpy galaxies at z $\sim$ 0.6, \citet{2010MNRAS.406..535P} 
suggests that cold flows are expected to feed z$>$1 clumpy galaxies, while
mergers are probably the dominant driver leading to the formation of
clumpy galaxies at z$<$1. For our sample with redshifts ranging from 0.024 to 0.04,
mergers may be one of the reasons for the appearance of irregularities and central star
formations. 

Figure~\ref{fig.mass.morphology} shows the histogram distributions of stellar masses
for regular and irregular LSB galaxies. The stellar masses are recovered from MPA-JHU
\footnote{http://www.mpa-garching.mpg.de/SDSS/DR7/} and estimated using fits to the photometry.
Irregular, star forming galaxies have lower stellar mass than others. This is probably due to the different mass-to-light ratios (M/L)
of these two subsamples. Such variations have been 
really presented when we found the differences in the fraction of young stars. In order to 
estimate this incompleteness, 
we determine the mean stellar mass of the 
galaxies which are located very near the luminosity cut (between -19 and -18.8) 
and consider that as a mass-completeness limit. In Figure~\ref{fig.mass.morphology}, the 
dot-dashed lines give the mass-completeness limits of irregular and regular LSB galaxies, 
which are 9.0 and 9.25, respectively. The LSB galaxies with stellar mass higher than these 
limits are considered to be complete in stellar mass. We focus in the following on 
irregular LSB galaxies with log$M_{\odot}$$>$9.5 (to compare to the regular ones) 
as well as the irregular ones with log$M_{\odot}$$<$9.0. 

\begin{figure*}
\centering
\includegraphics [width=3cm, height=3cm] {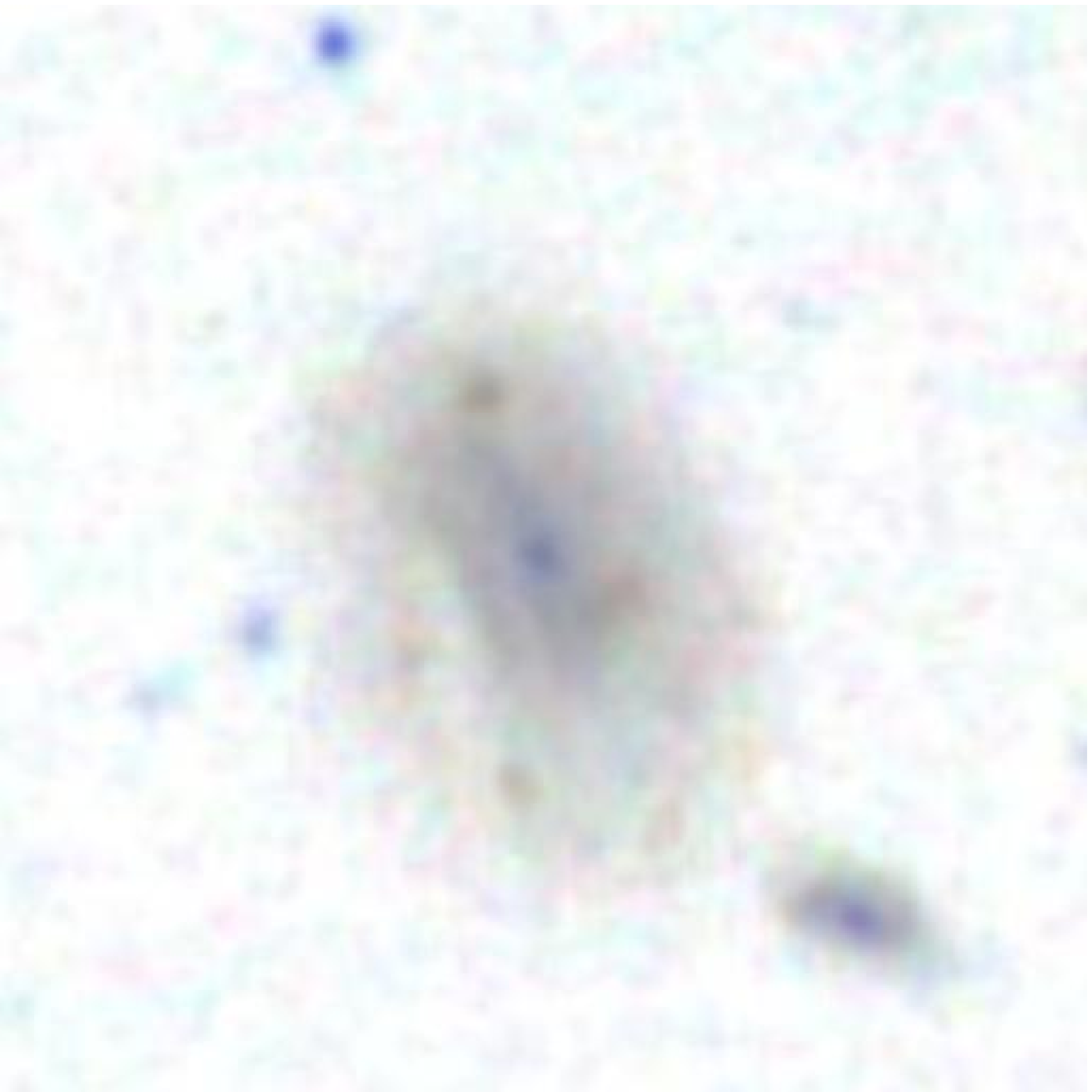}
\includegraphics [width=3cm, height=3cm] {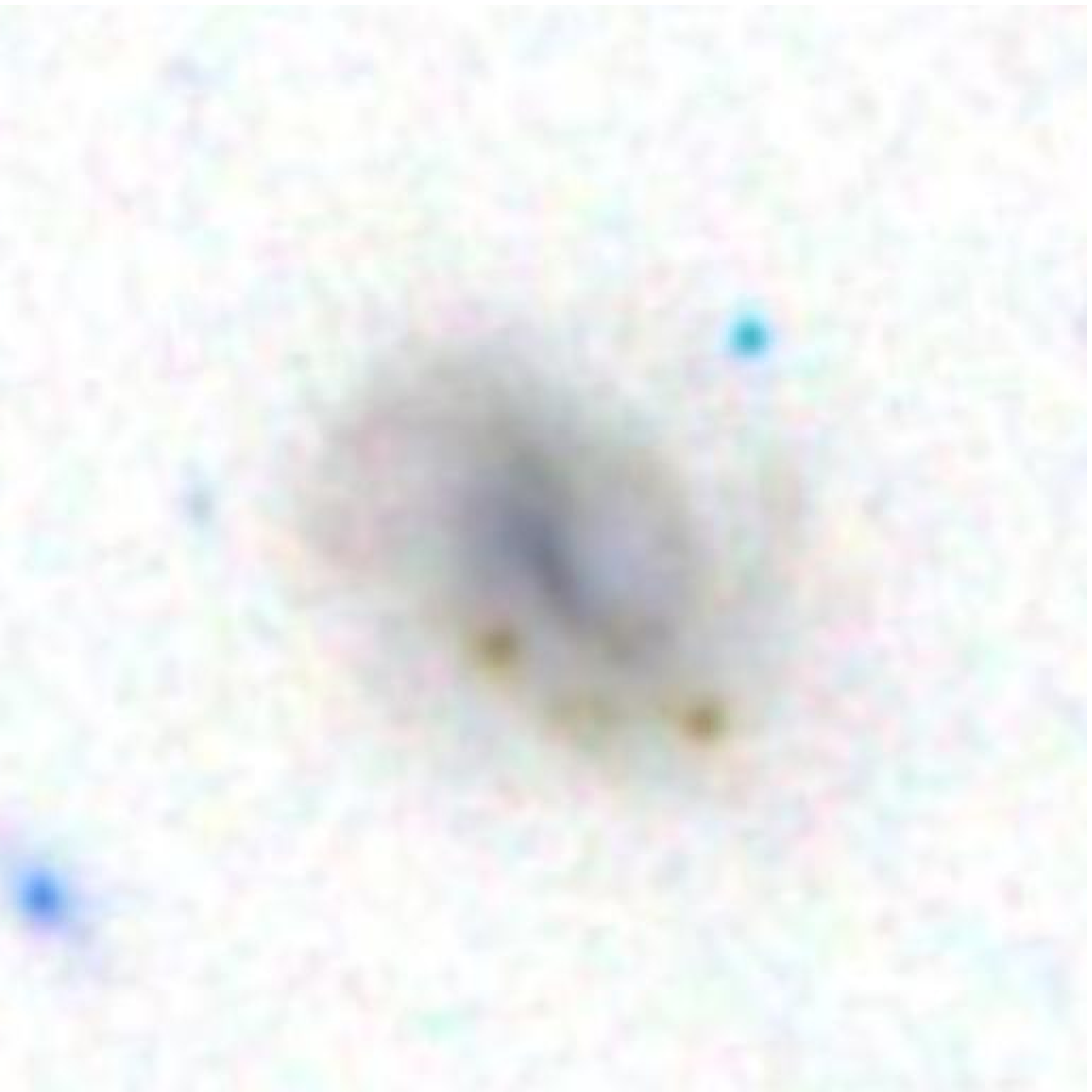}
\includegraphics [width=3cm, height=3cm] {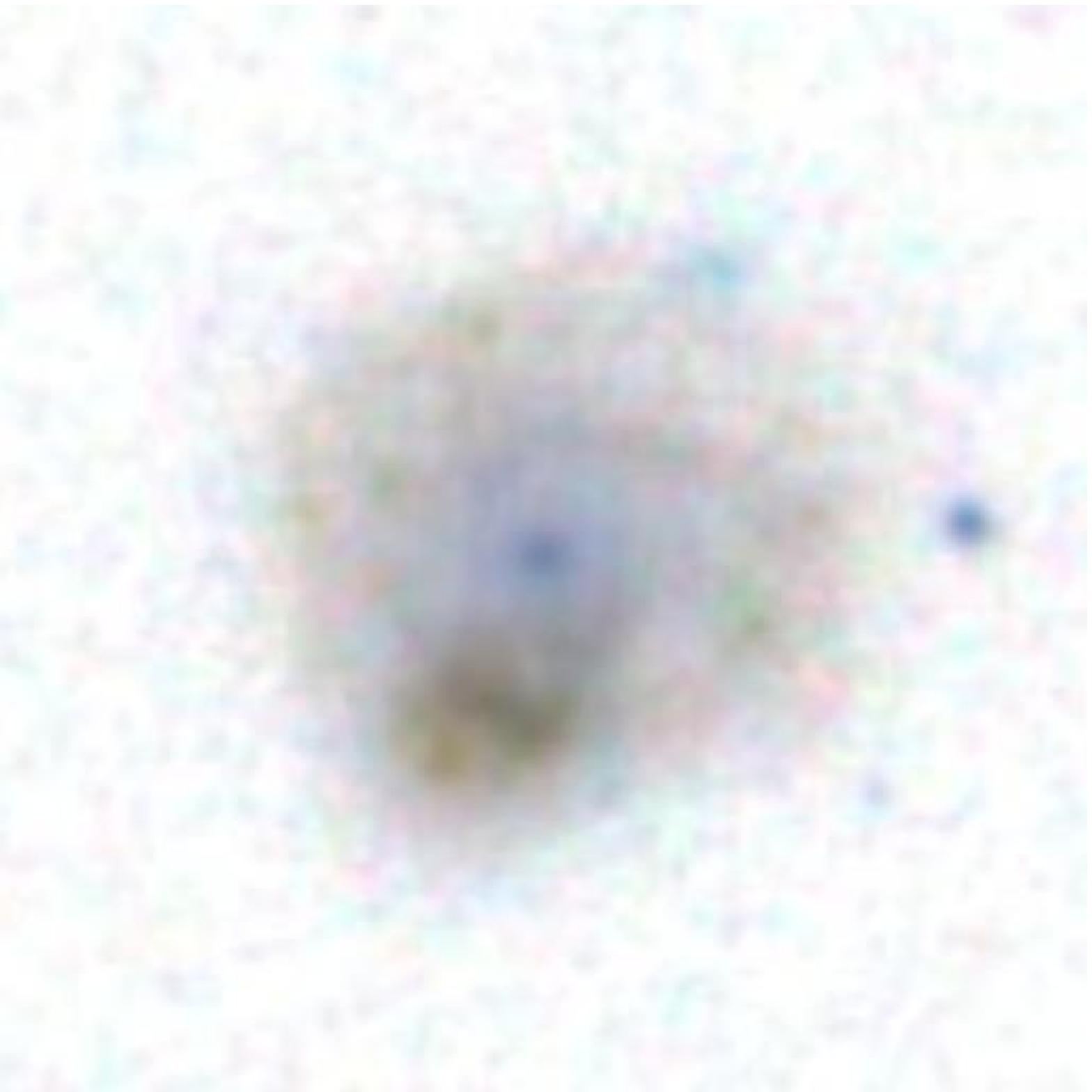}
\includegraphics [width=3cm, height=3cm] {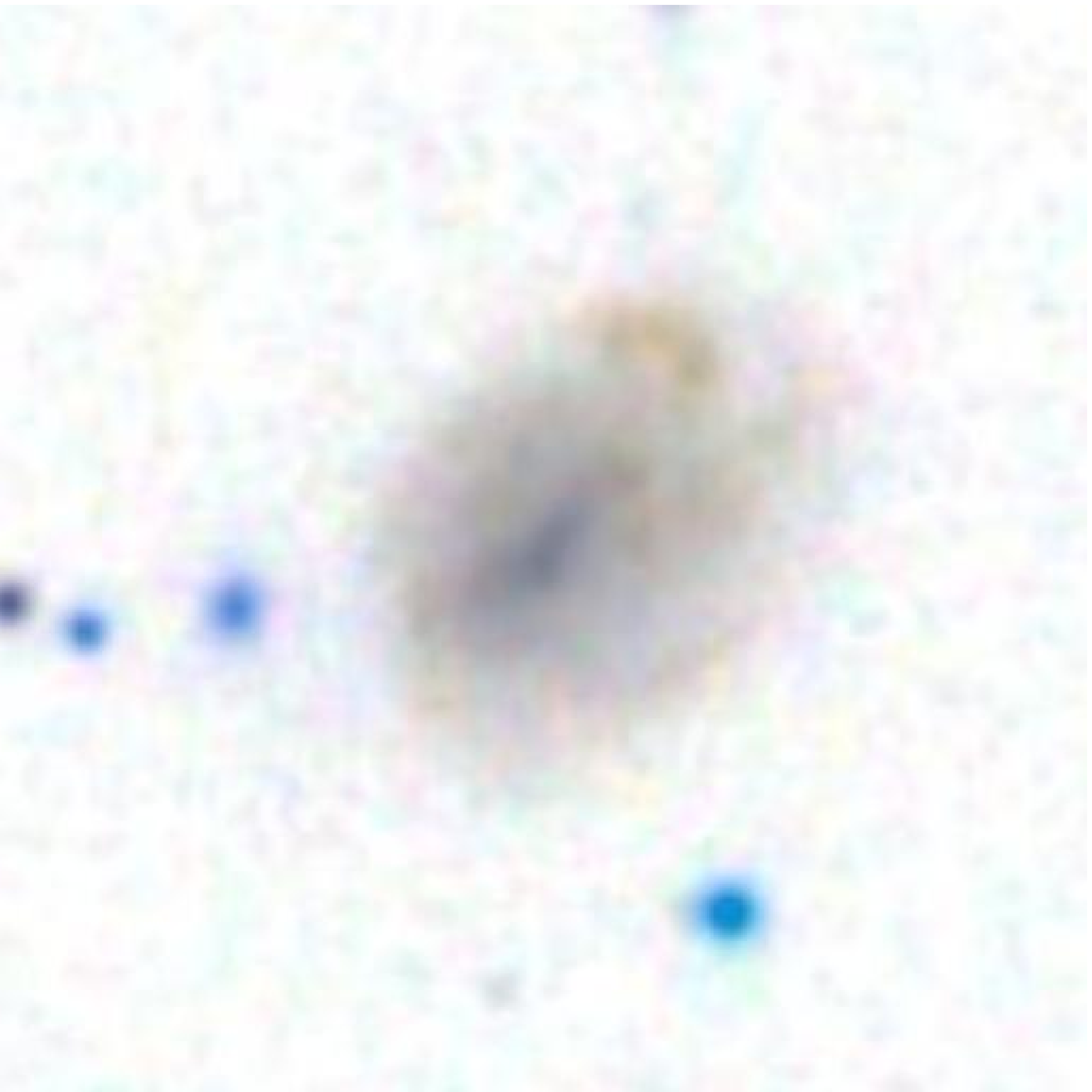}
\includegraphics [width=3cm, height=3cm] {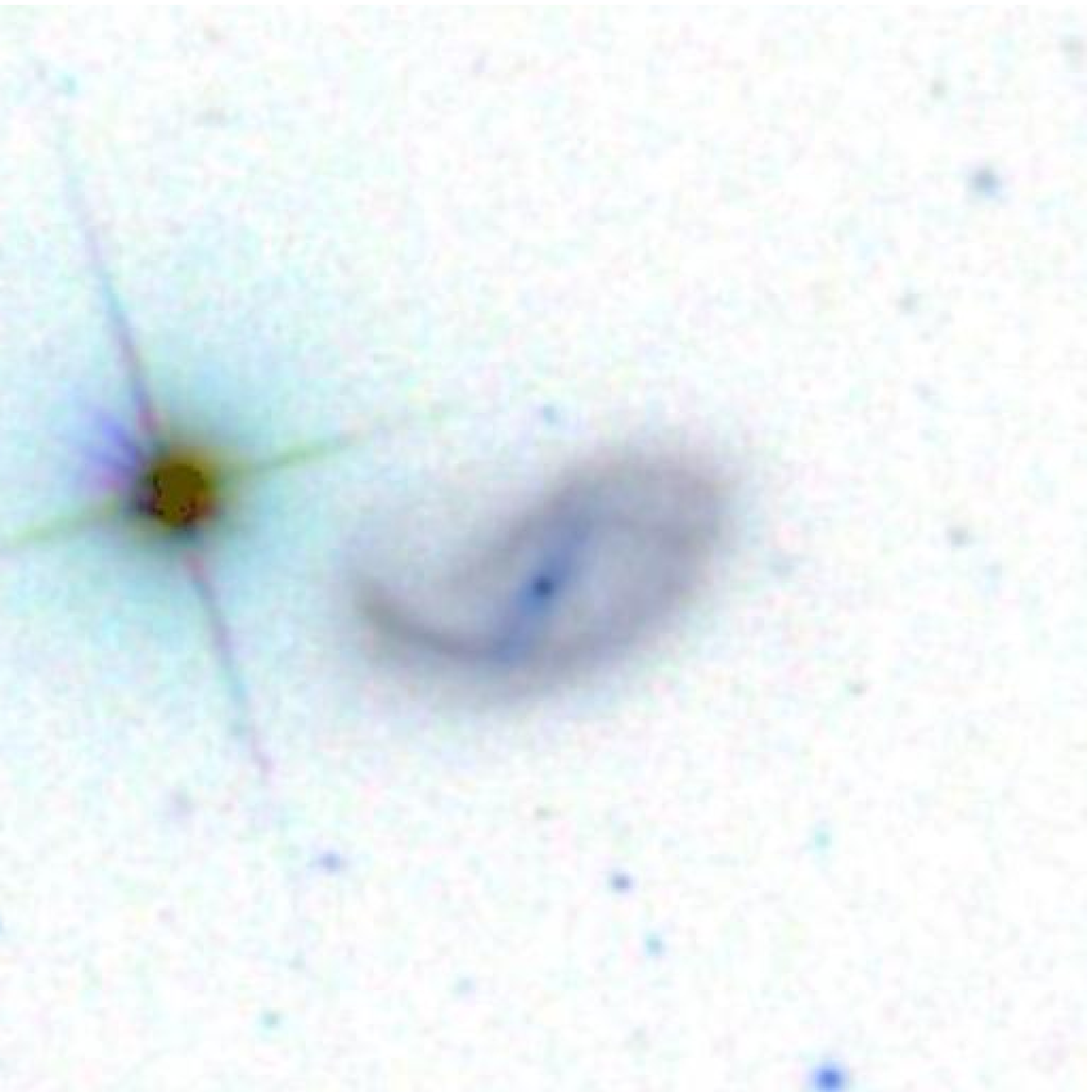}
\includegraphics [width=3cm, height=3cm] {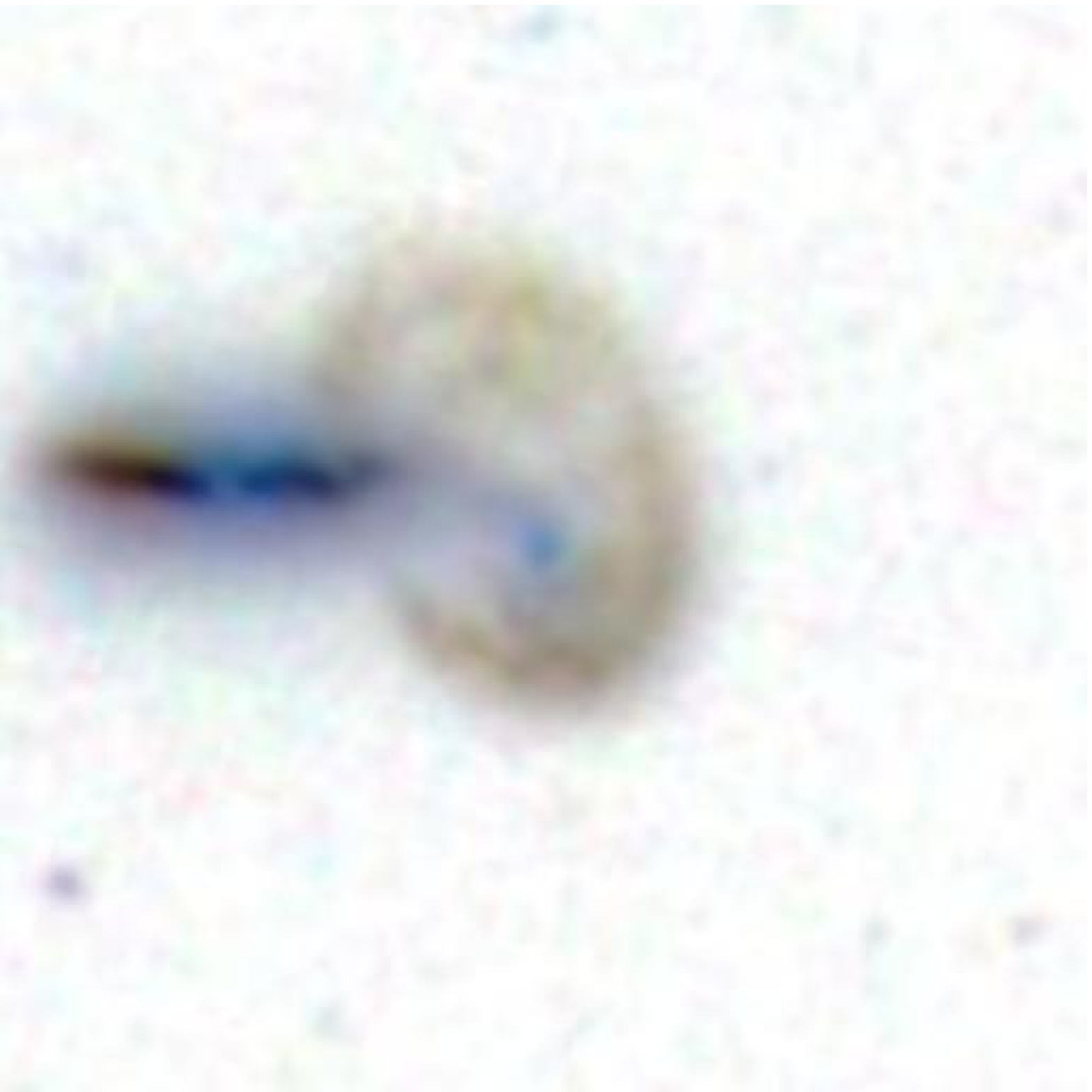}
\includegraphics [width=3cm, height=3cm] {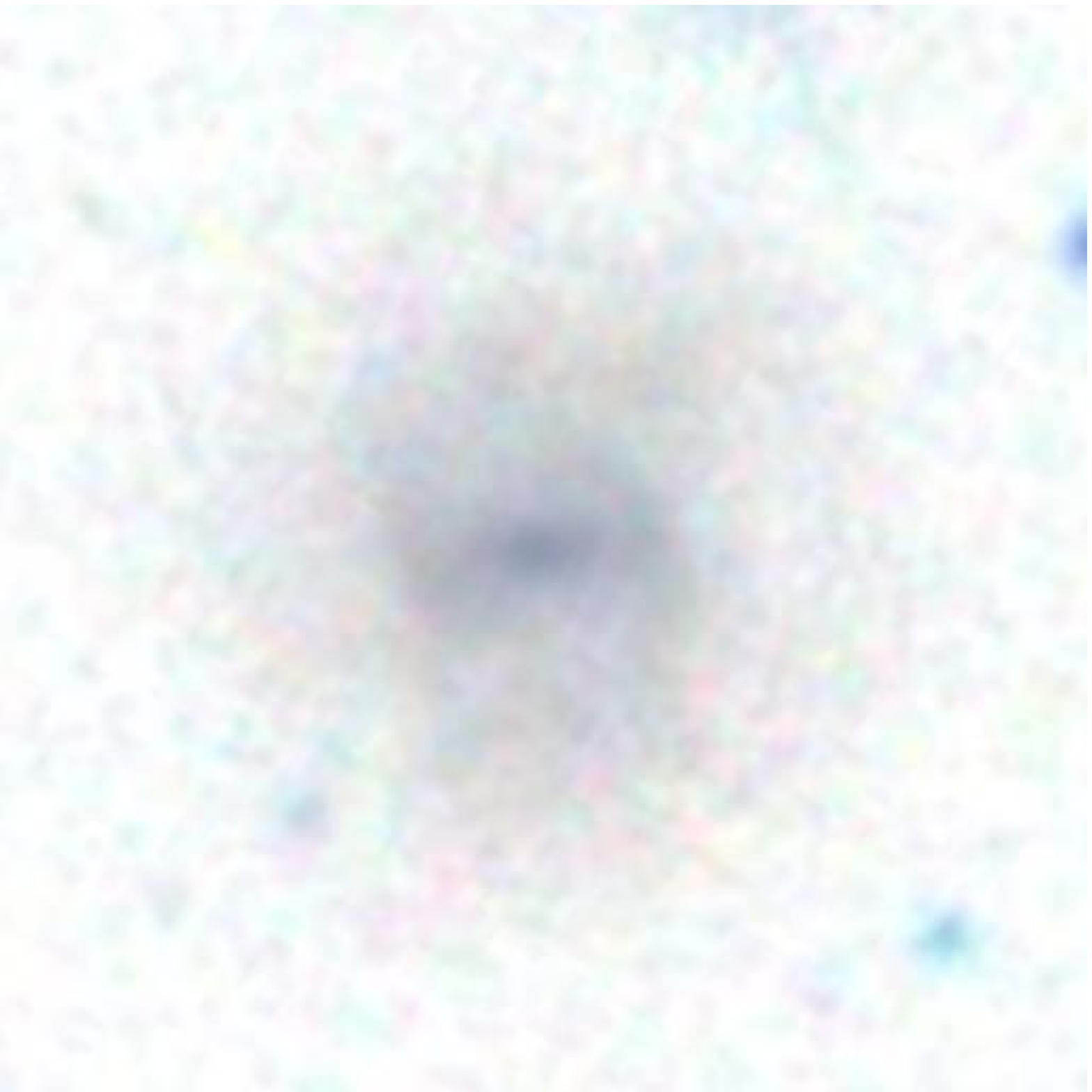}
\includegraphics [width=3cm, height=3cm] {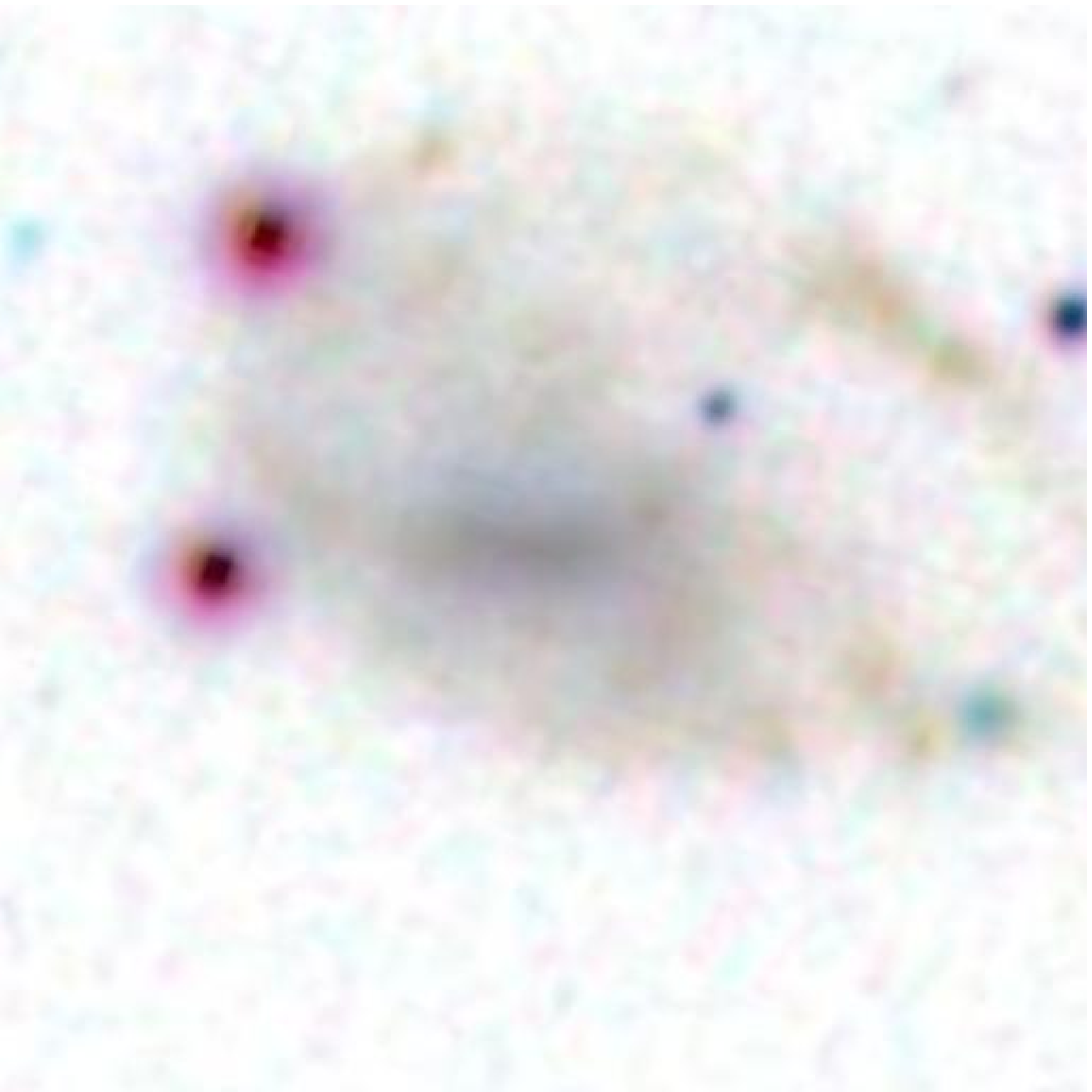}
\includegraphics [width=3cm, height=3cm] {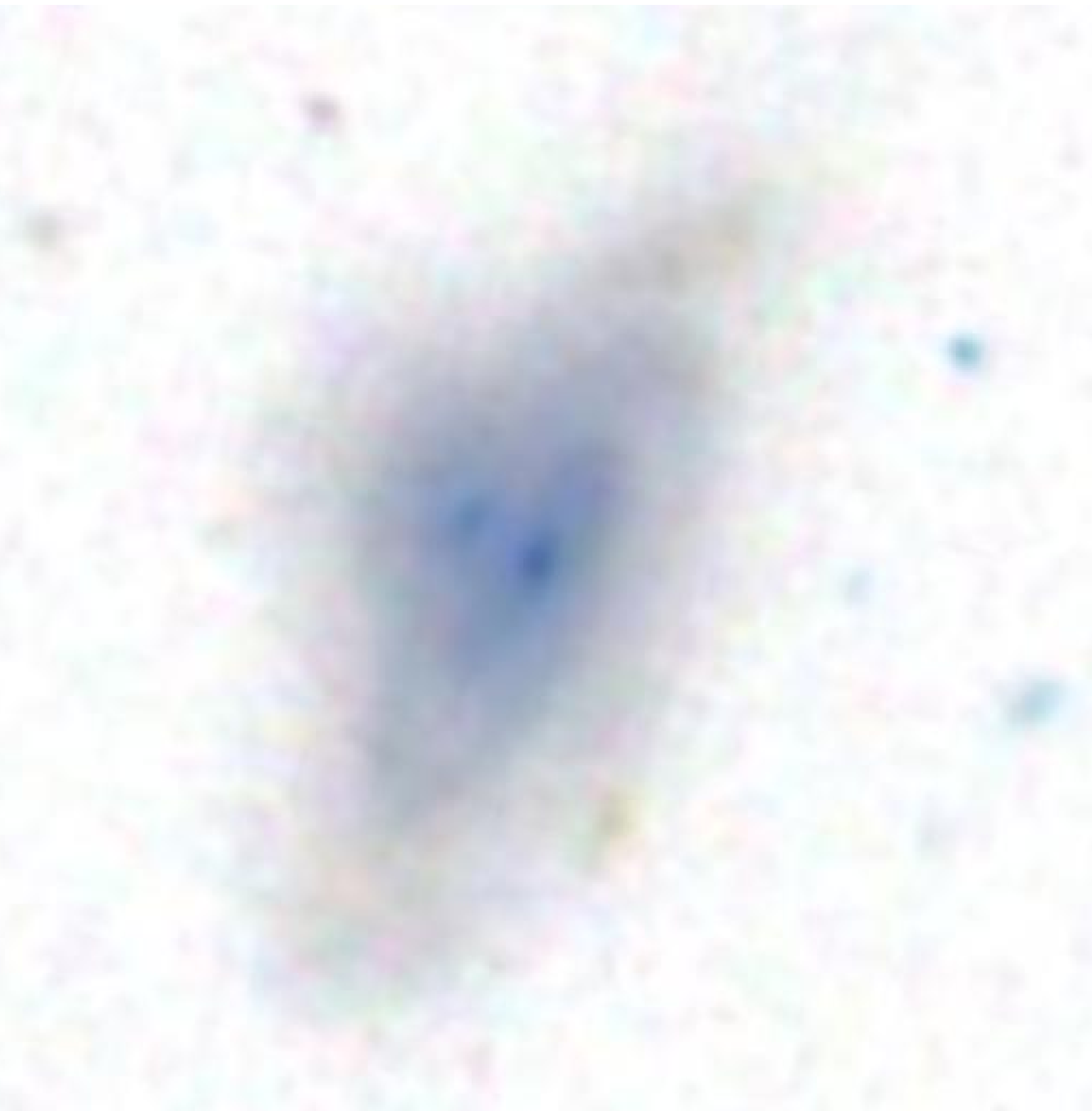}
\includegraphics [width=3cm, height=3cm] {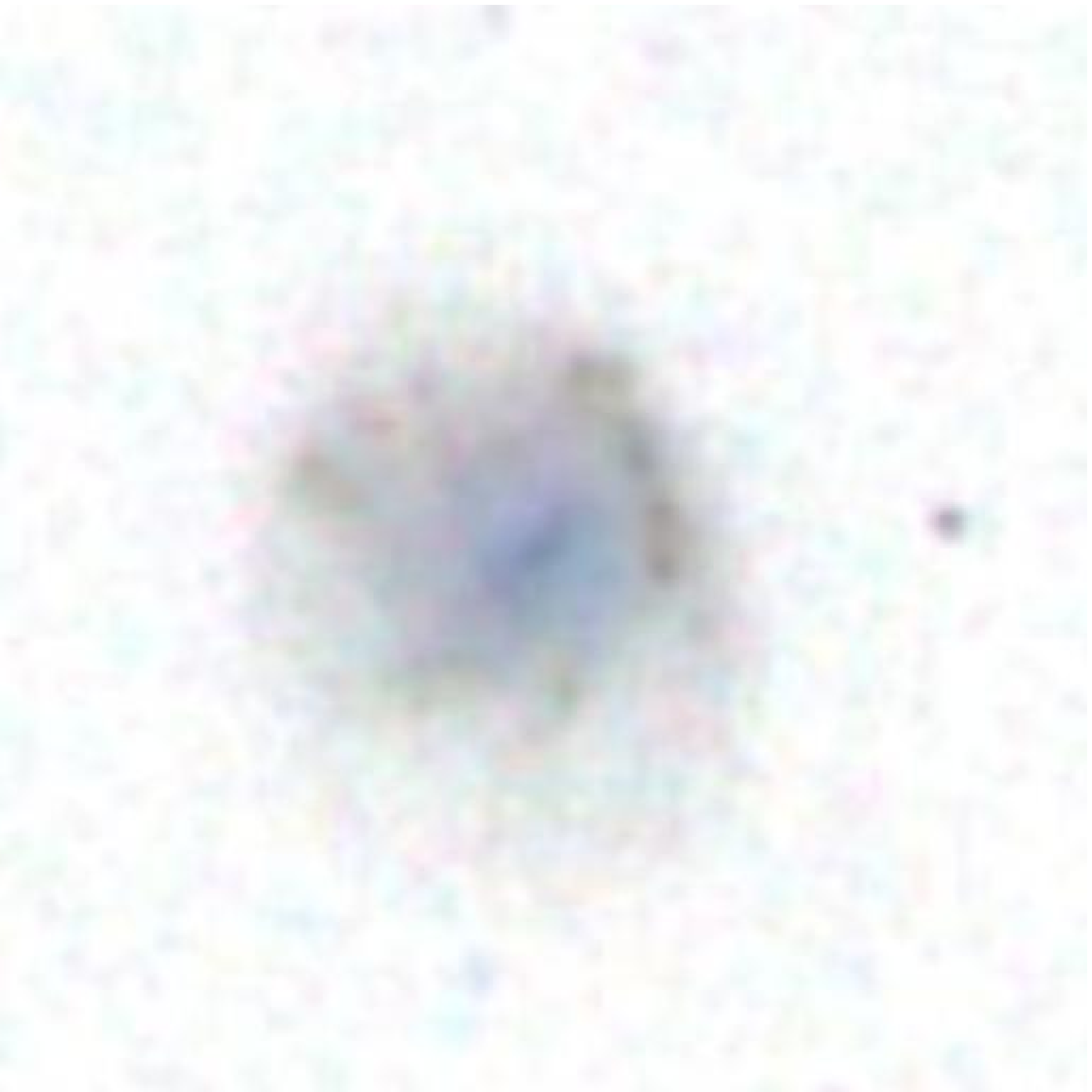}
\includegraphics [width=3cm, height=3cm] {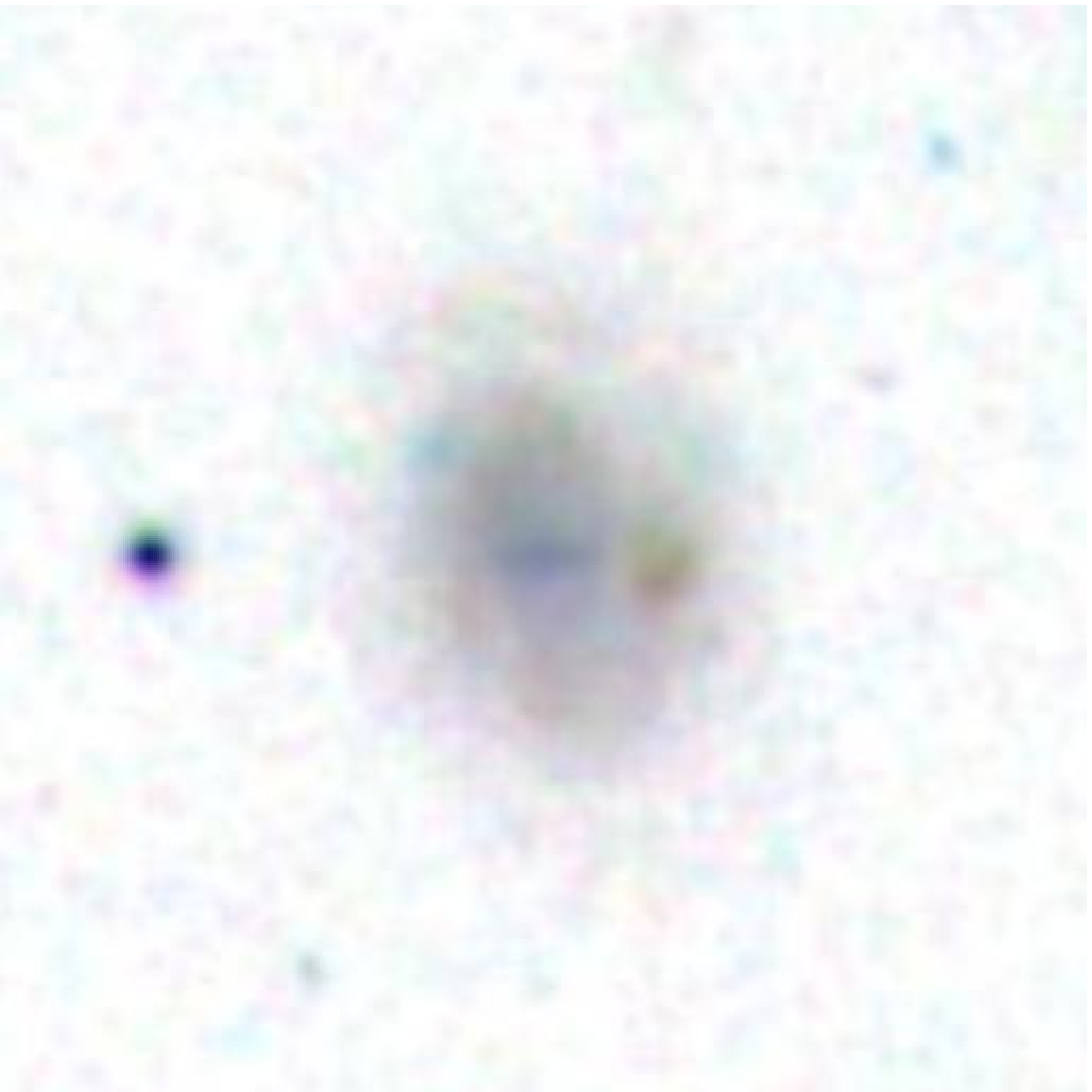}
\includegraphics [width=3cm, height=3cm] {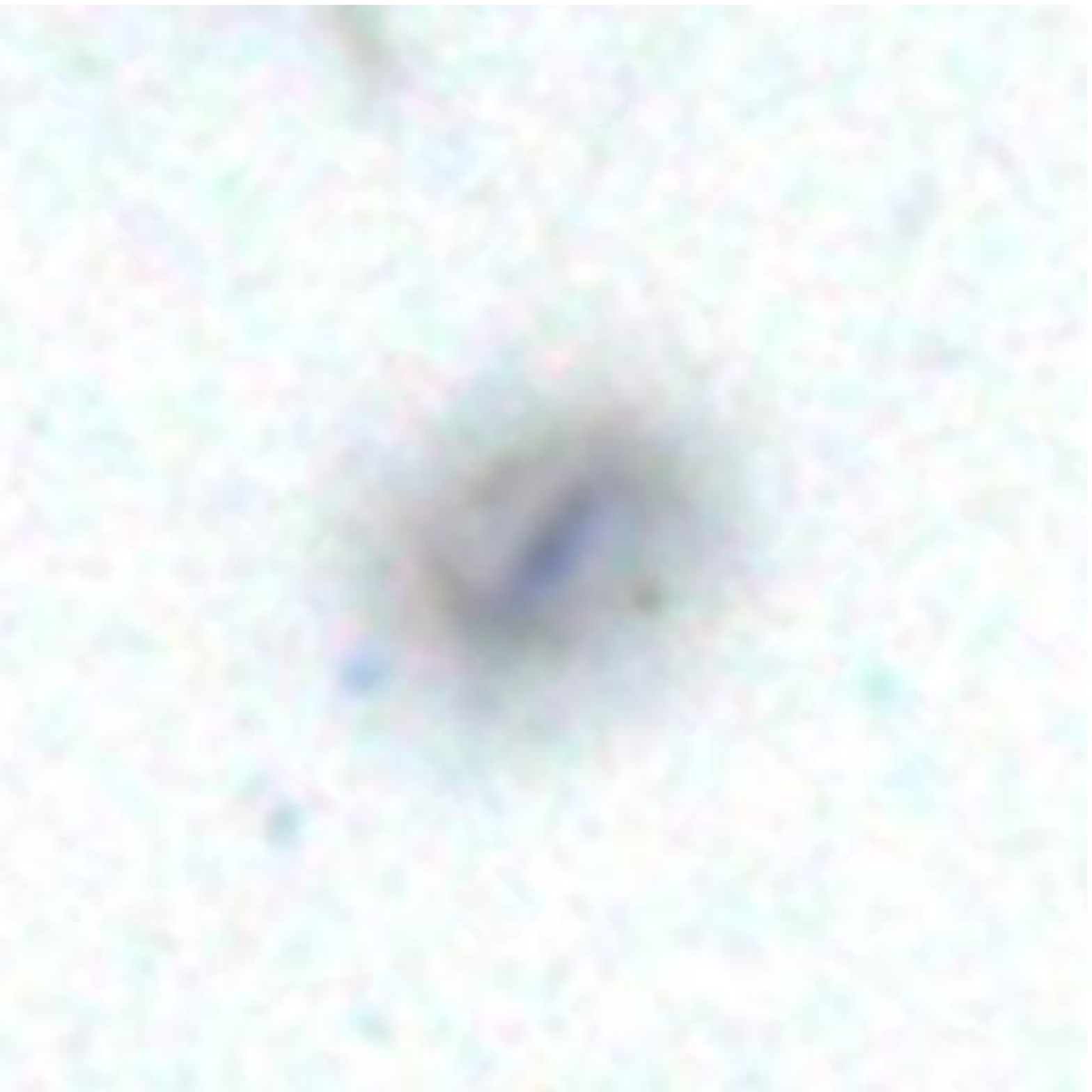}
\includegraphics [width=3cm, height=3cm] {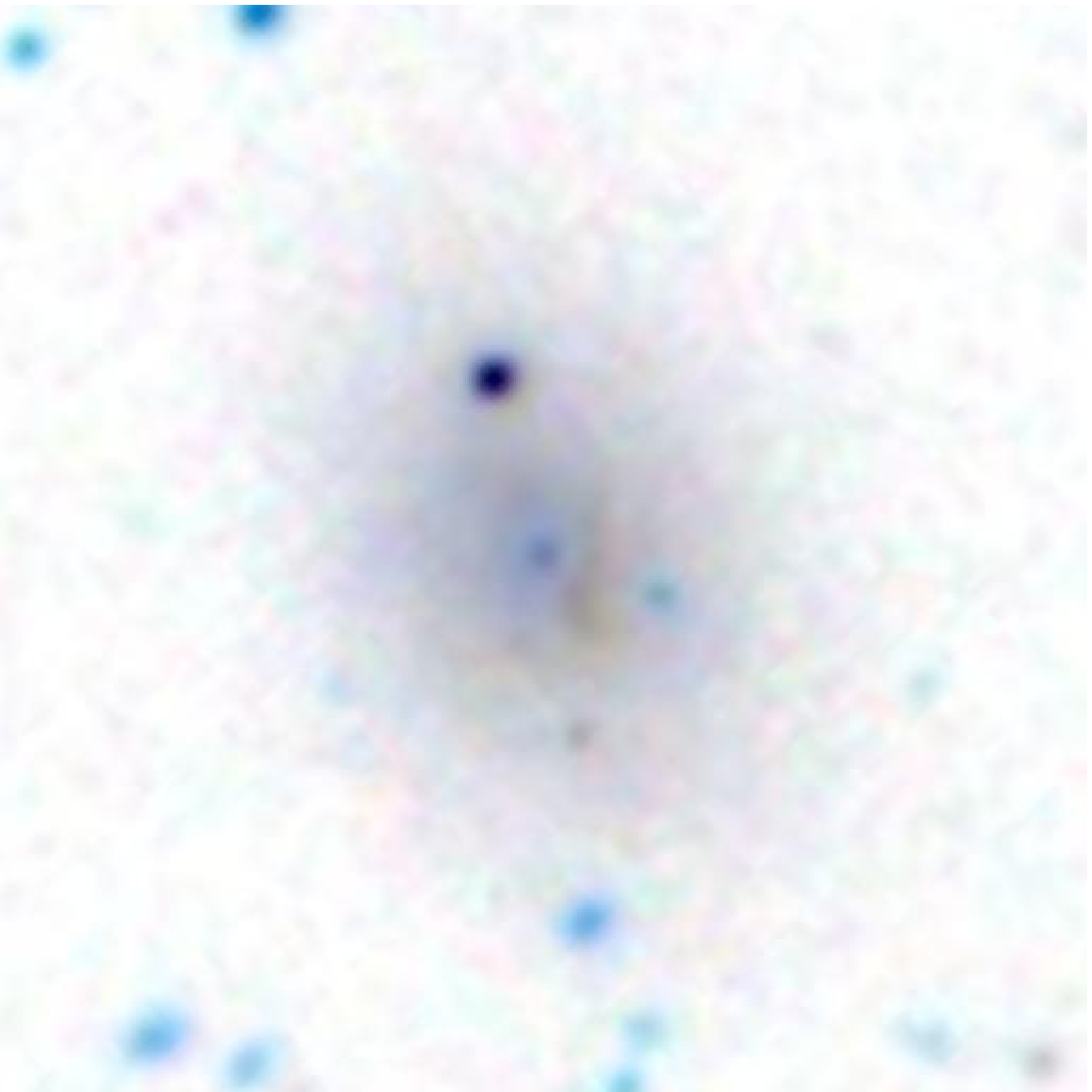}
\includegraphics [width=3cm, height=3cm] {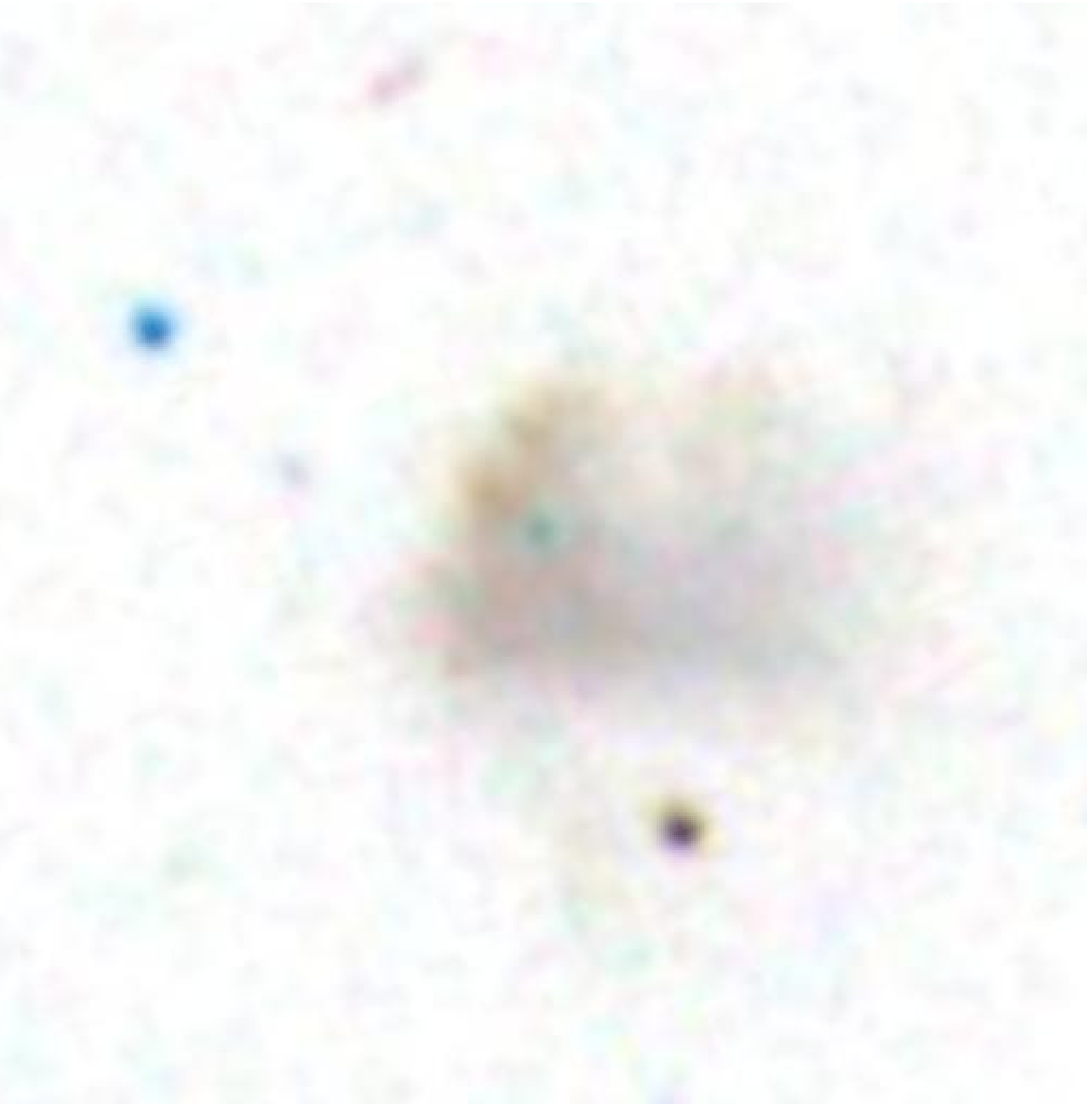}
\includegraphics [width=3cm, height=3cm] {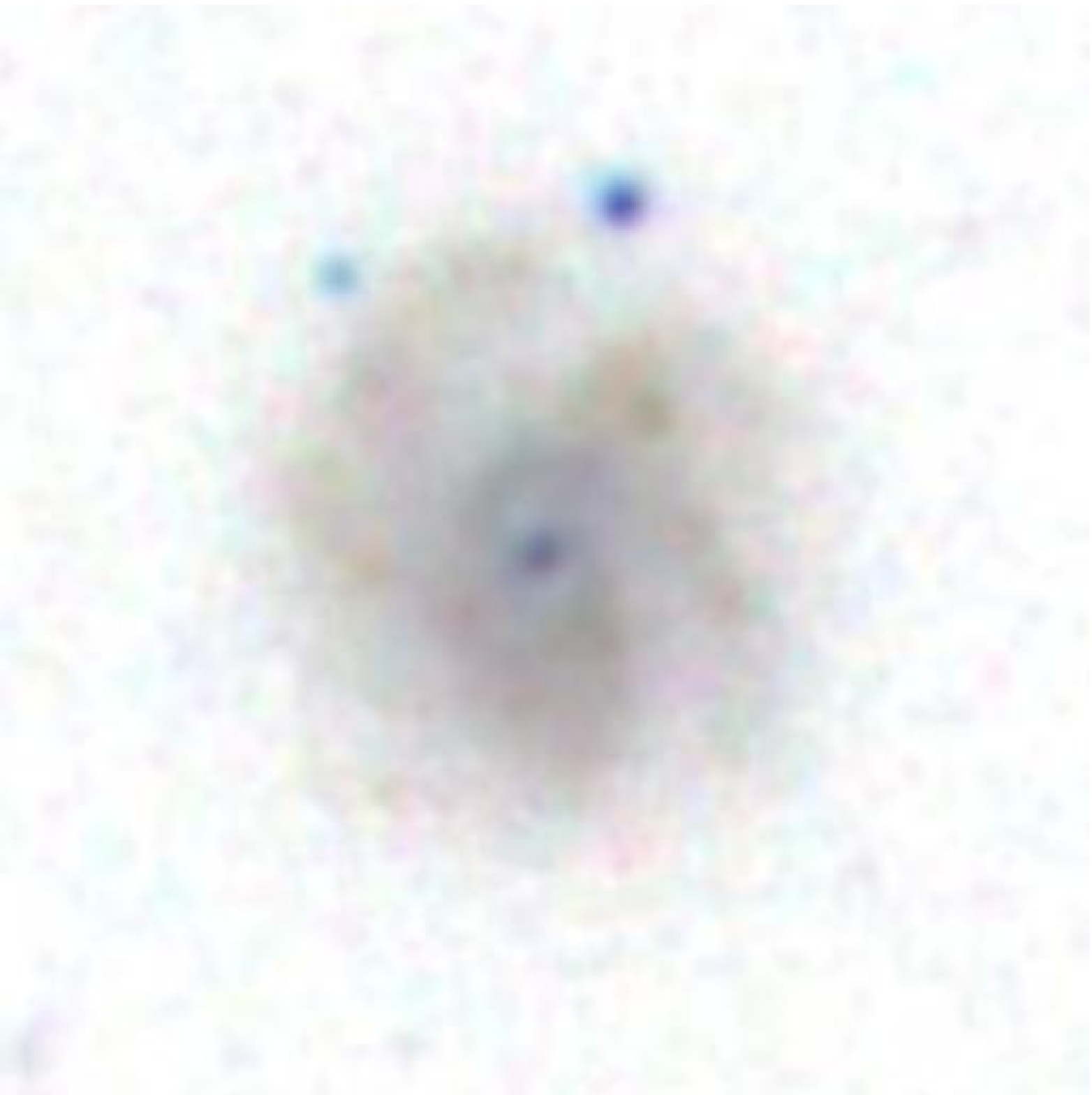} \vspace{1.5cm}

\includegraphics [width=3cm, height=3cm] {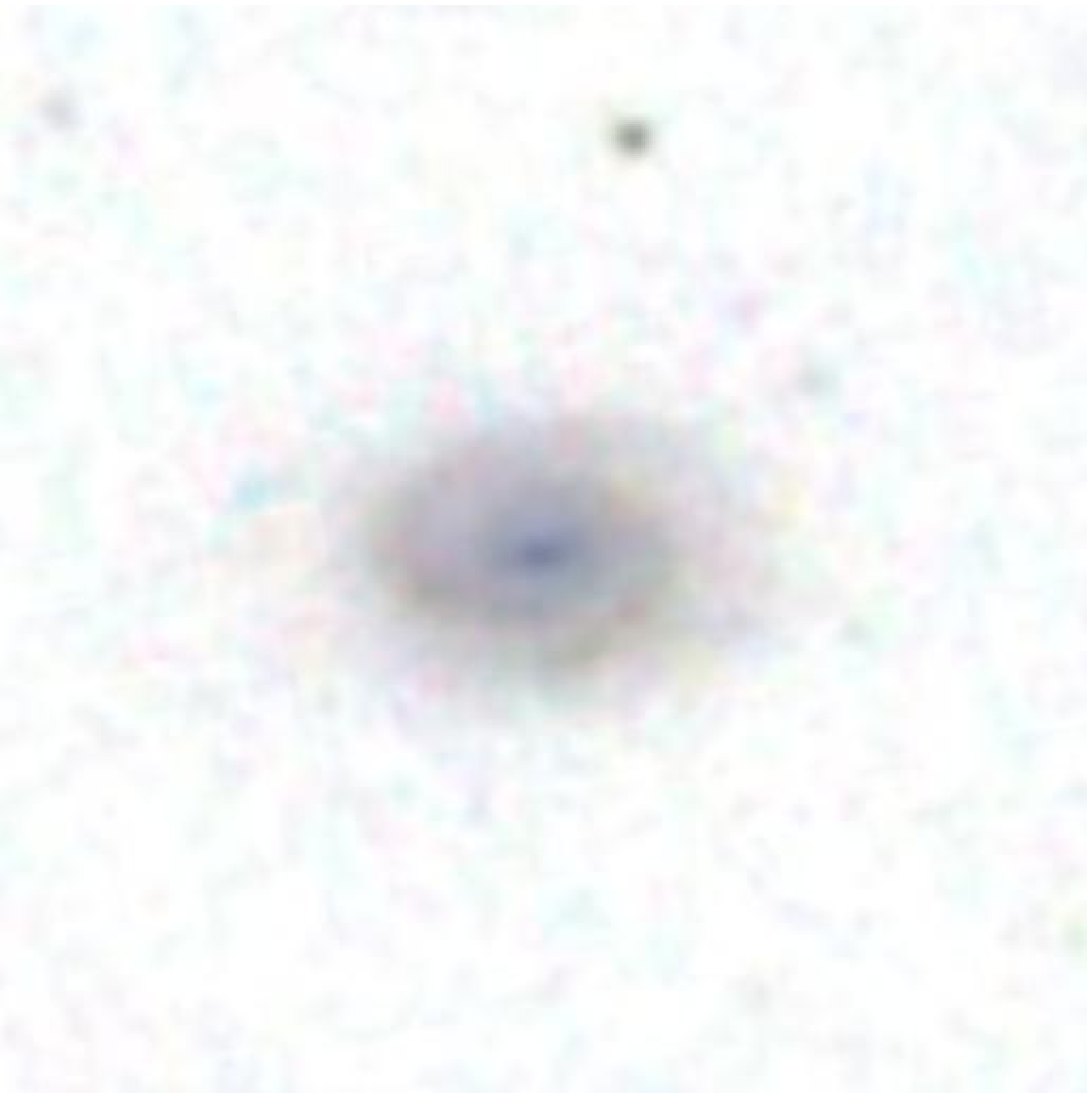}
\includegraphics [width=3cm, height=3cm] {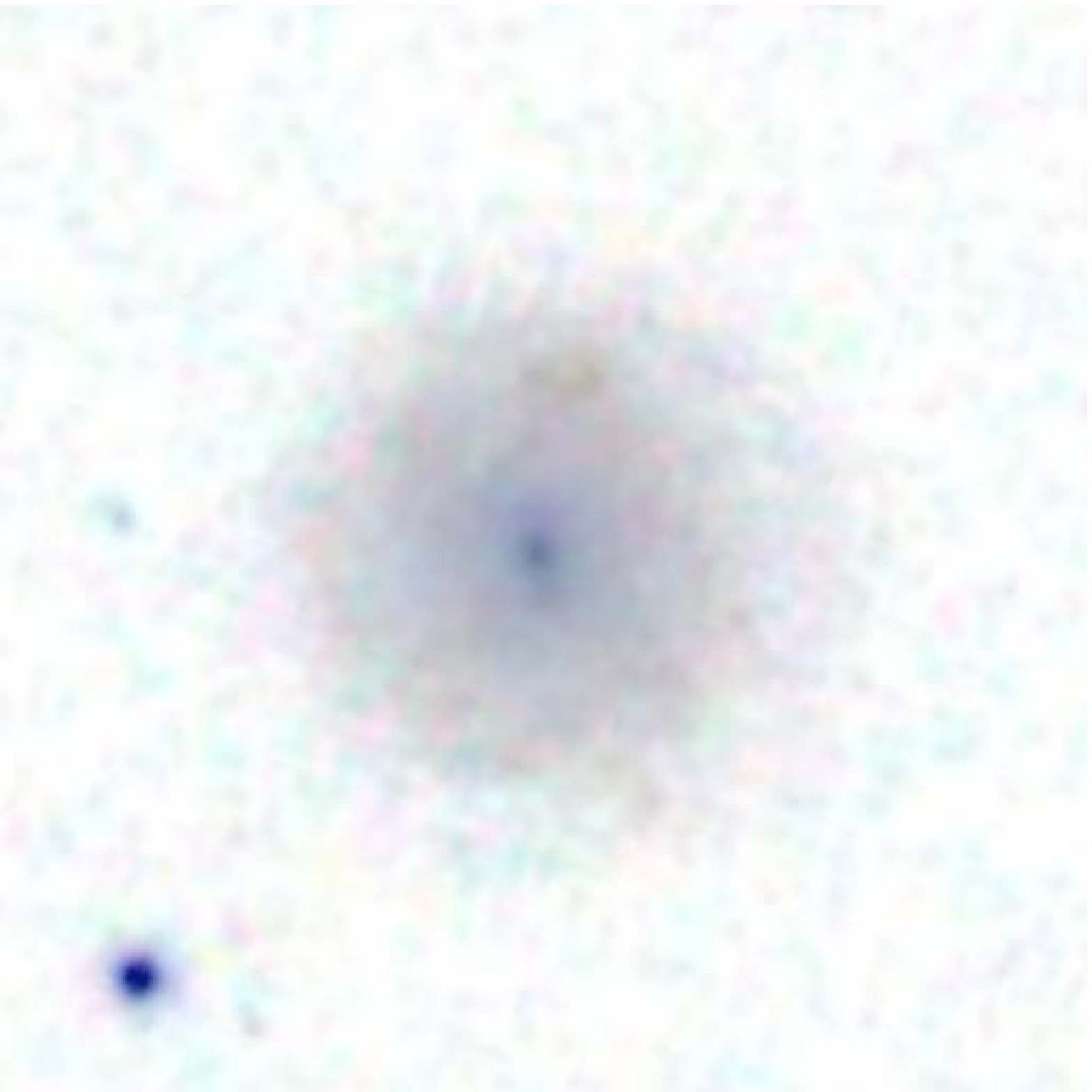}
\includegraphics [width=3cm, height=3cm] {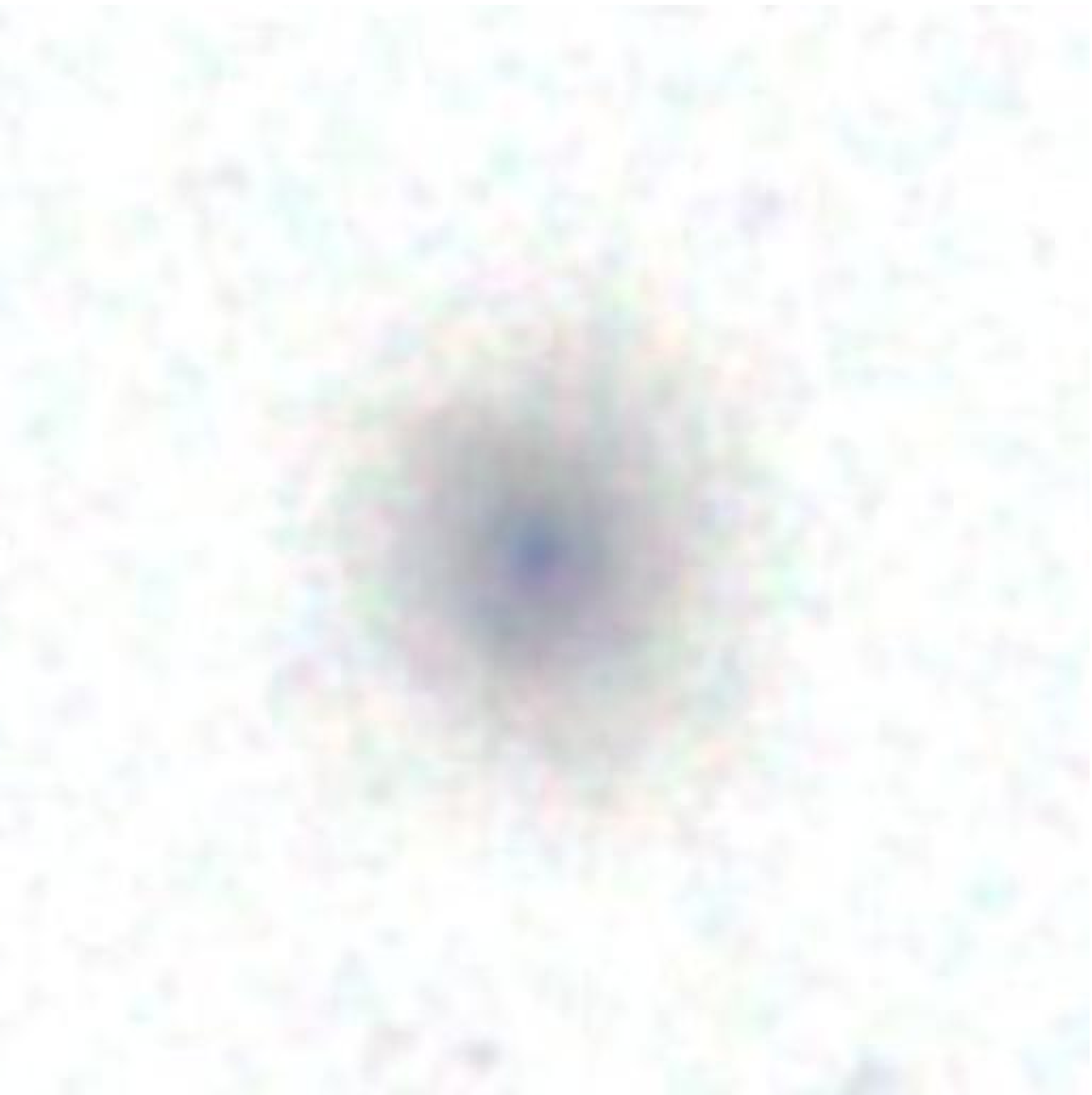}
\includegraphics [width=3cm, height=3cm] {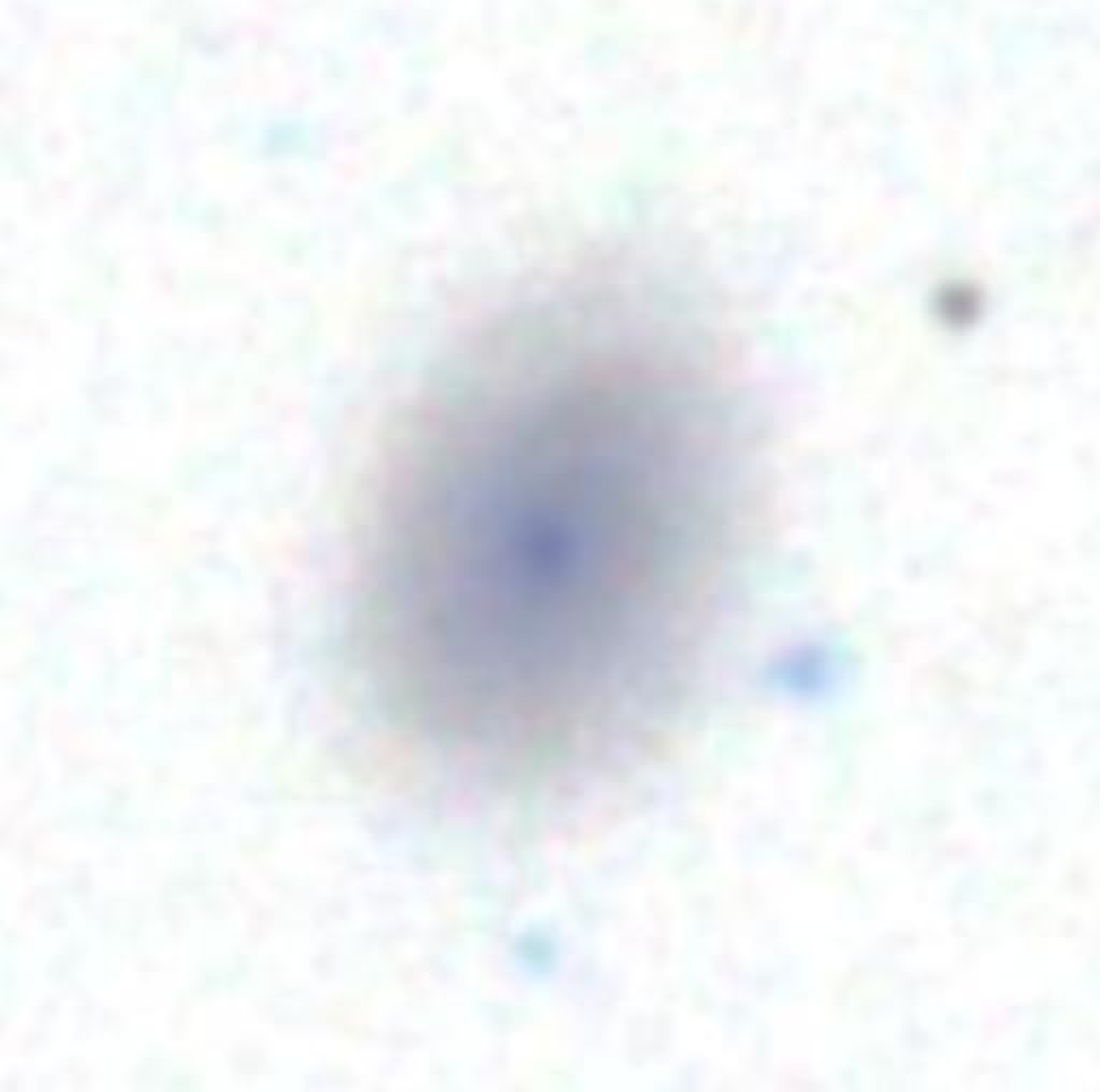}
\includegraphics [width=3cm, height=3cm] {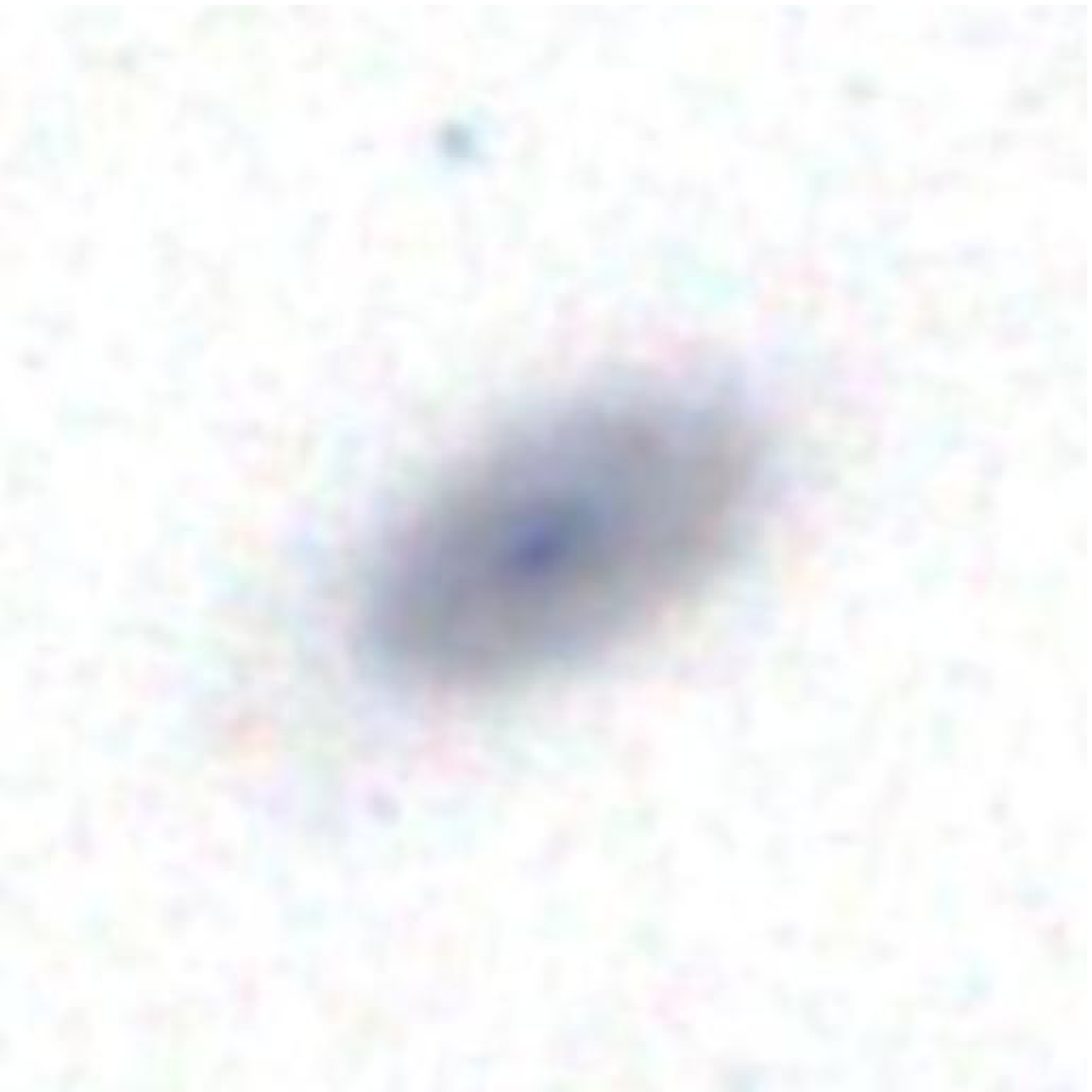}
\includegraphics [width=3cm, height=3cm] {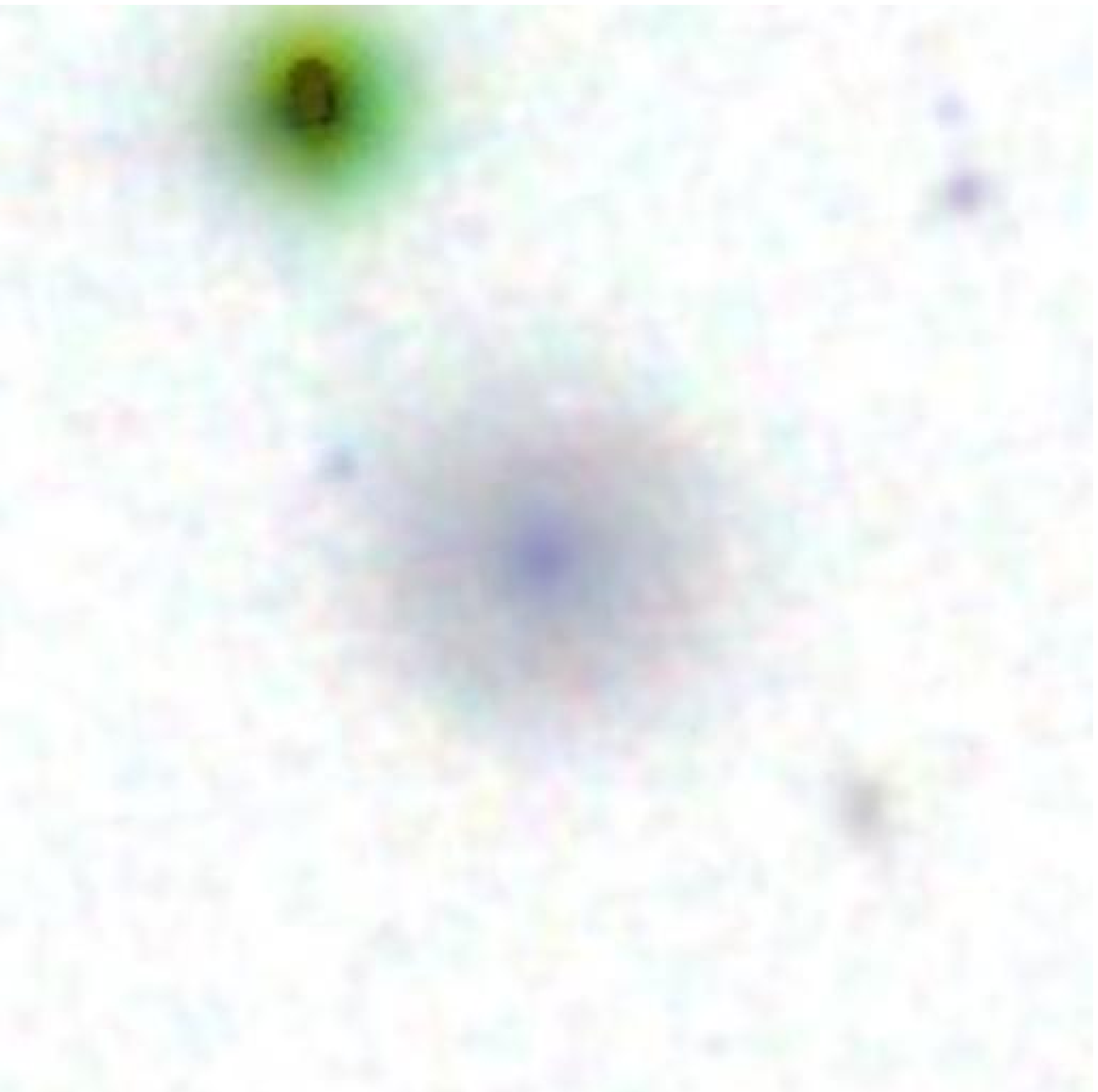}
\includegraphics [width=3cm, height=3cm] {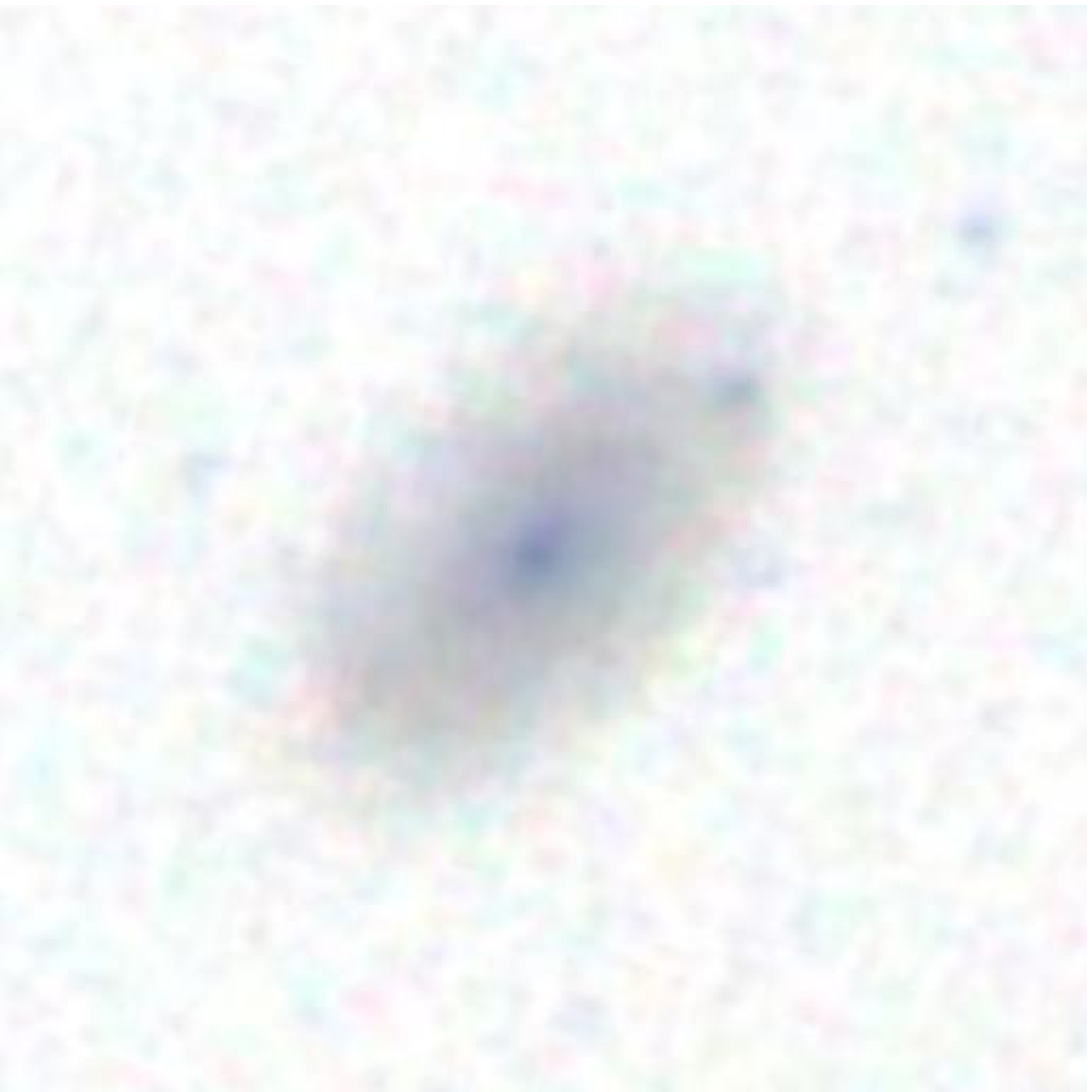}
\includegraphics [width=3cm, height=3cm] {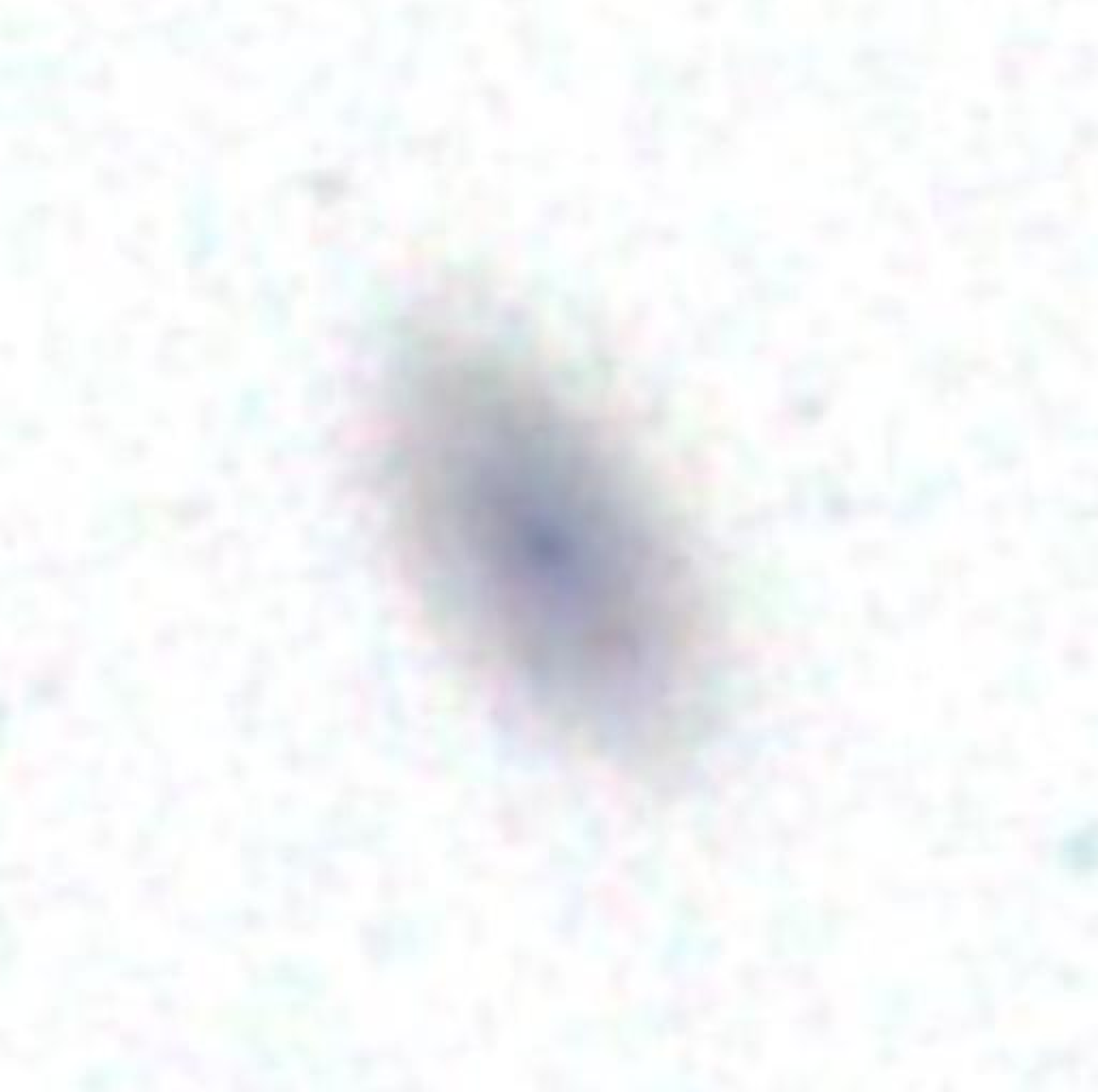}
\includegraphics [width=3cm, height=3cm] {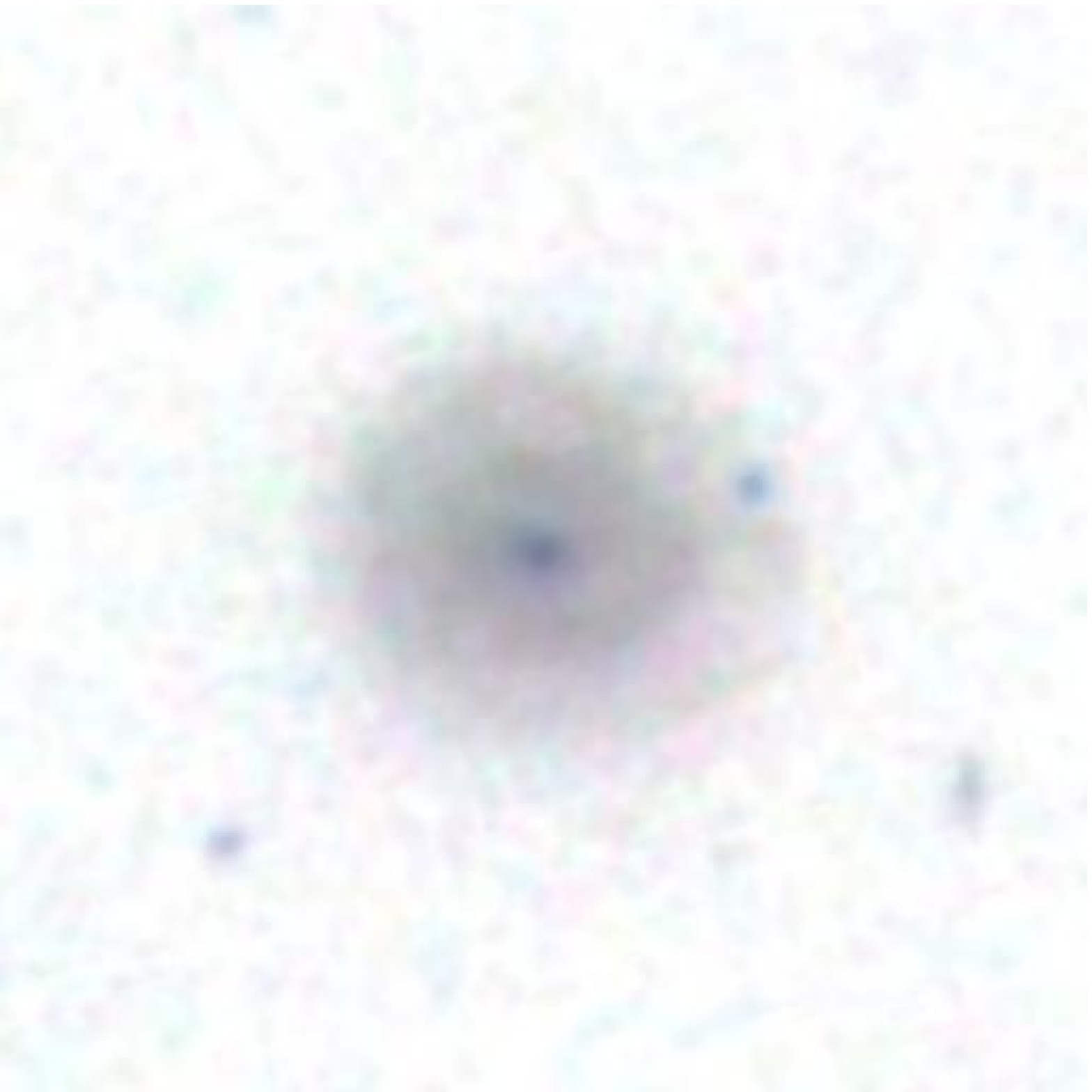}
\includegraphics [width=3cm, height=3cm] {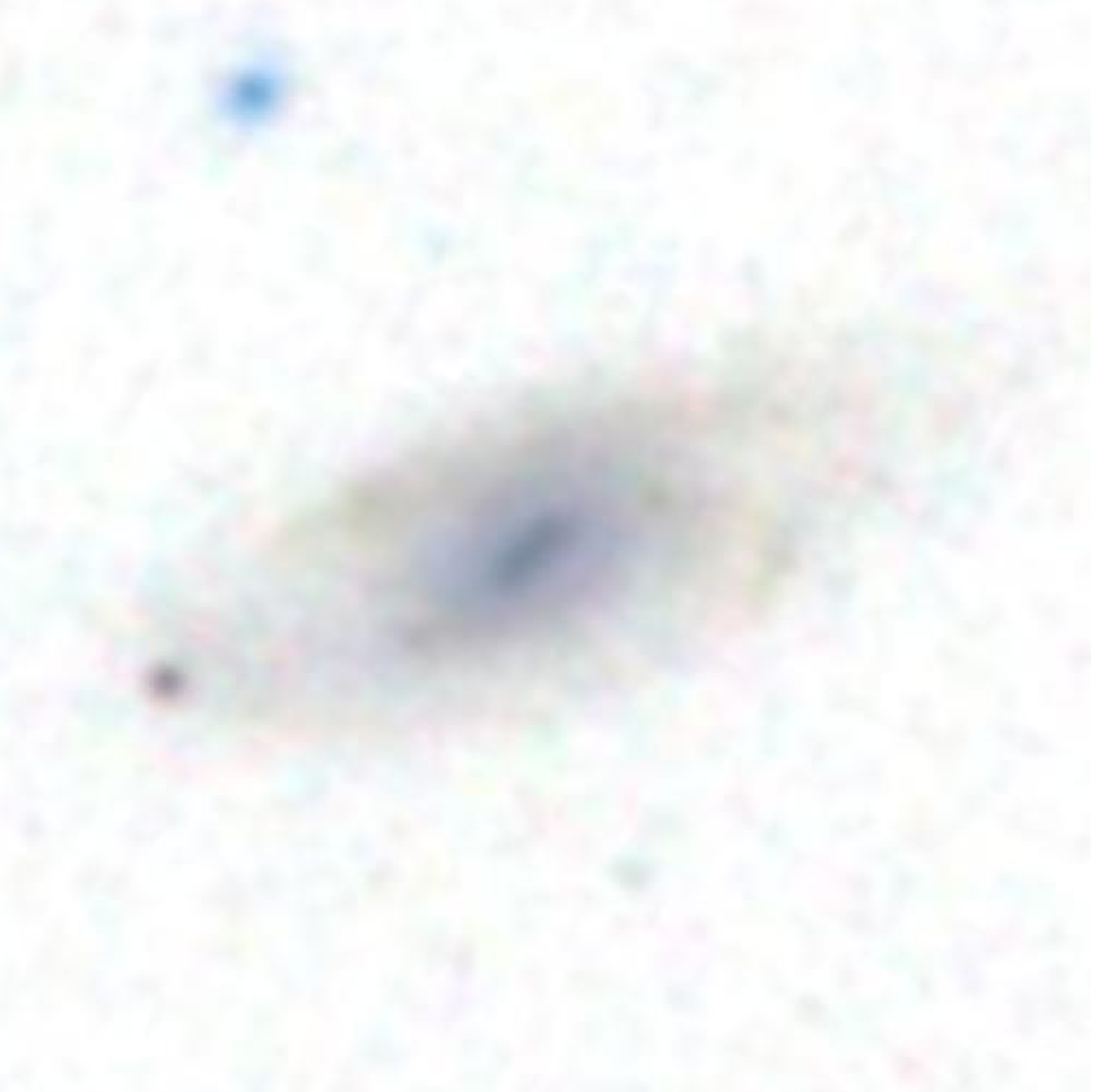}
\includegraphics [width=3cm, height=3cm] {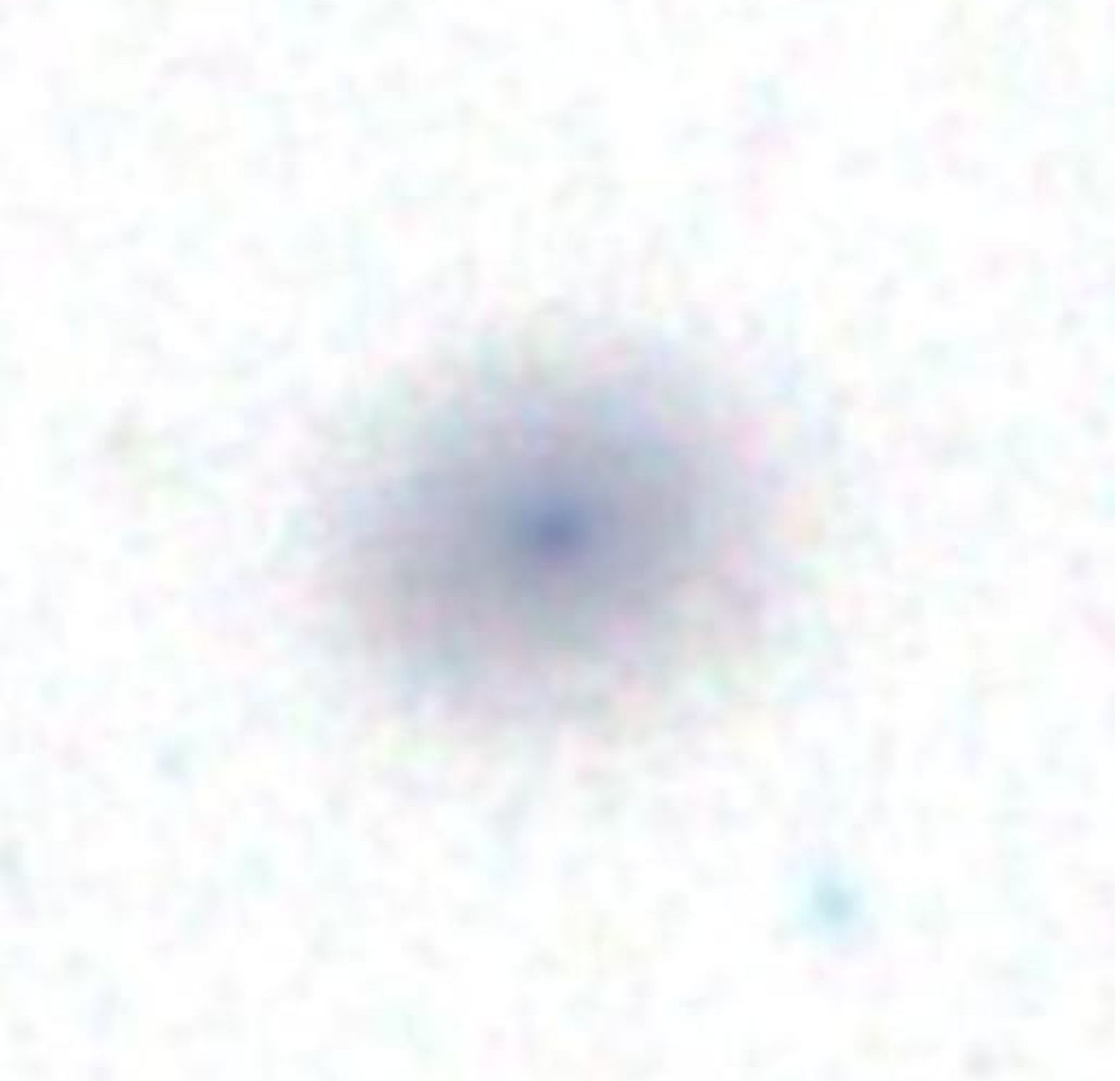}
\includegraphics [width=3cm, height=3cm] {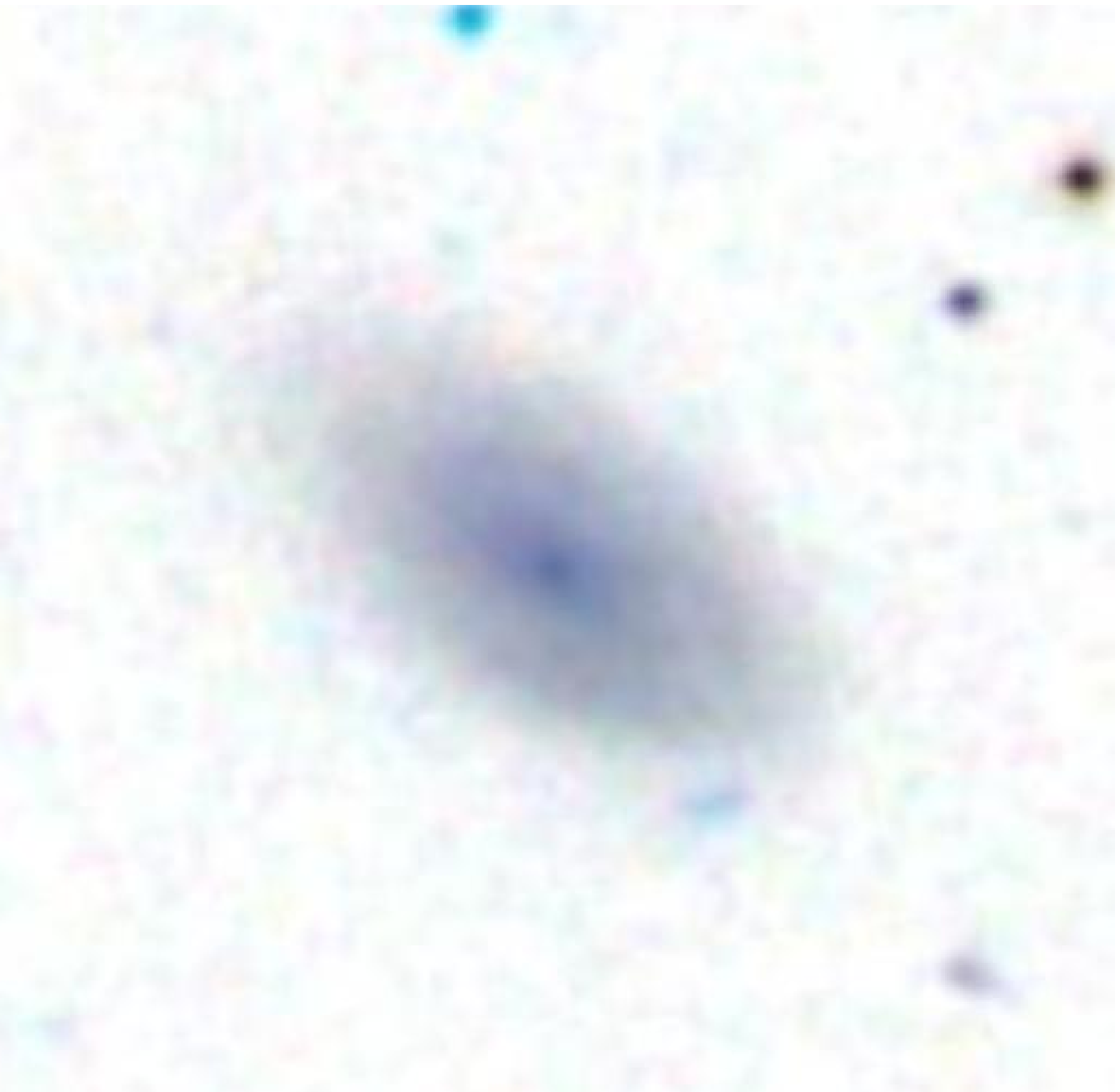}
\includegraphics [width=3cm, height=3cm] {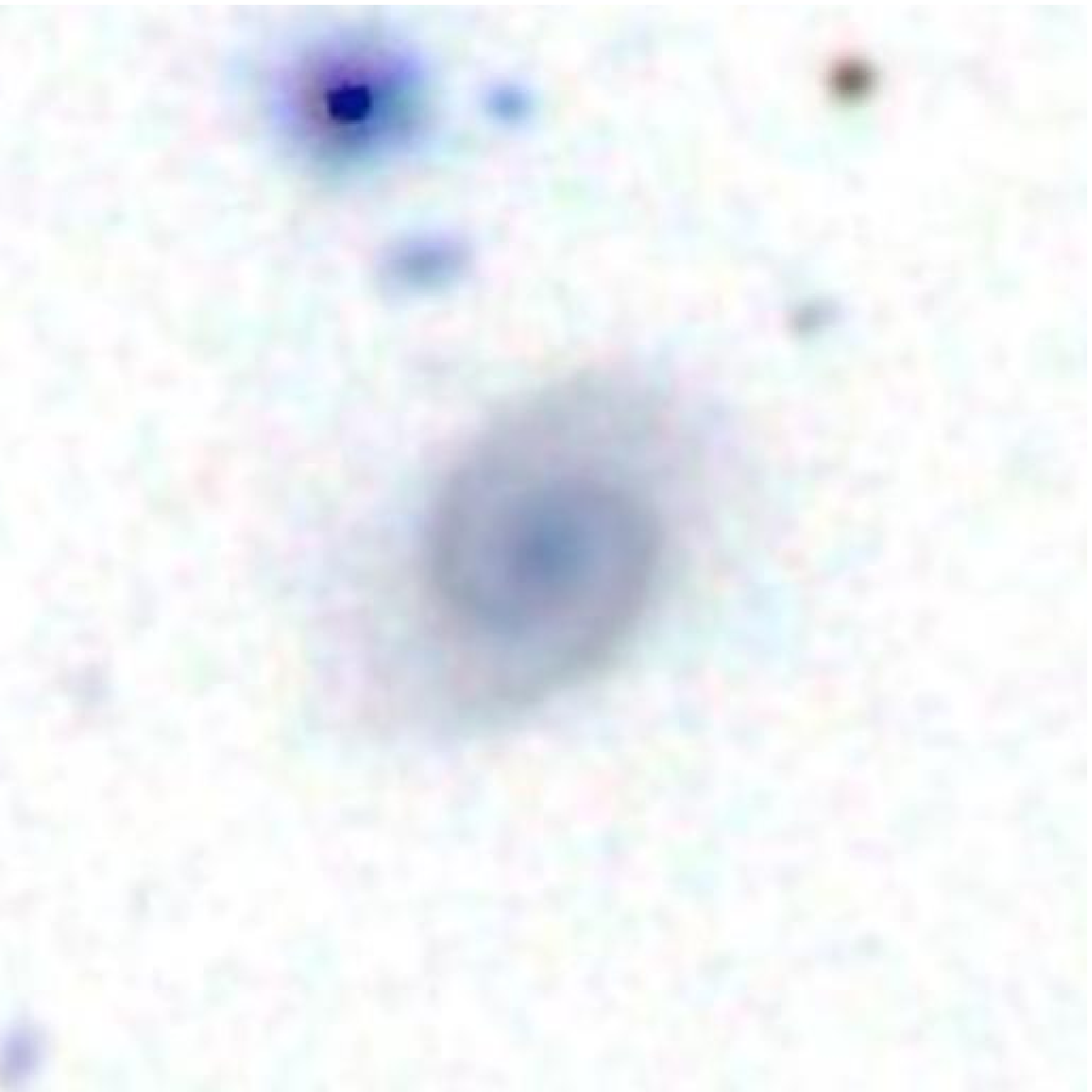}
\includegraphics [width=3cm, height=3cm] {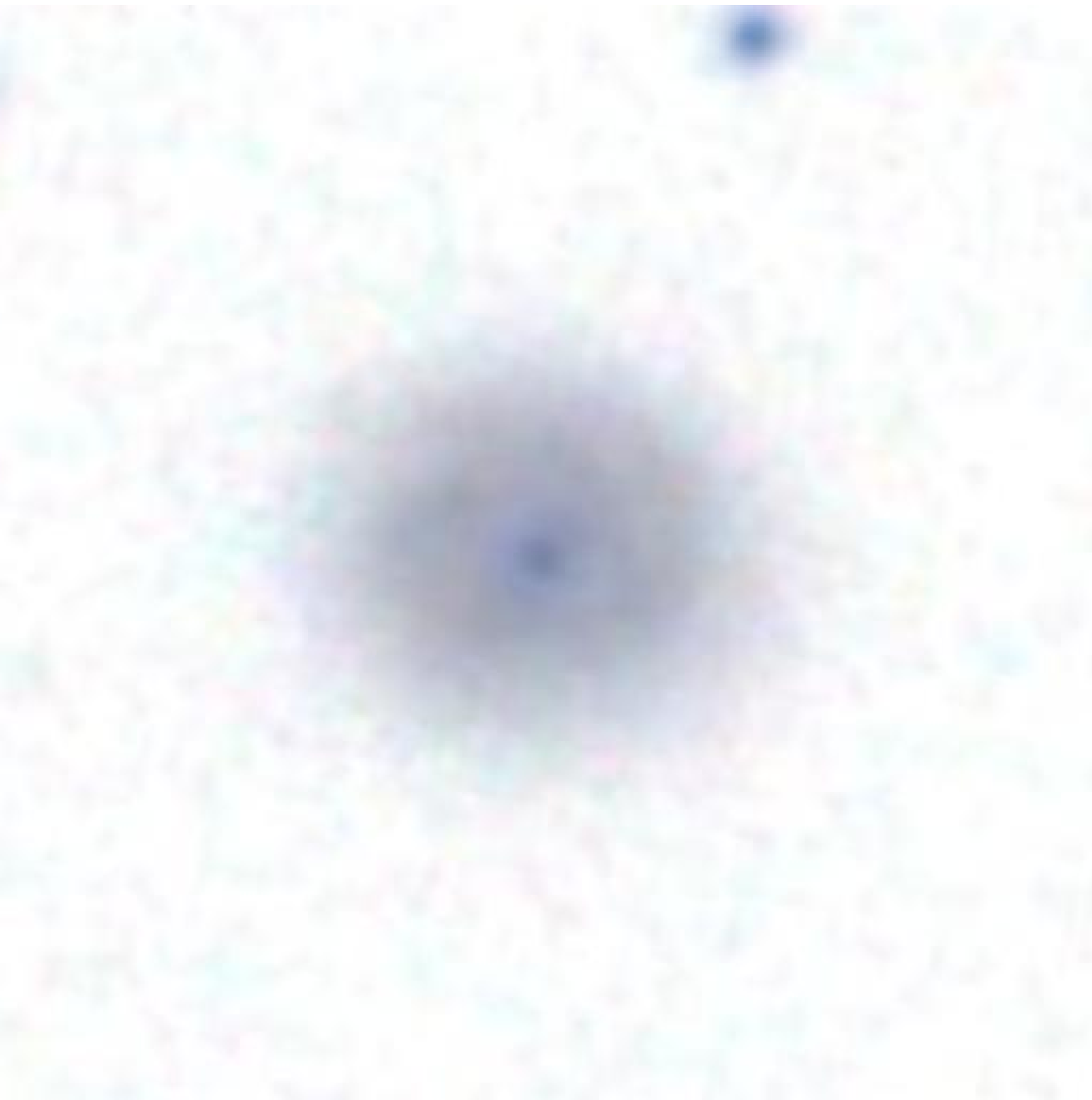}
\includegraphics [width=3cm, height=3cm] {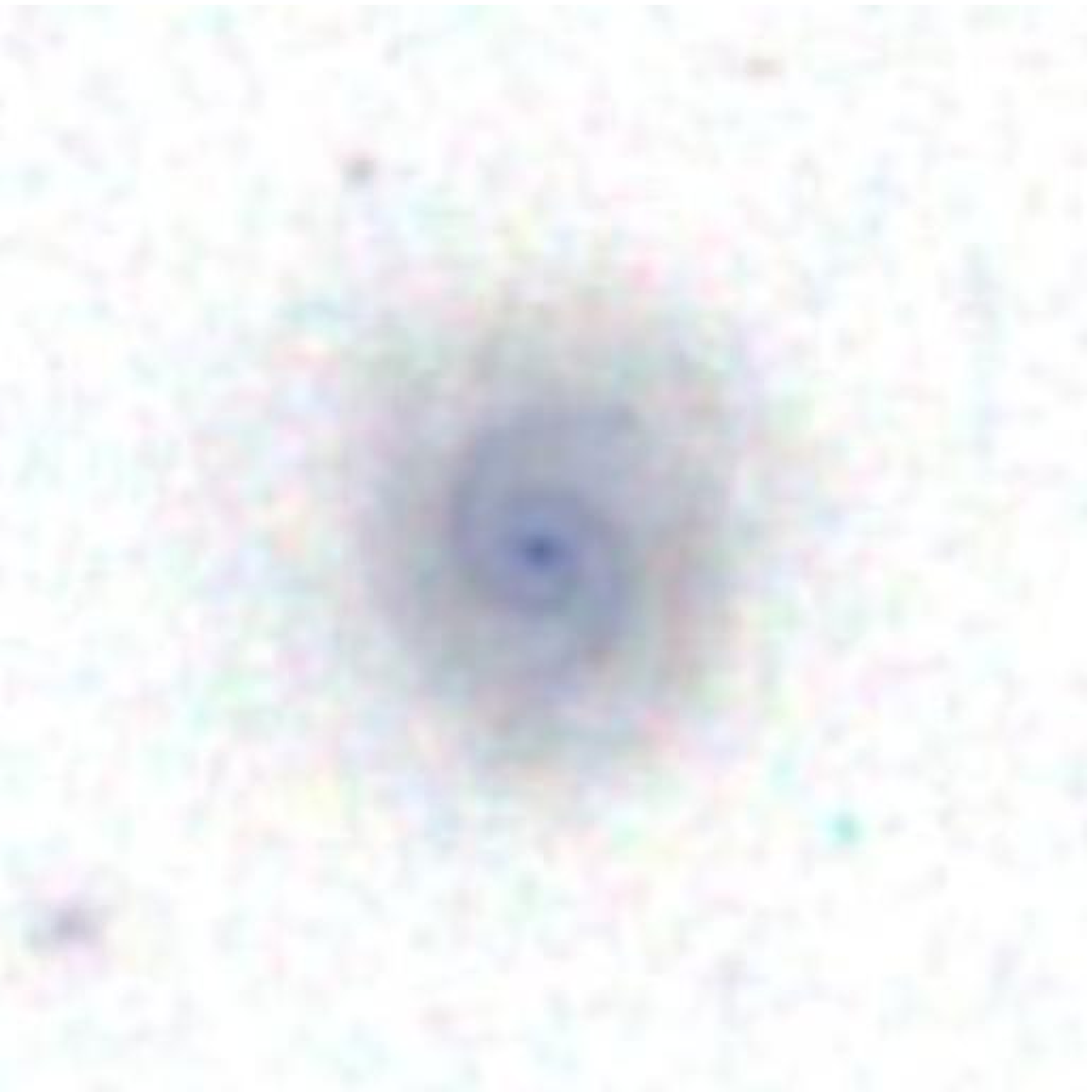}
\caption {The images of high mass irregular LSB galaxies (log$M_{\odot}$$>$9.5, top panel) 
and high mass regular LSB galaxies (log$M_{\odot}$$>$9.5, bottom panel). 
From top-left to bottom-right in the top panel, the 3rd, 5th, 6th, 8th, 9th, 10th, 12th and 15th 
show evidences for either multiple components, double nuclei 
and tidal features evidencing merging.} \label{fig.highmass}
\end{figure*}

For massive LSB galaxies(log$M_{\odot}$$>$9.5), normally it is difficult for the gas to fall into the
center and form stars. One possible mechanism to form
young stars in the central region is through a gas-rich merger that could transport gas
from the outskirts of the galaxy into the center \citep{2009A&A...507.1313H, 2008A&A...492..371R, 2012MNRAS.421.2888R}.
For more than half of irregular LSB galaxies in Figure~\ref{fig.highmass}, we see a merger happening within them when checking their images. 
For the irregular LSB galaxies with log$M_{\odot}$$>$9.5
in our sample, we also find seven with bars appearing in
the center, a phenomenon that can be enhanced by merging in peculiar galaxies. The presence of a bar is believed to be an efficient way
to drive gas towards the central region of a galaxy \citep{1996ApJ...462..114N}, which may
trigger strong star formation \citep{2009ApJ...692.1623H}. 

\begin{figure*}
\centering
\includegraphics [width=3cm, height=3cm] {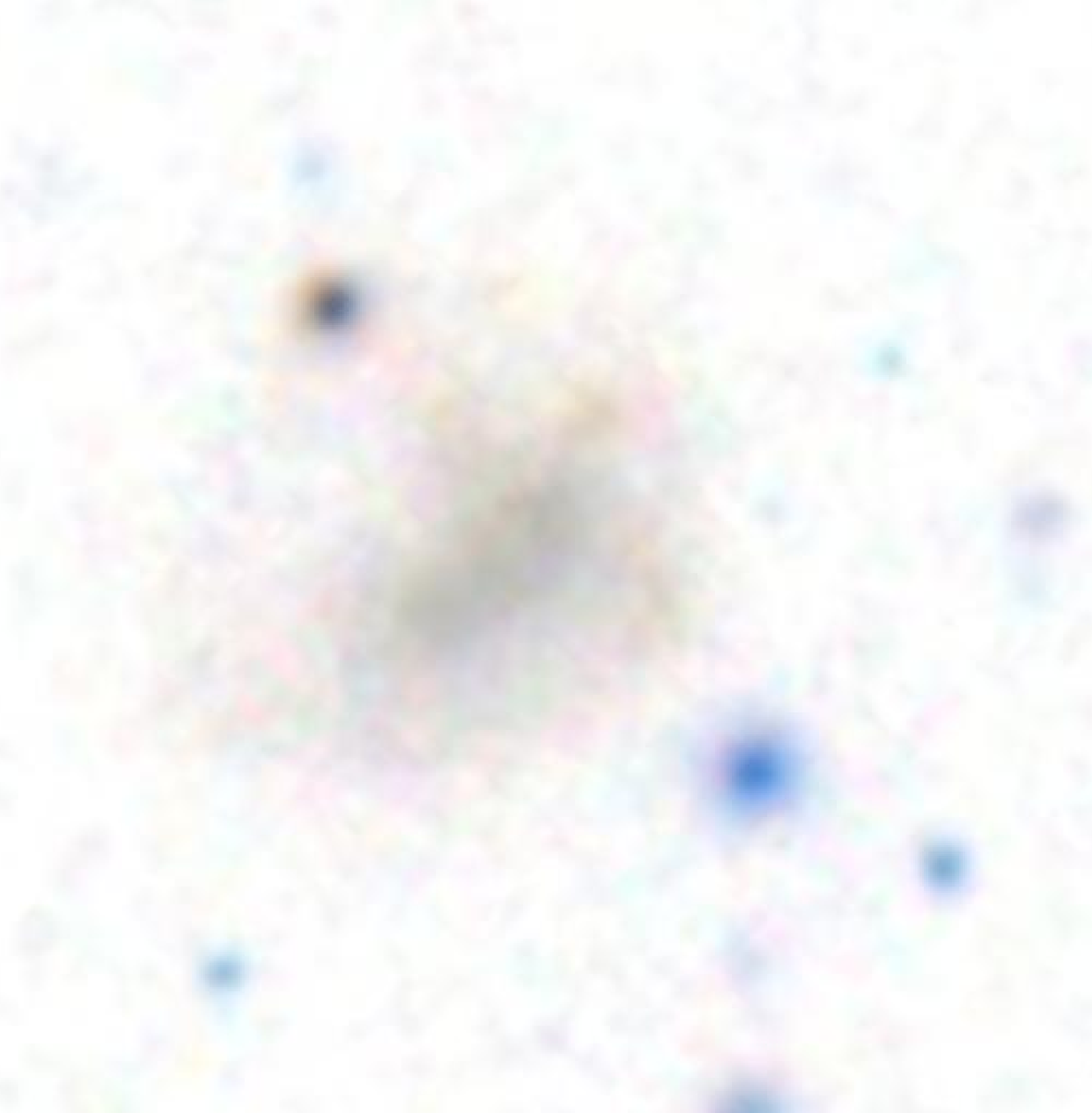}
\includegraphics [width=3cm, height=3cm] {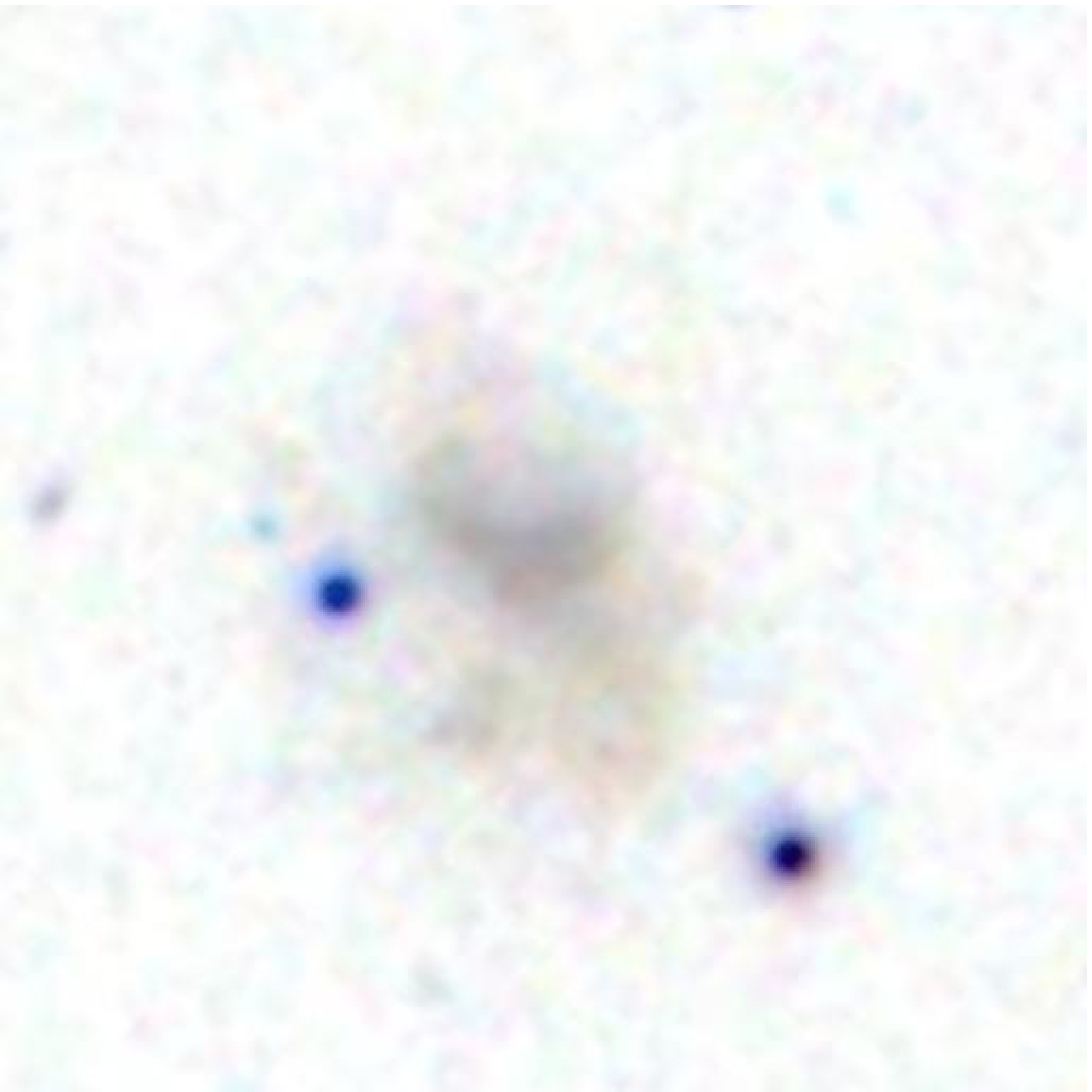}
\includegraphics [width=3cm, height=3cm] {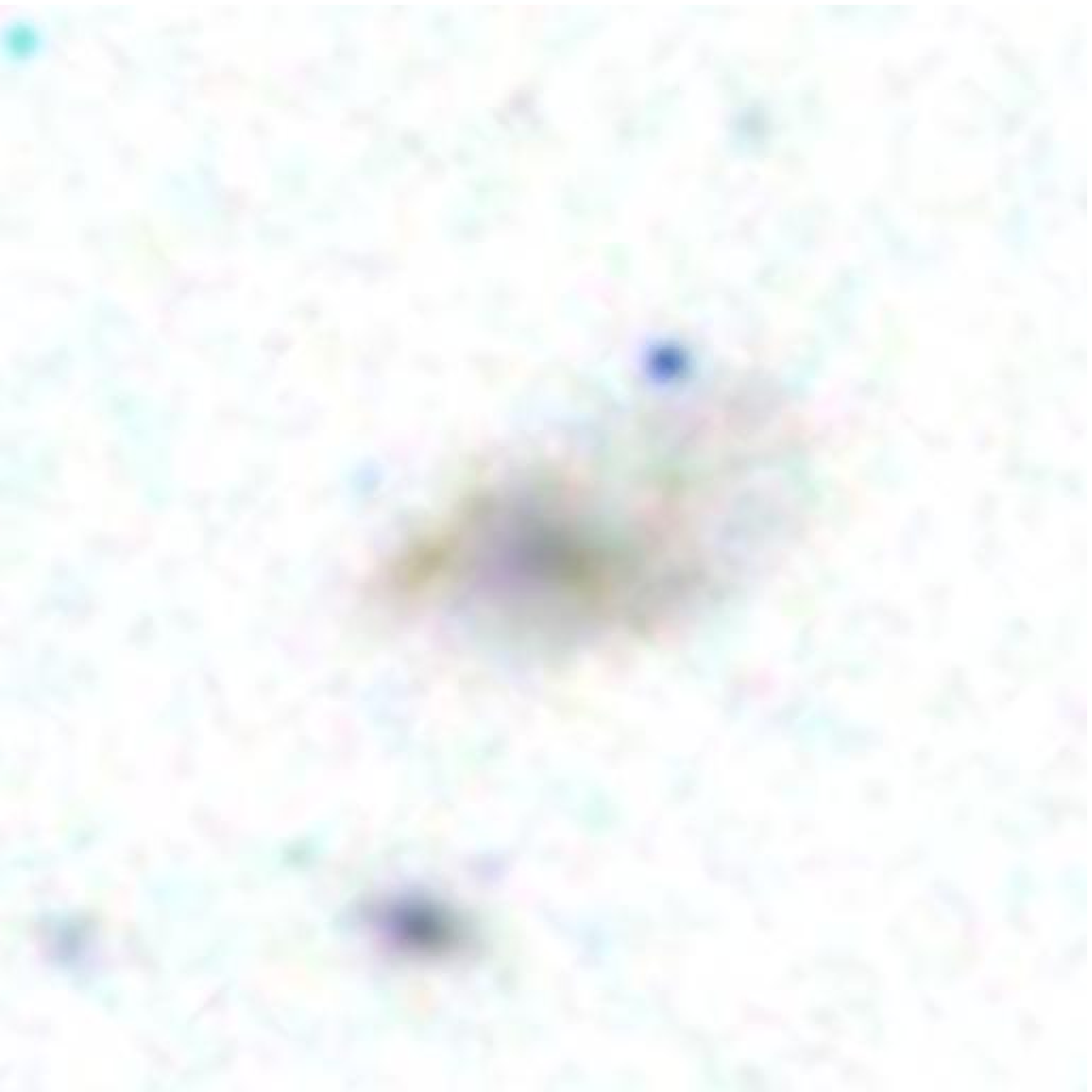}
\includegraphics [width=3cm, height=3cm] {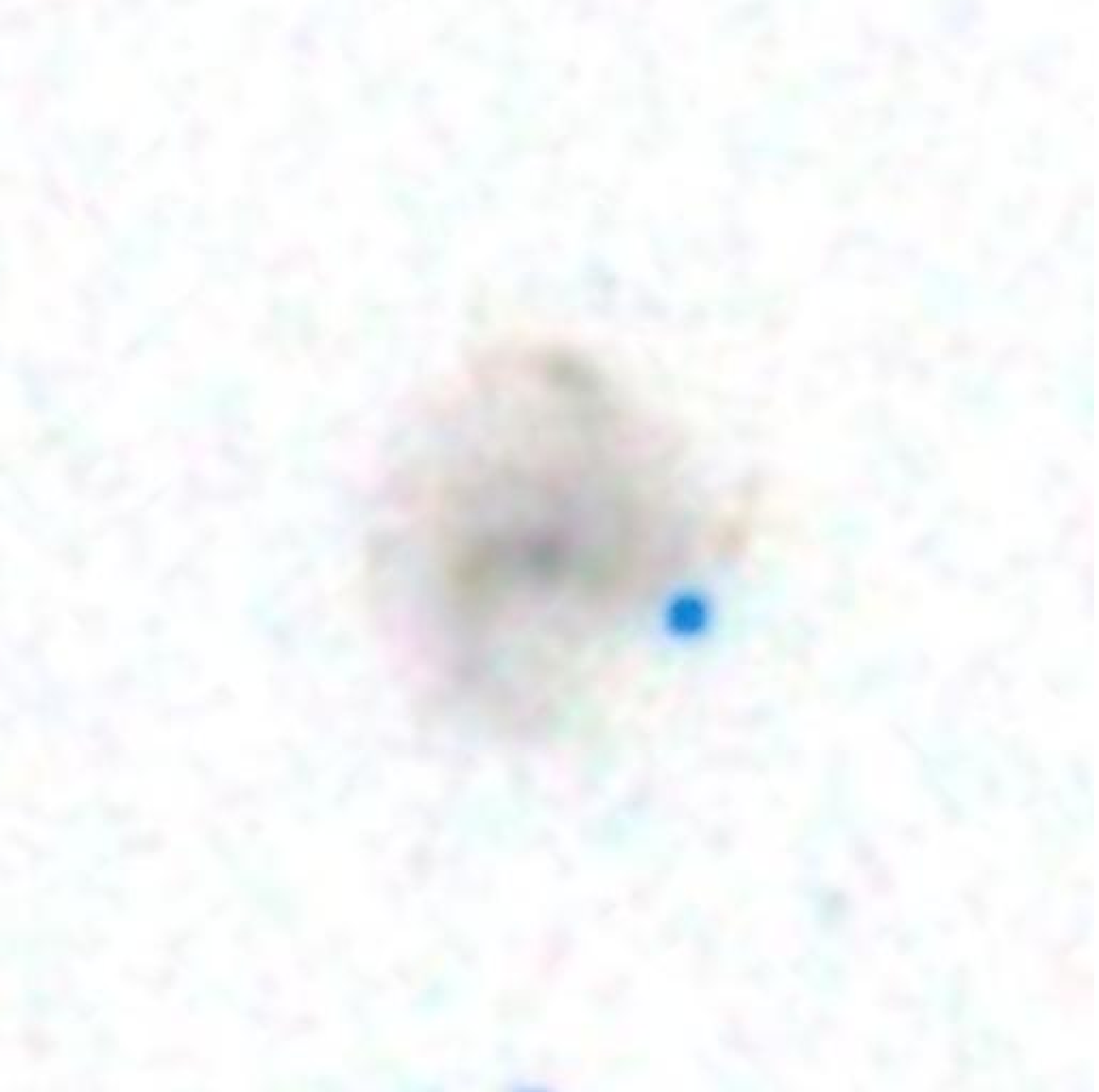}
\includegraphics [width=3cm, height=3cm] {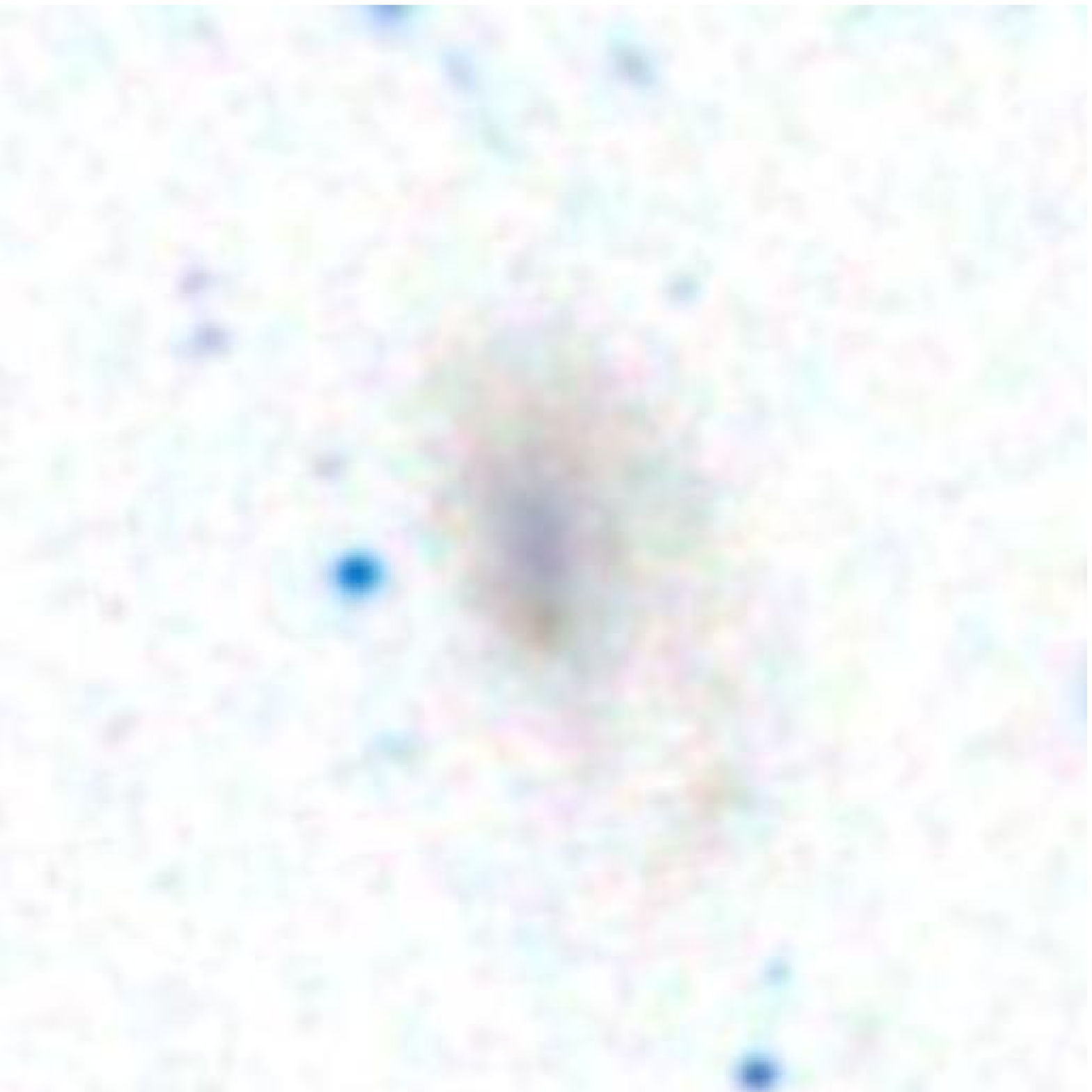}
\includegraphics [width=3cm, height=3cm] {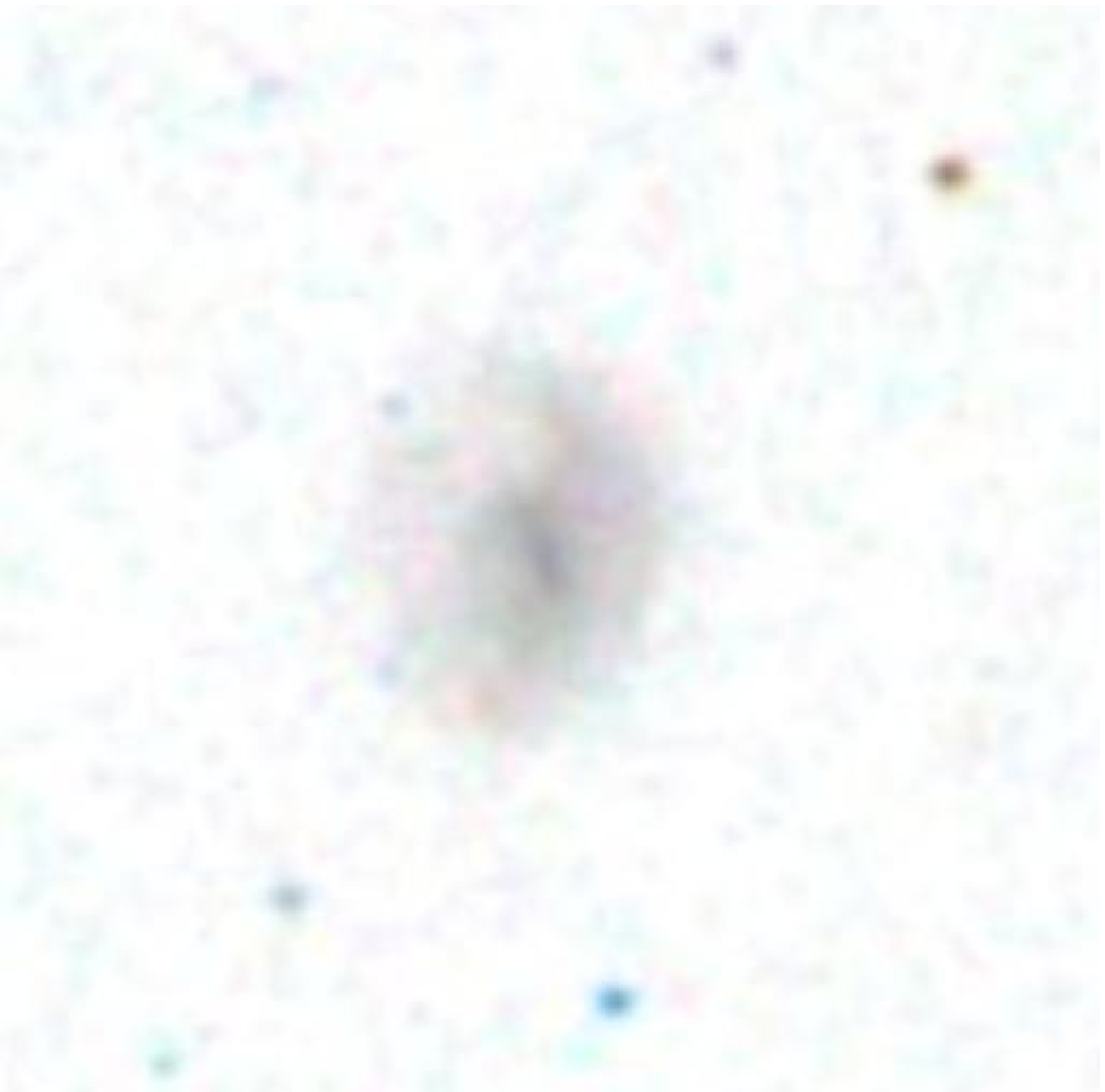}
\includegraphics [width=3cm, height=3cm] {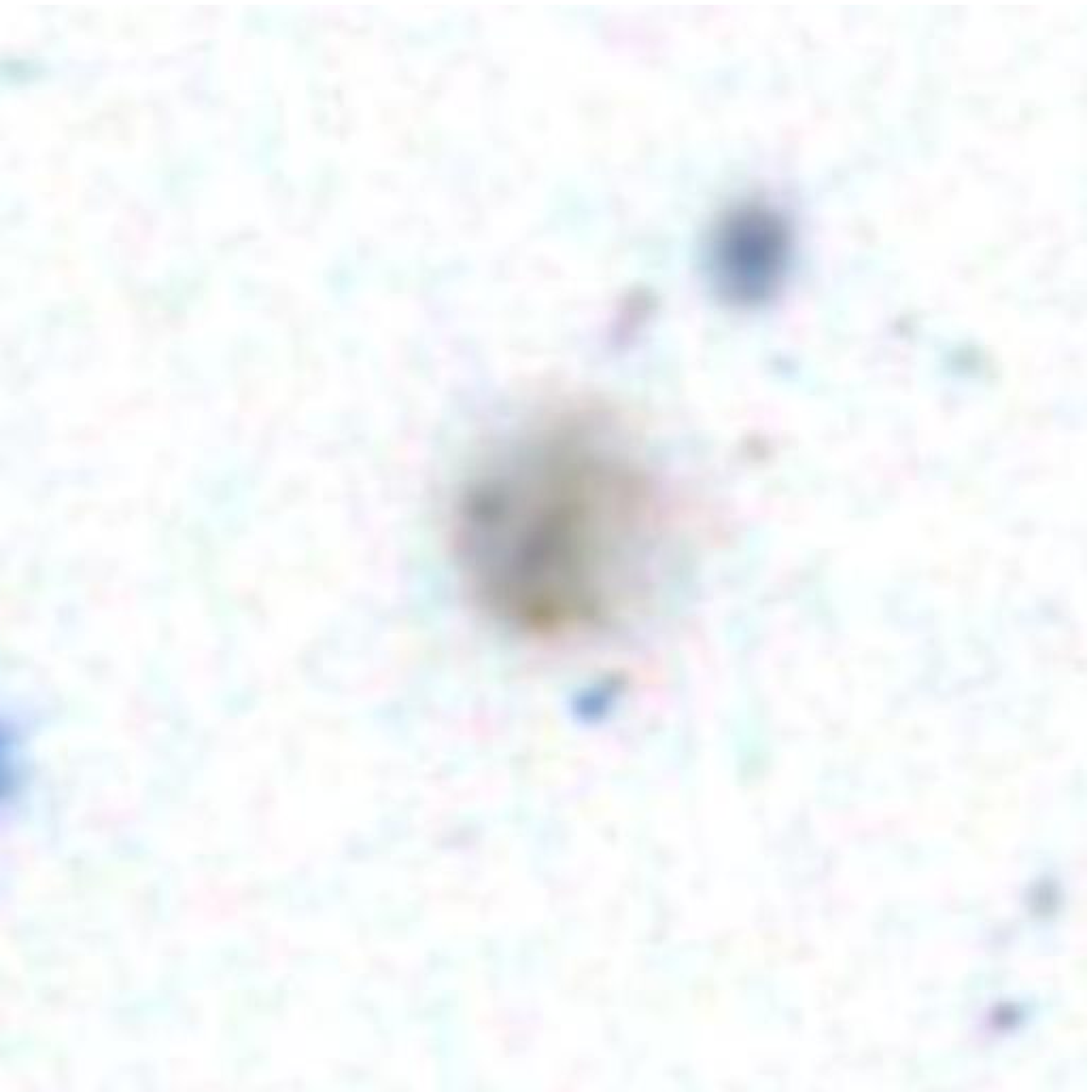}
\includegraphics [width=3cm, height=3cm] {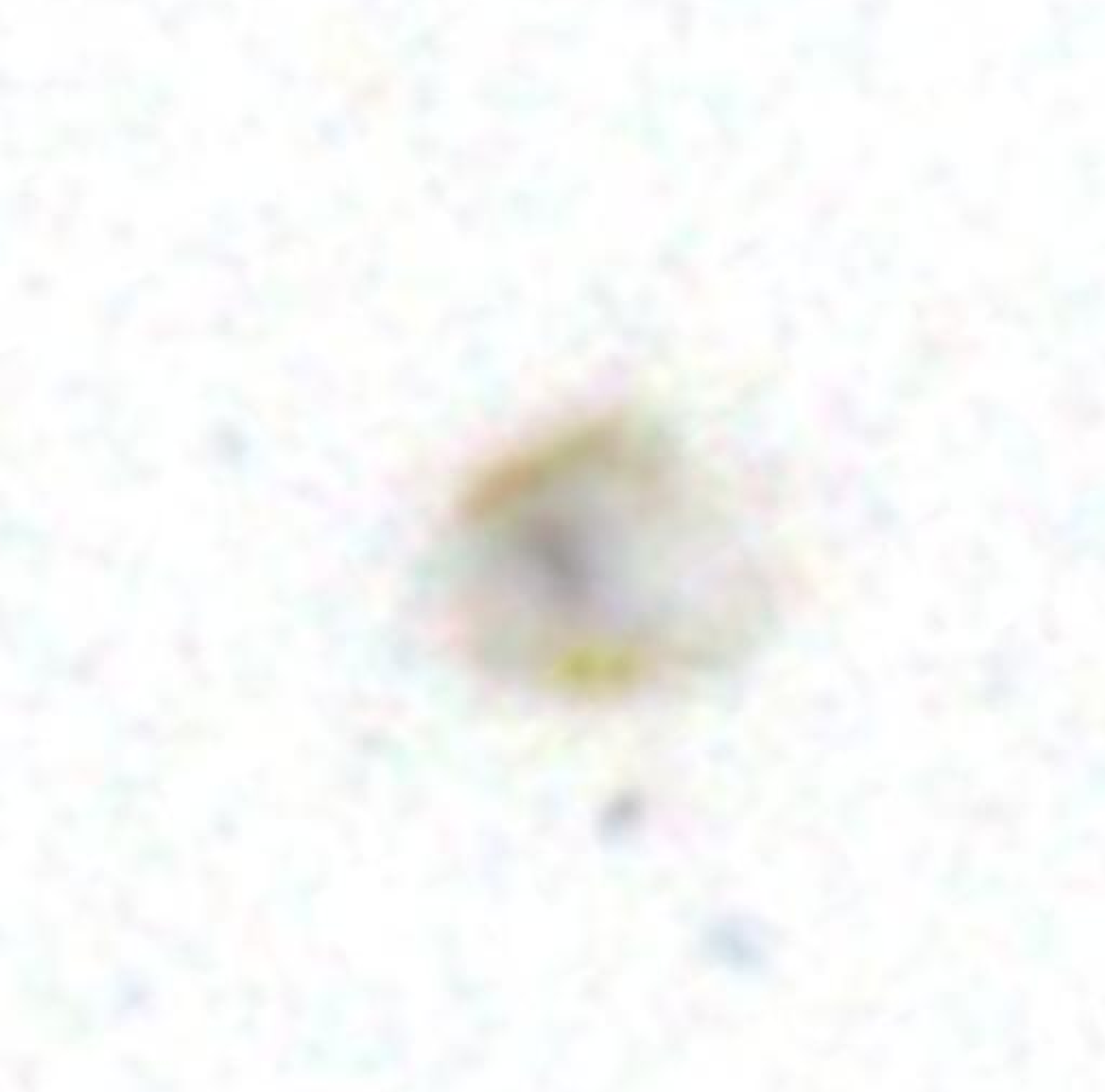}
\includegraphics [width=3cm, height=3cm] {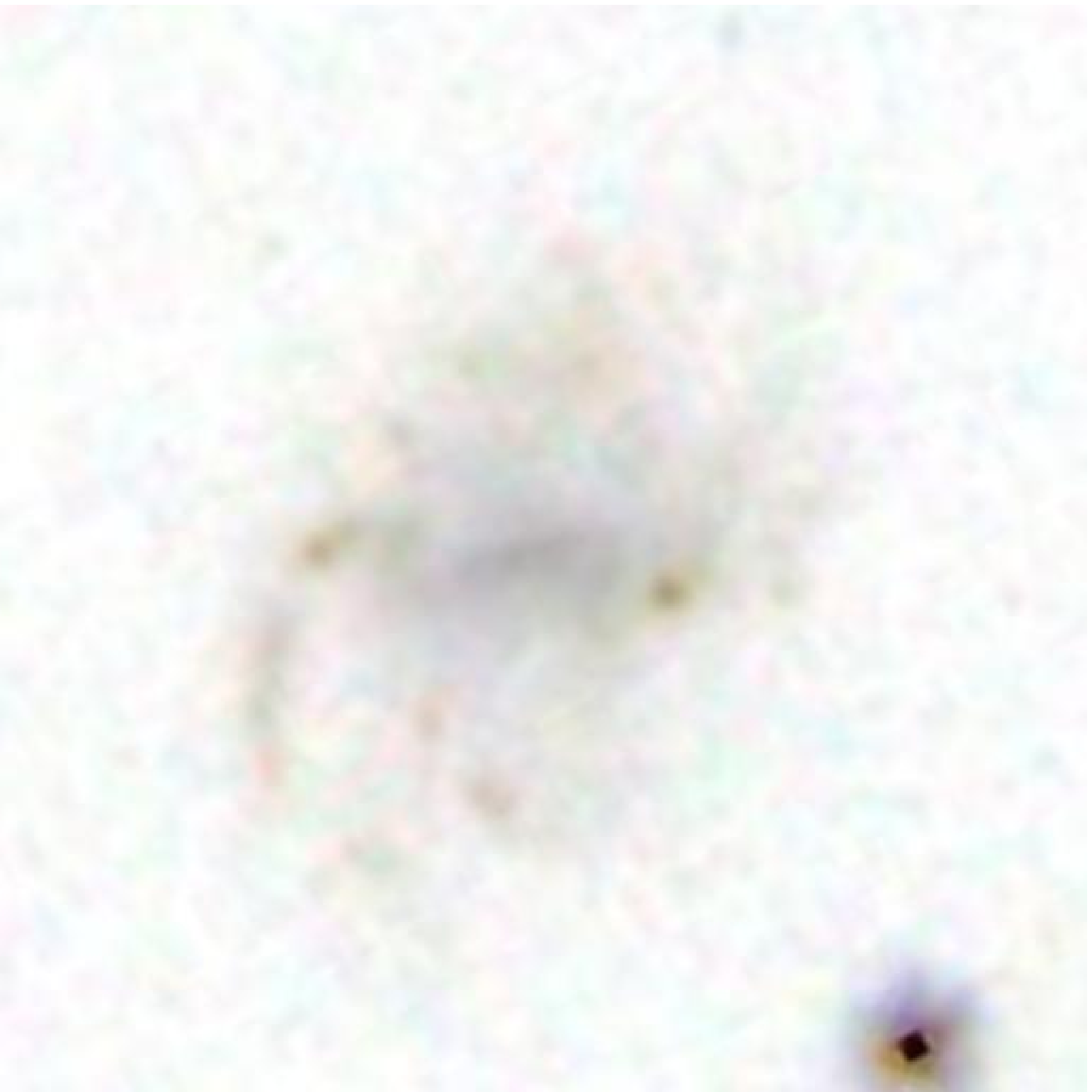}
\includegraphics [width=3cm, height=3cm] {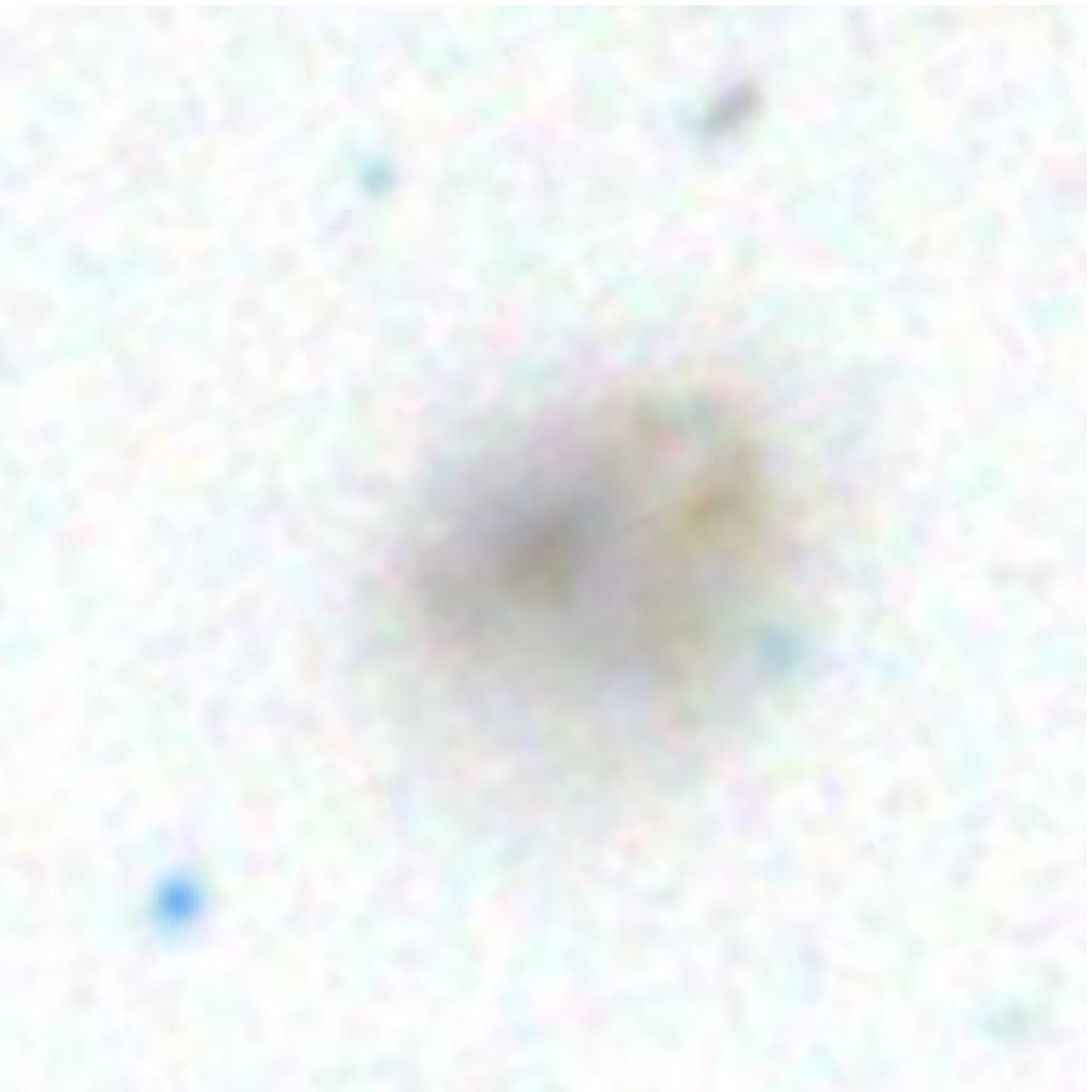}
\includegraphics [width=3cm, height=3cm] {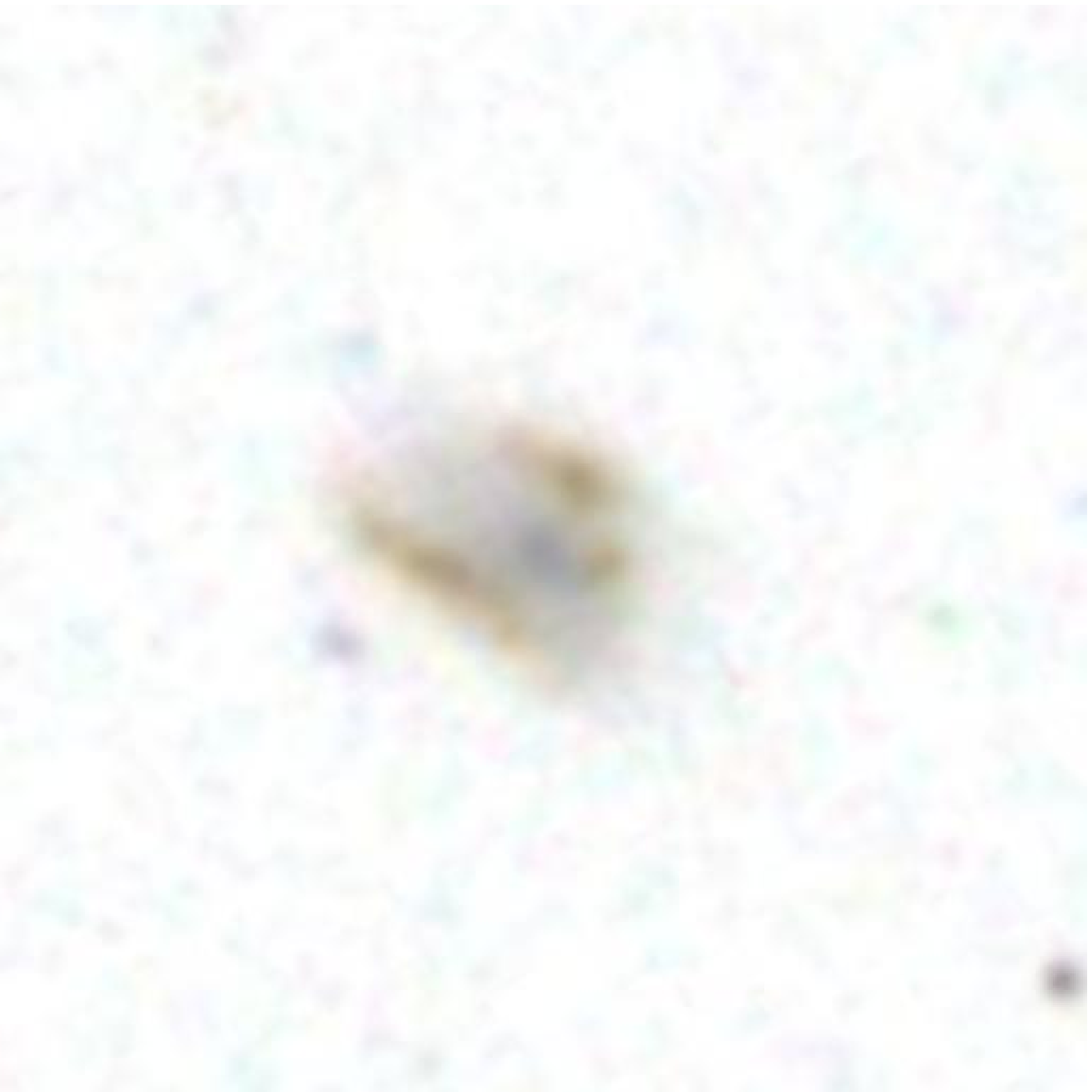}
\includegraphics [width=3cm, height=3cm] {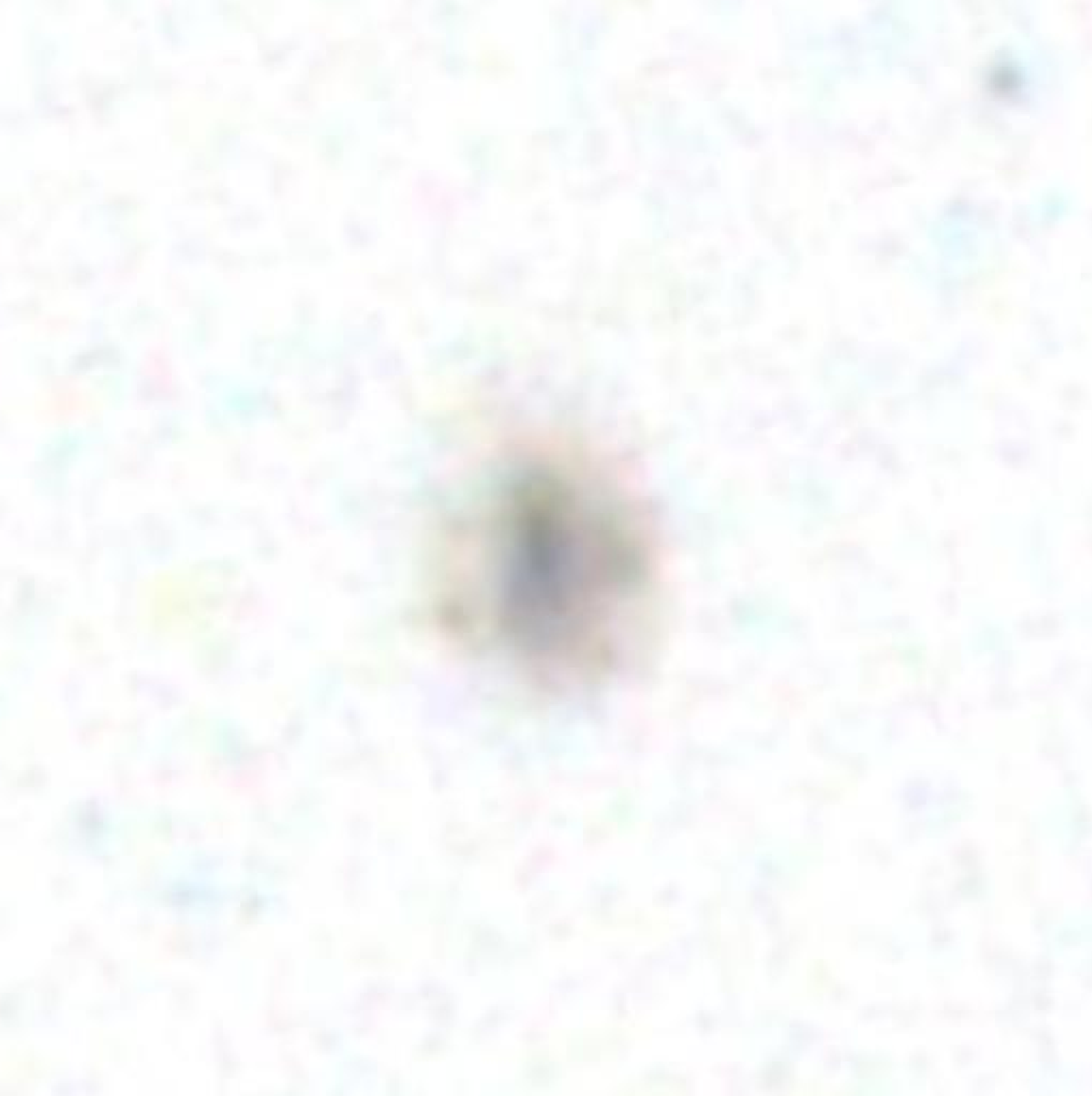}
\includegraphics [width=3cm, height=3cm] {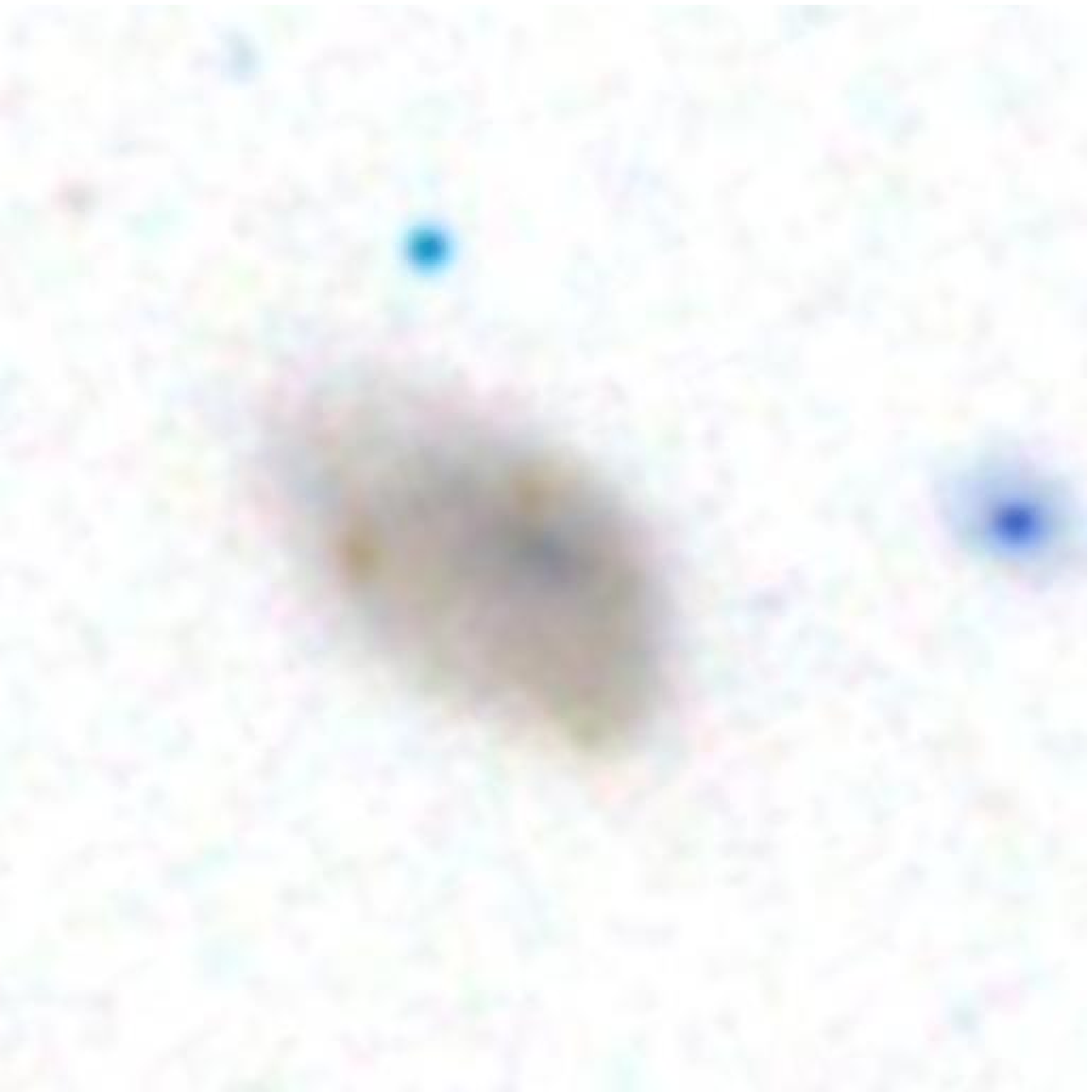}
\includegraphics [width=3cm, height=3cm] {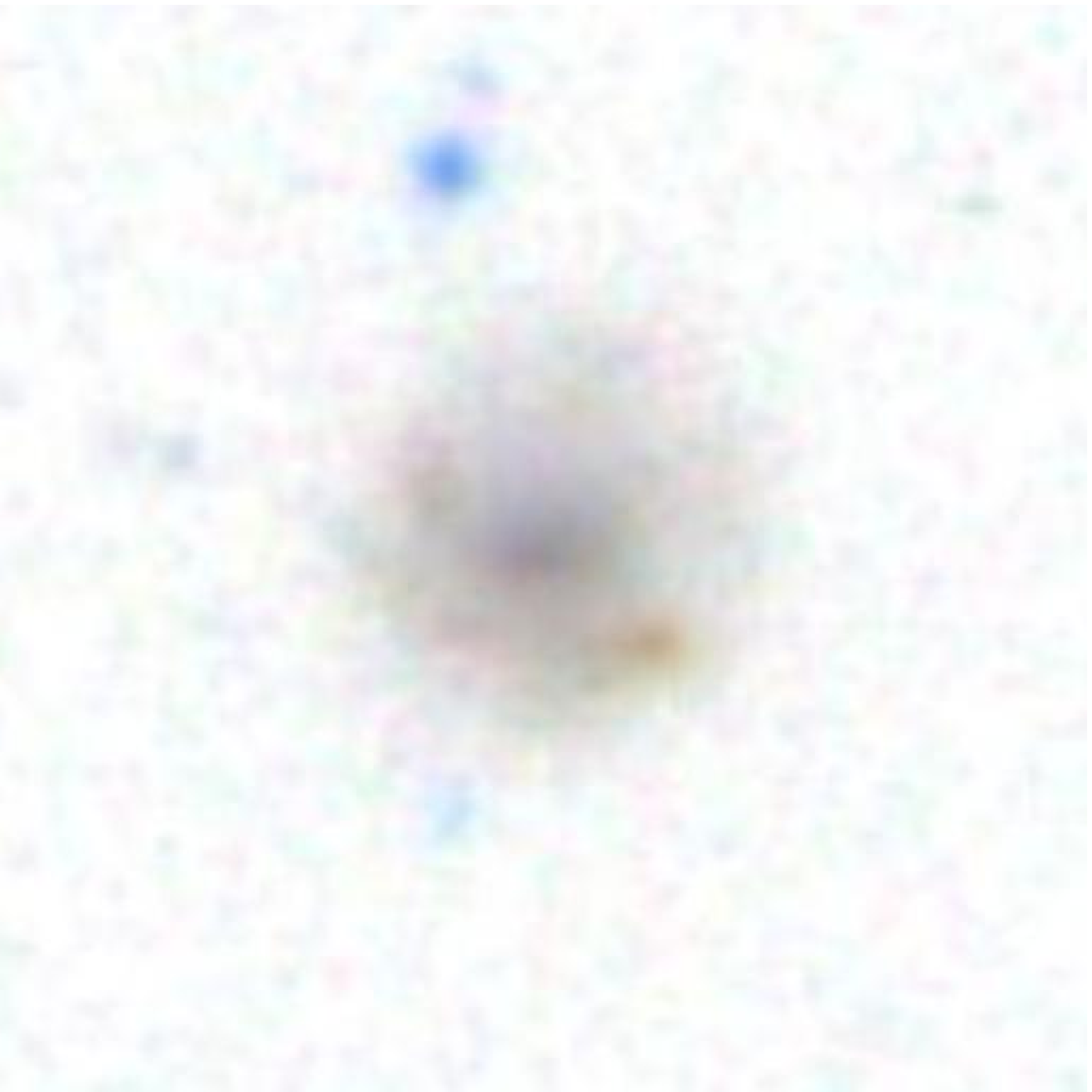}
\includegraphics [width=3cm, height=3cm] {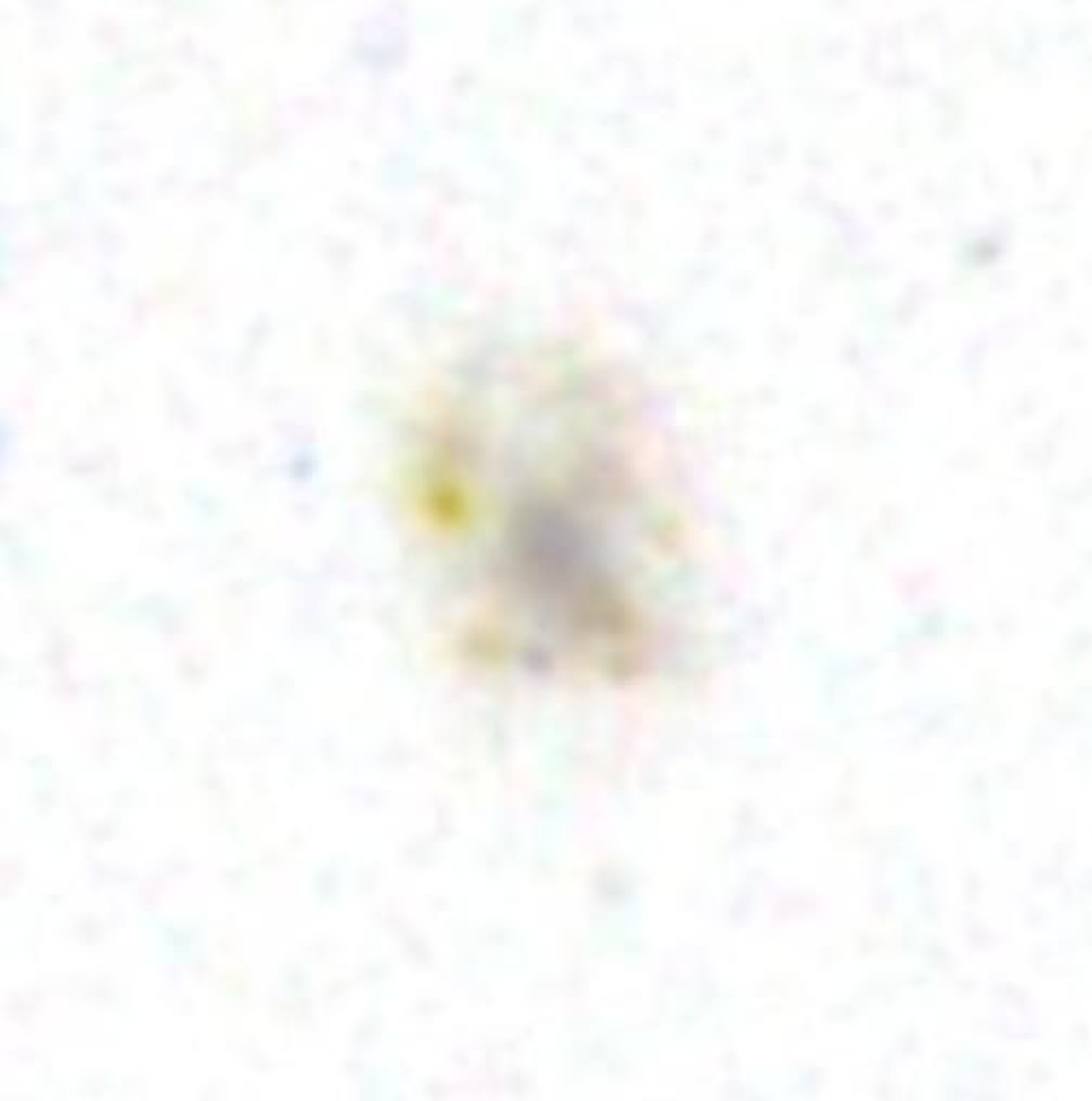}
\includegraphics [width=3cm, height=3cm] {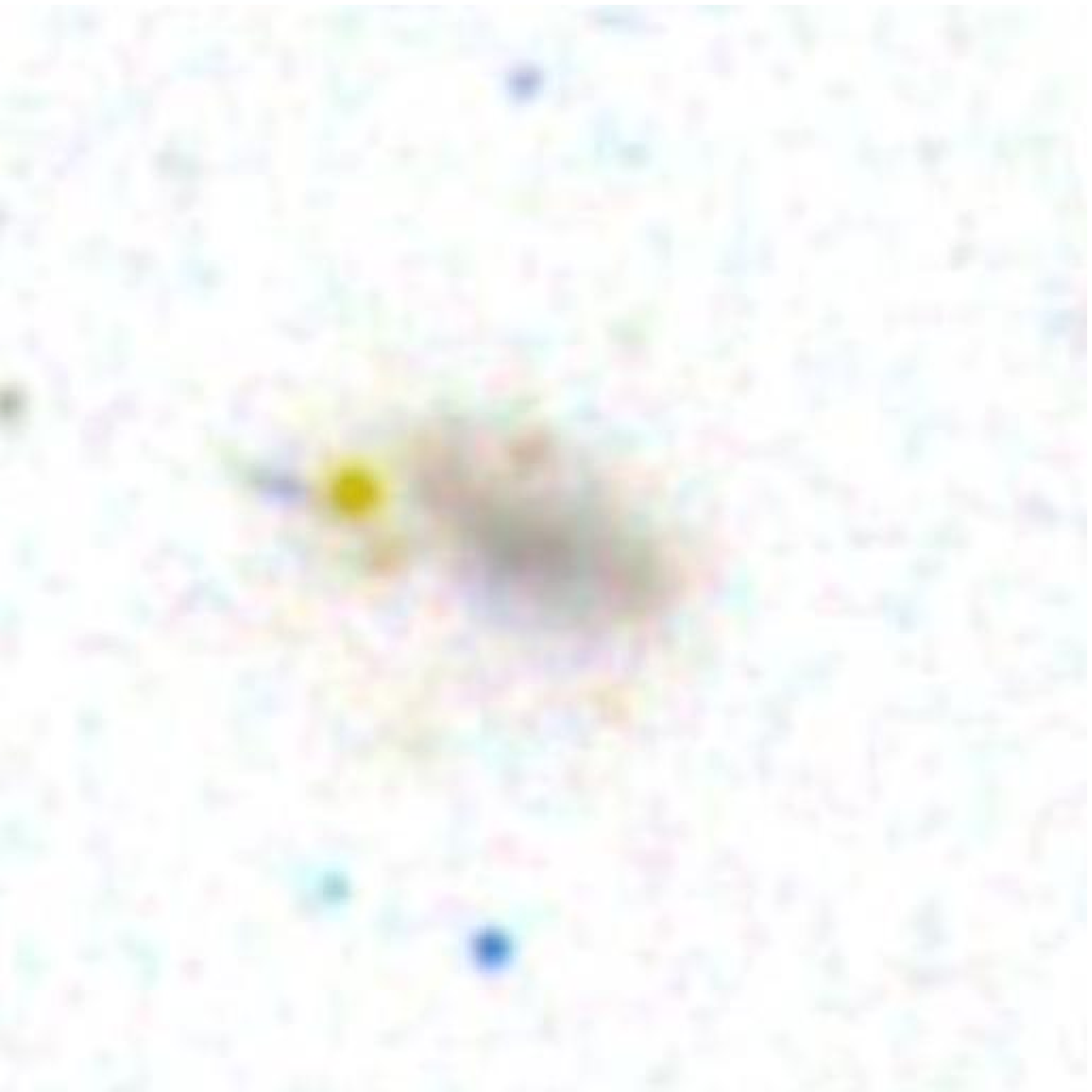}
\includegraphics [width=3cm, height=3cm] {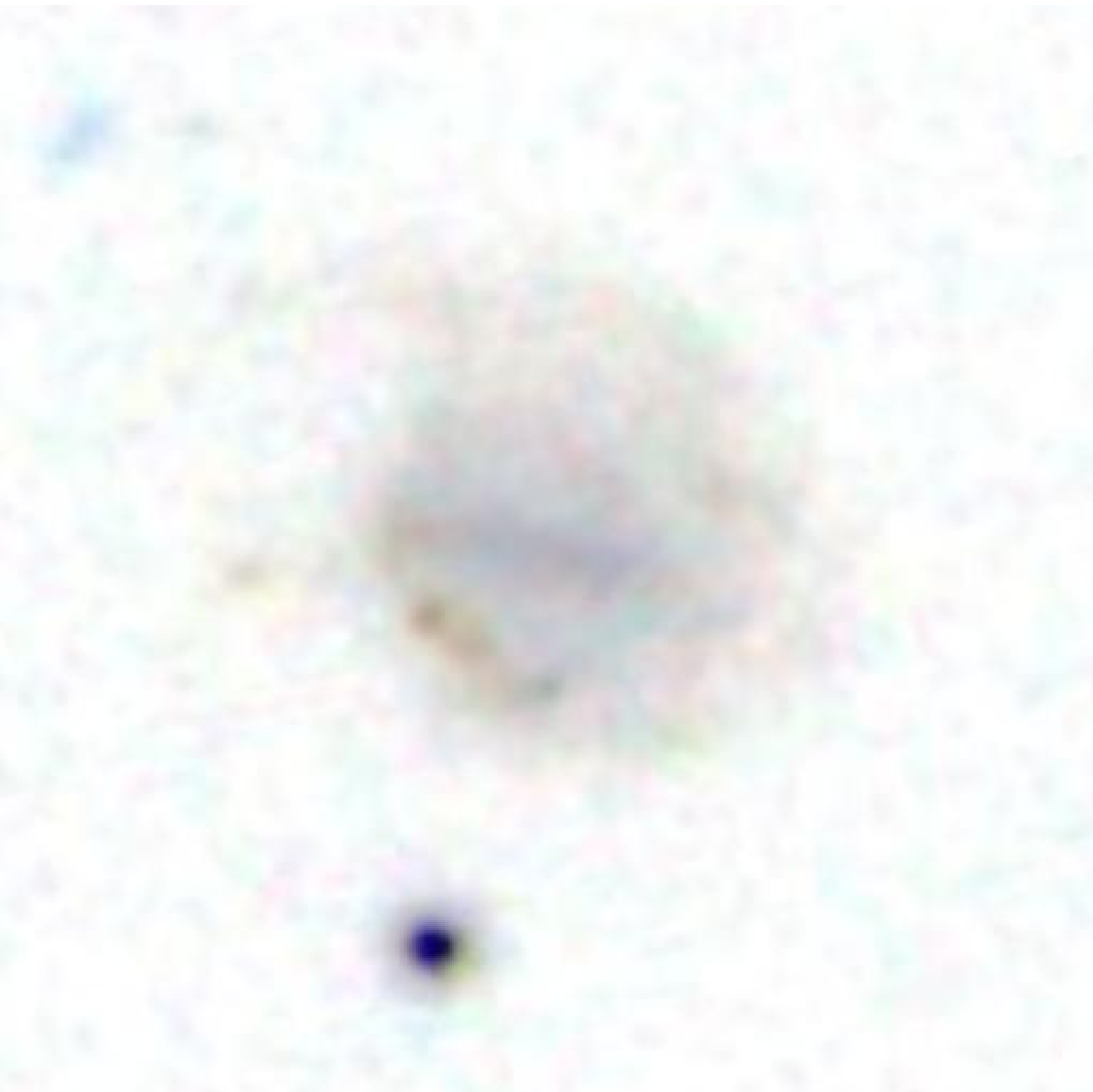}
\includegraphics [width=3cm, height=3cm] {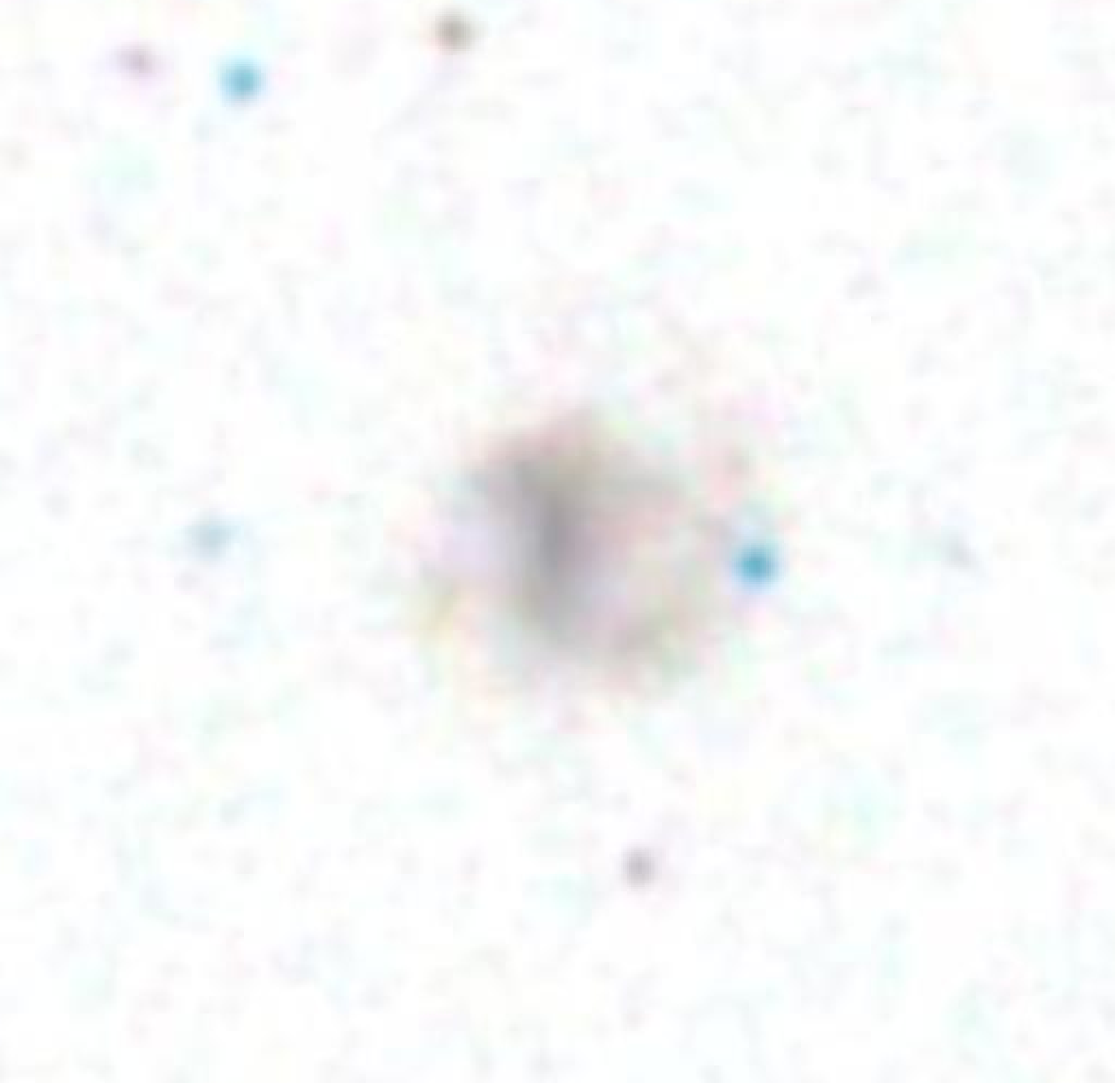}
\includegraphics [width=3cm, height=3cm] {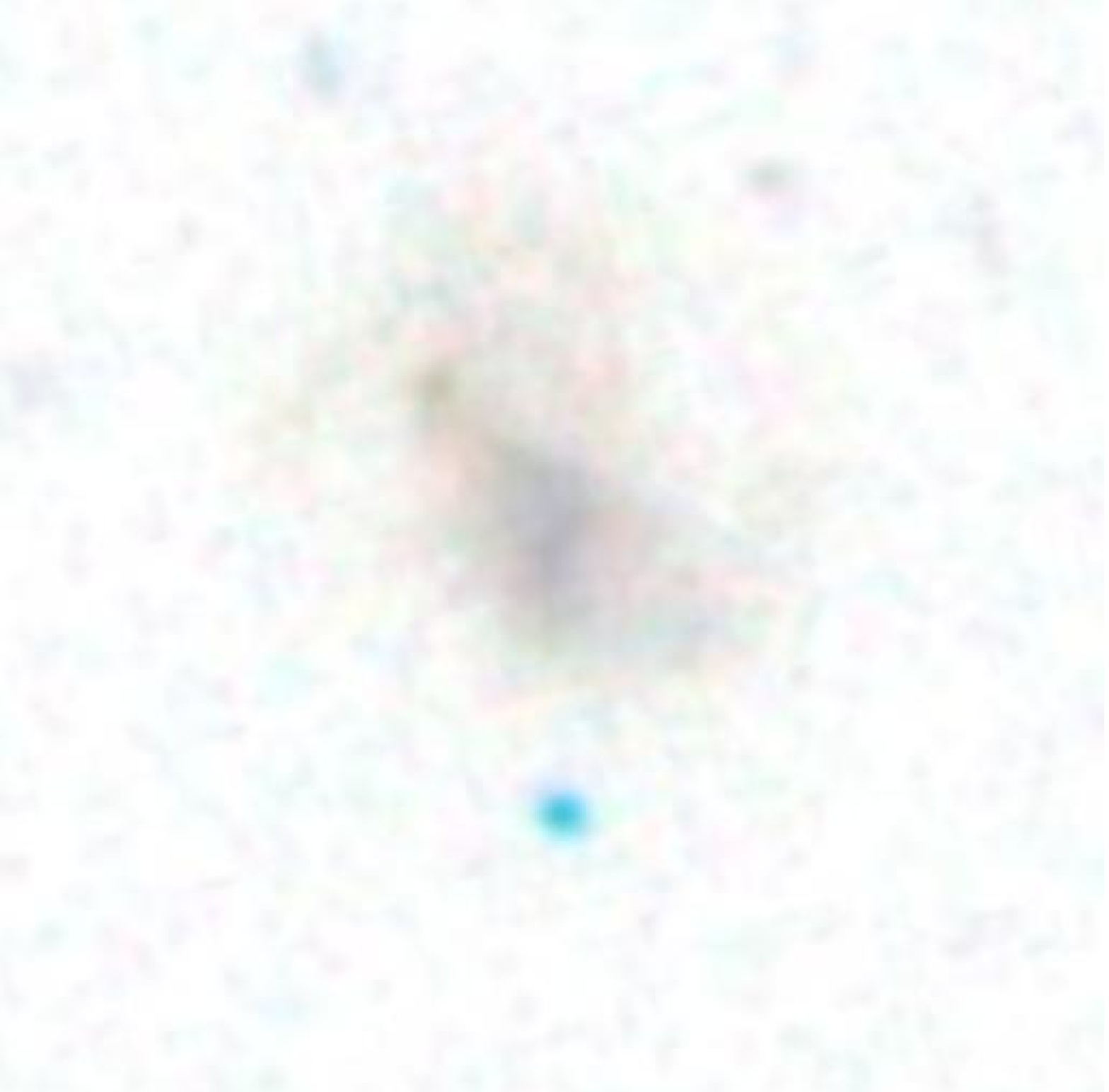}
\includegraphics [width=3cm, height=3cm] {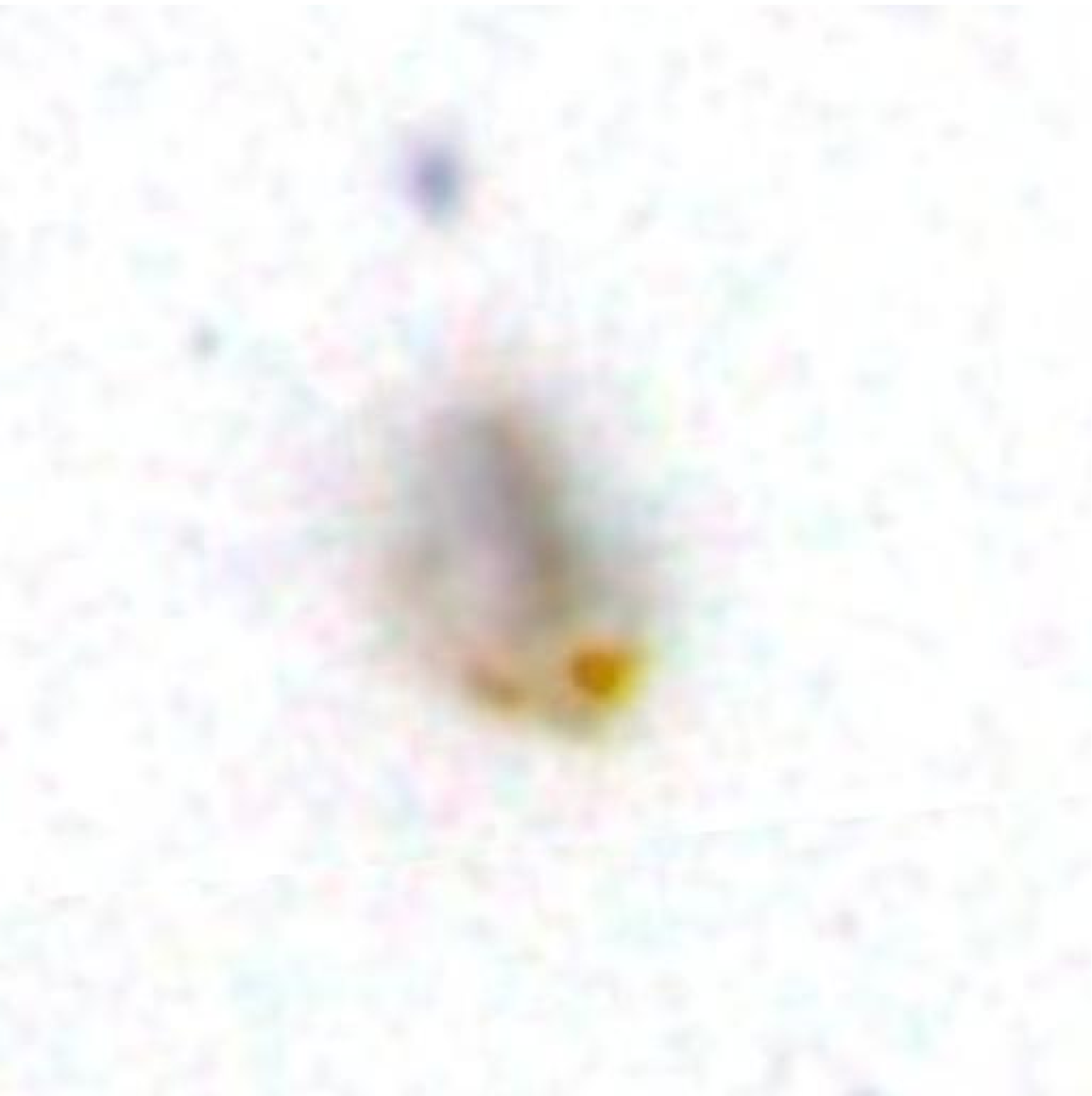}
\includegraphics [width=3cm, height=3cm] {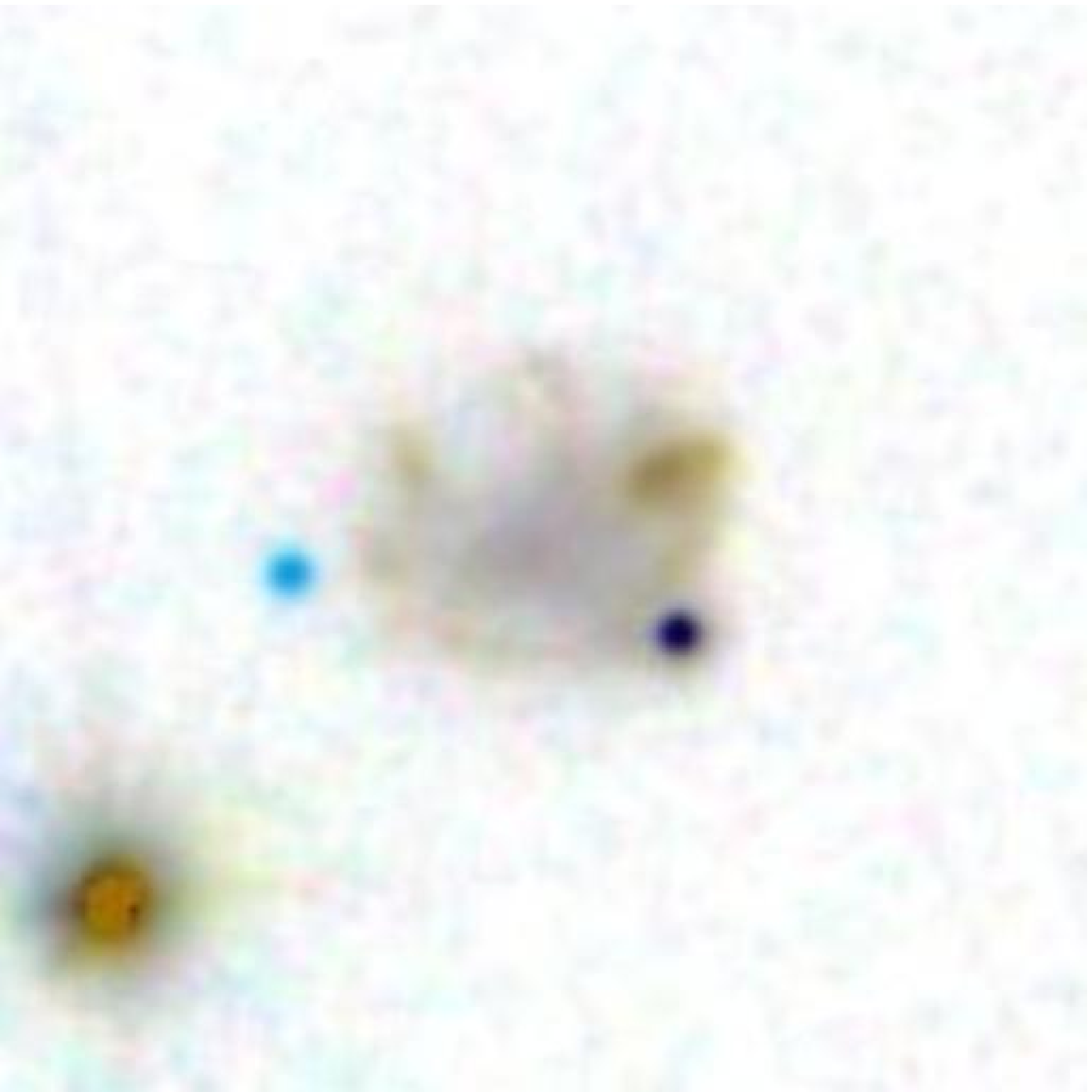}
\includegraphics [width=3cm, height=3cm] {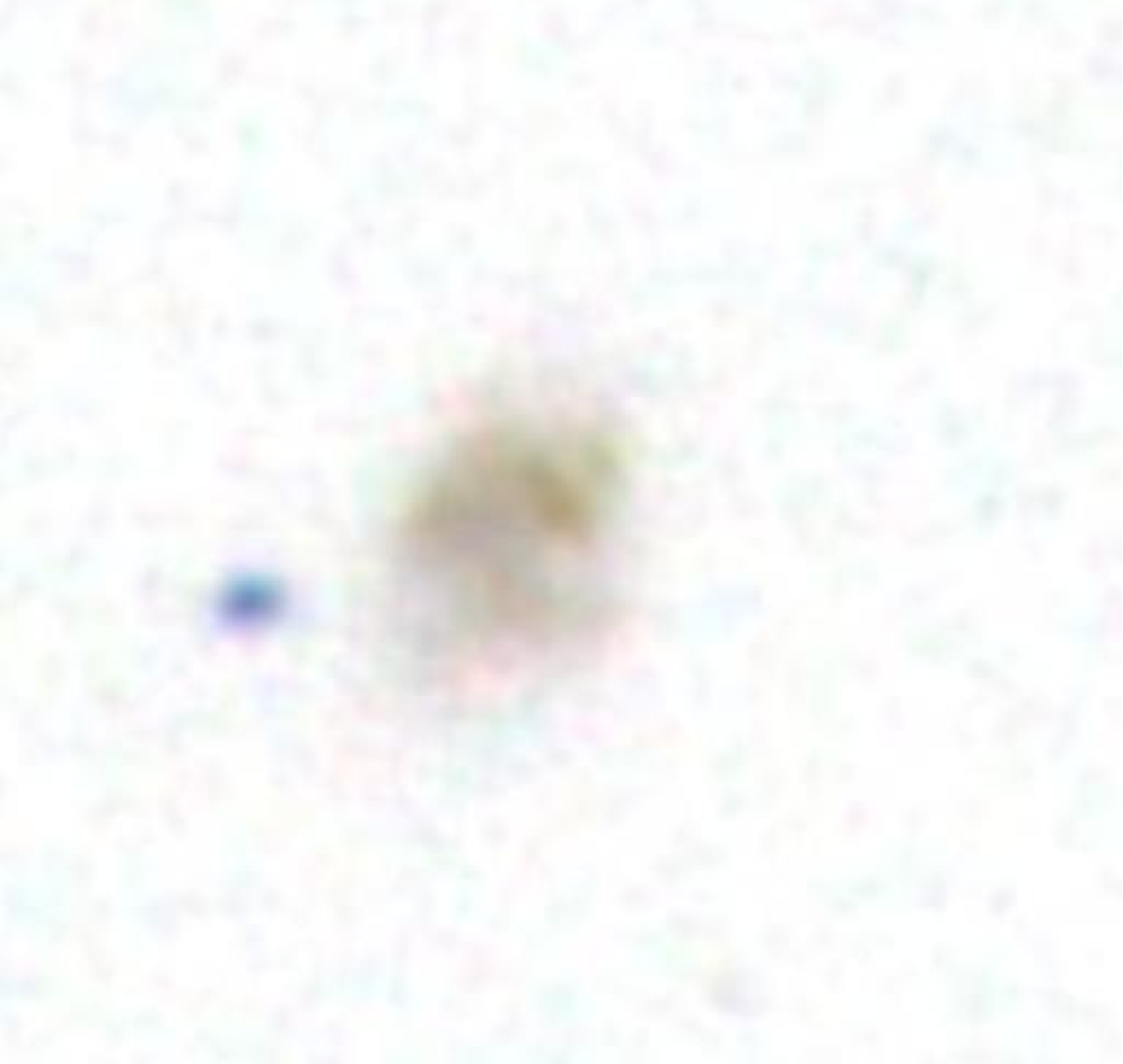}
\includegraphics [width=3cm, height=3cm] {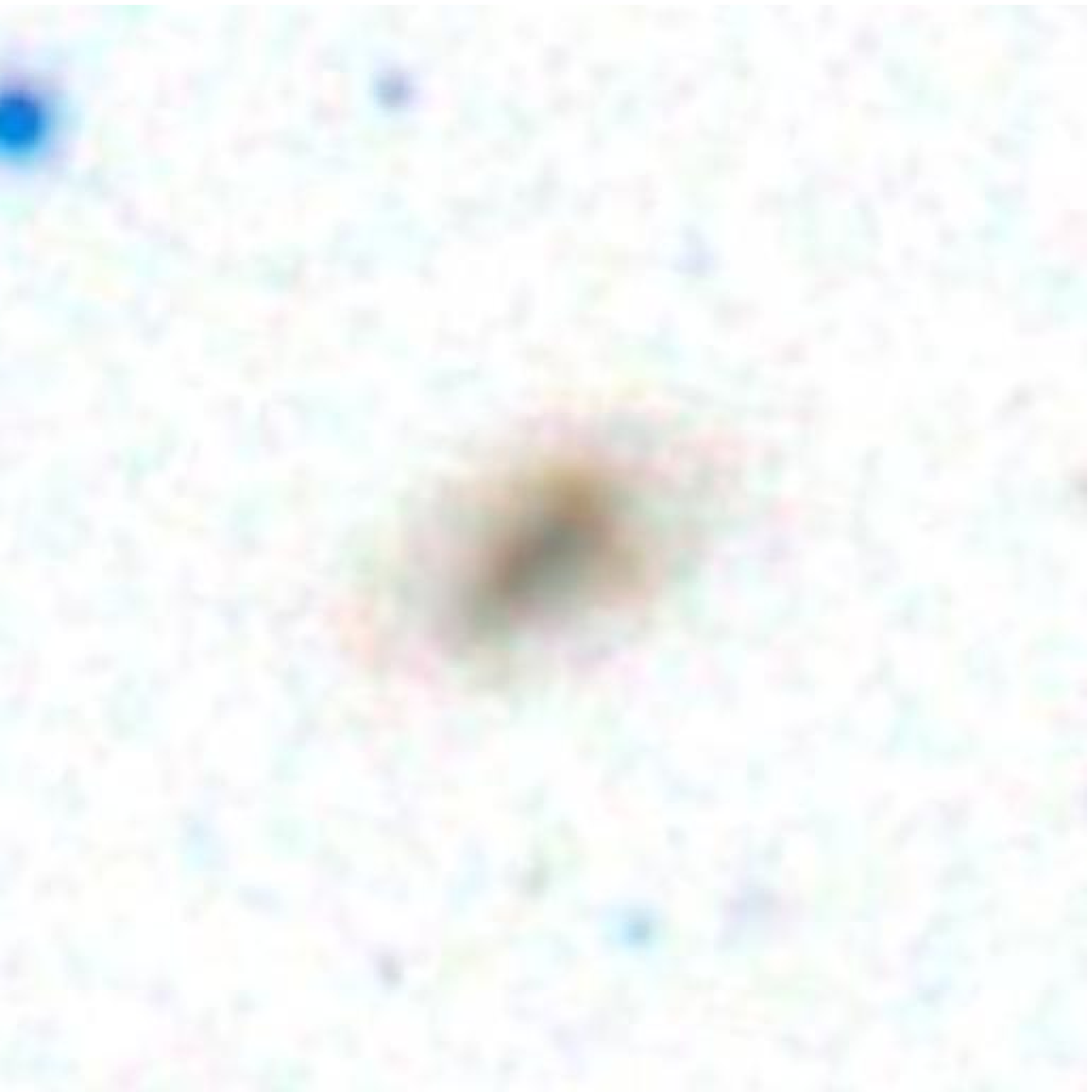}
\includegraphics [width=3cm, height=3cm] {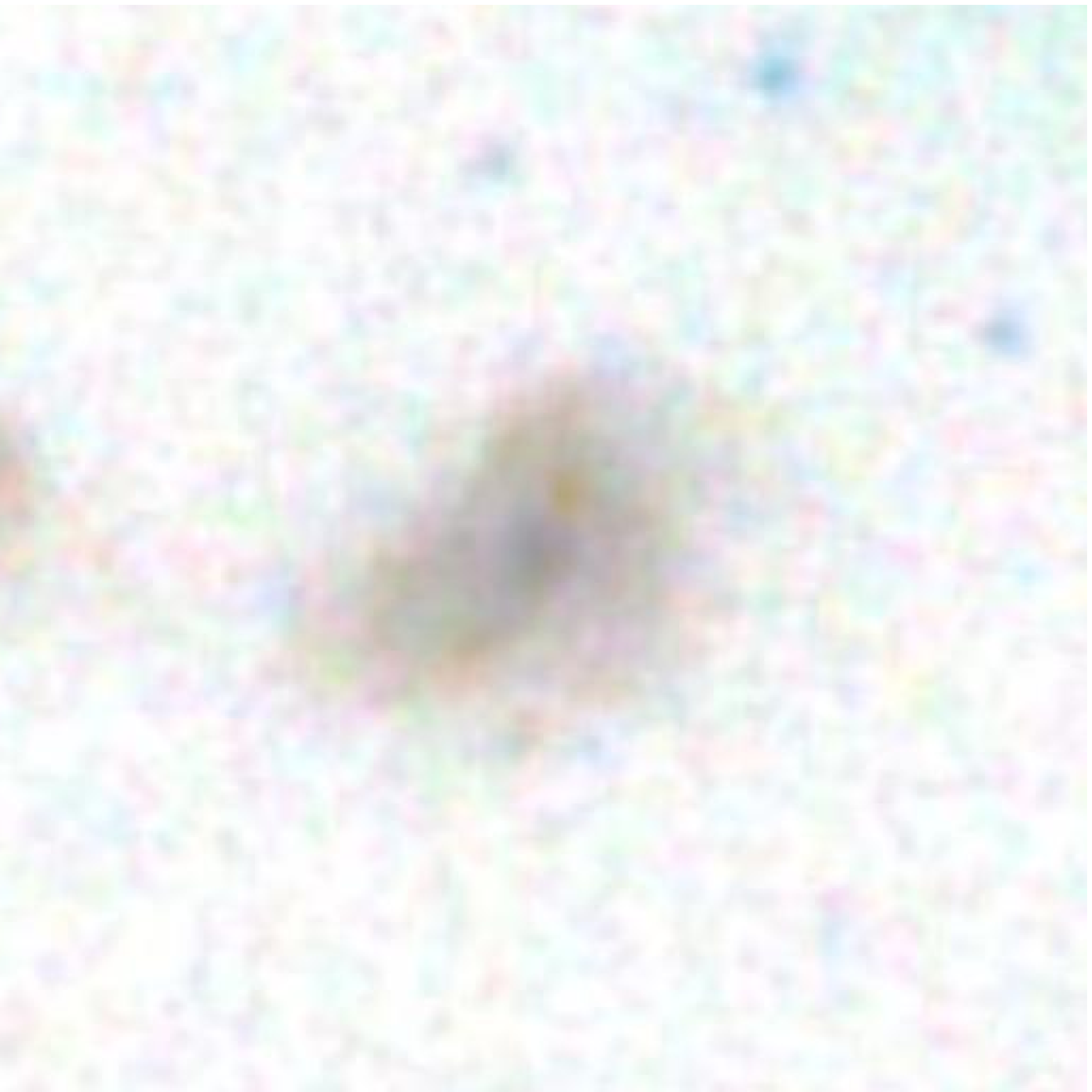}
\includegraphics [width=3cm, height=3cm] {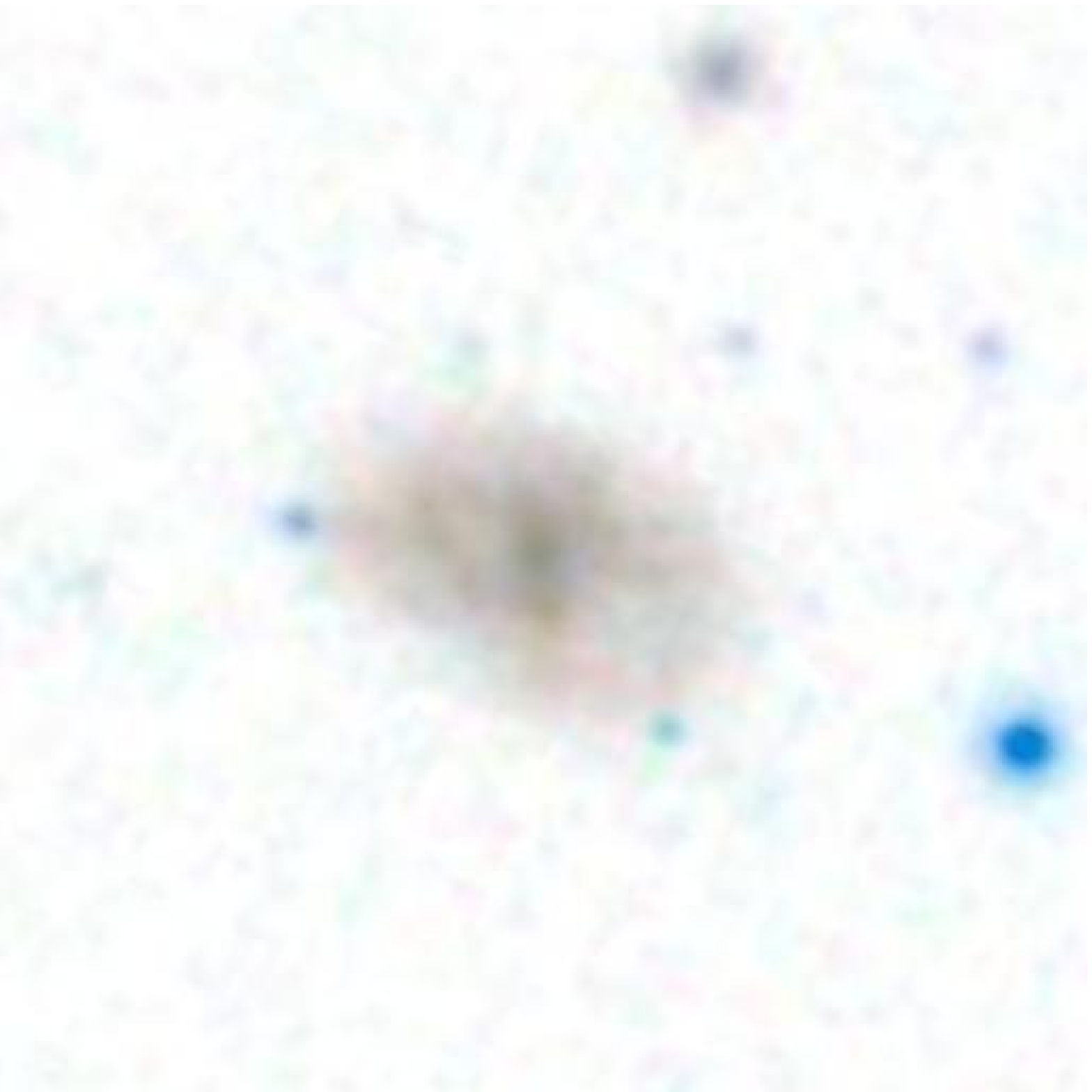}
\includegraphics [width=3cm, height=3cm] {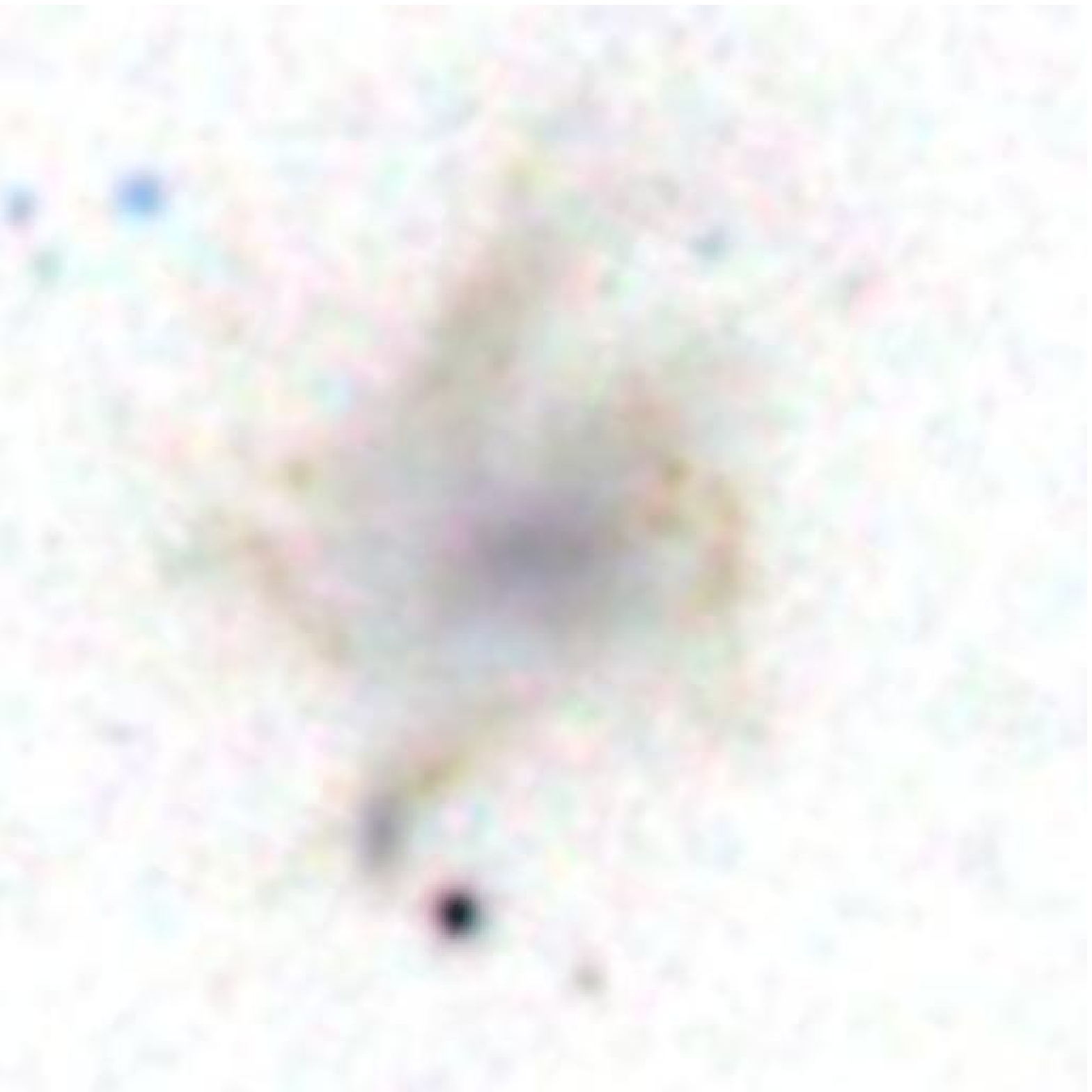}
\includegraphics [width=3cm, height=3cm] {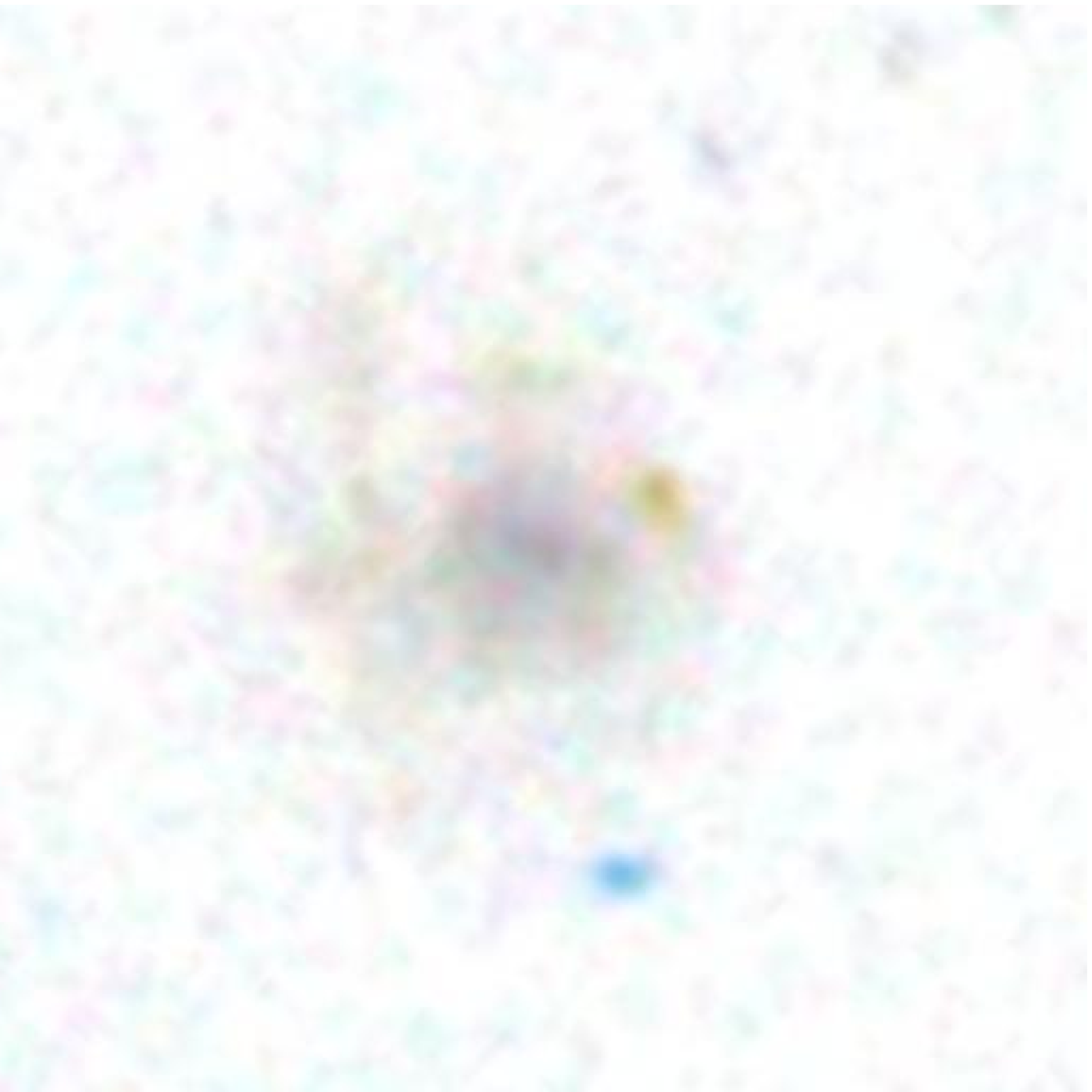}
\includegraphics [width=3cm, height=3cm] {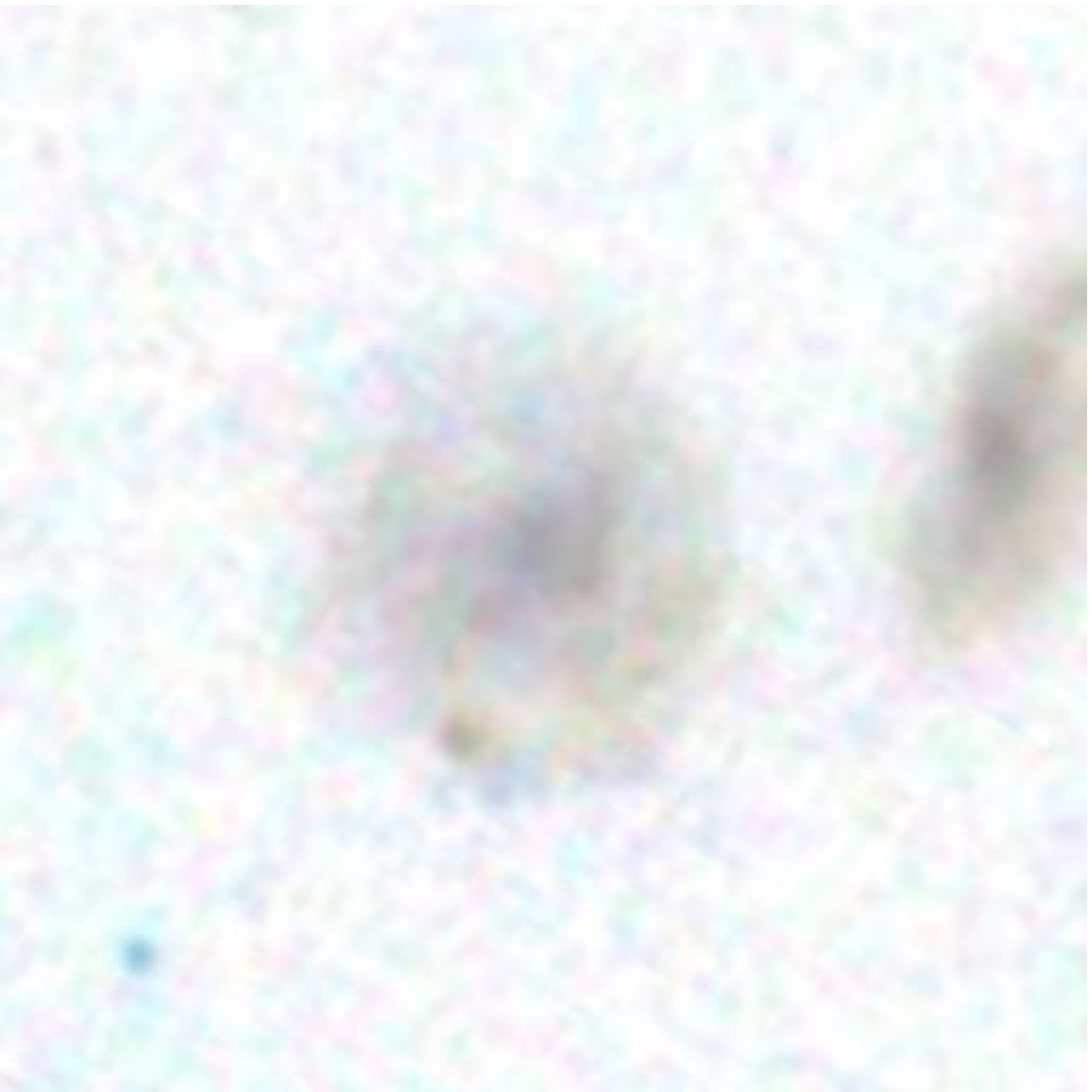}
\caption {The images of low mass irregular LSB galaxies 
(log$M_{\odot}$$<$9.0).} \label{fig.lowmass}
\end{figure*}

\citet{2012AJ....144....4M} reviewed nearly 100 nearby dwarf galaxies and found that
most of them have stellar mass less than 10$^{9}$$M_{\odot}$.
When checking the images of irregular LSB galaxies with log$M_{\odot}$$<$9.0 in Figure~\ref{fig.lowmass},
we find that most of them are dwarf irregulars (dIrrs).
Several studies have revealed that recent star formation in dIrrs
takes place at their center \citep{1998AJ....116.2363W, 2001ApJ...562..713M, 2009A&A...502.1015G, 2013A&A...549A..47L, Fouquet et al.2015}.
These star formation episodes might be triggered by the collapse
of high density gas. Indeed \citet{2005A&A...439..111M} found that young stars are
located near regions with high HI column density in the Sagittarius dwarf irregular galaxy.
There is still the possibility that the low-mass irregular
LSB galaxies are tidal dwarf galaxies (TDGs), which can form young stars in their
center \citep{2013MNRAS.431.3543H, Fouquet et al.2015}. For irregular LSB galaxies
with 9.0$<$log$M_{\odot}$$<$9.5, the situation is more complicated.
Kinematics need to be done in detail to see why they have a large fraction of young stars in
their central region. \\

From this quite crude analysis in absence of support from kinematics, it seems that
LSB galaxies share most of their properties with HSB galaxies. Mostly depending on their
masses, the star formation history of the most massive ones are affected by
past mergers, while the least massive ones are gradually more affected 
by stellar phenomena (e.g., outflows) and minor mergers. 

\section{Conclusion}

The environments in which bulgeless LSB galaxies and bulgeless HSB galaxies are located are compared and the
stellar populations of bulgeless LSB galaxies are investigated in this work. We selected
a large sample of 1235 bulgeless low inclination LSB galaxies in a volume-limited sample with redshifts
ranging from 0.024 to 0.04 and $M_r$ $\leq$ $-18.8$.
The local density parameter $\Sigma_5$ is calculated to trace their environment. This parameter
gives a hint as to how far the nearest neighbours are away from the galaxy, but it does not provide any insight into
what interactive histories the galaxies might have had.
For bulgeless galaxies, we find that their surface brightness does not depend on the environment.
Compared with previous studies, only selecting bulgeless
galaxies avoids significant bias when comparing their environments.

The stellar populations of bulgeless LSB galaxies in low density regions are similar to those of bulgeless LSB
galaxies in high density regions, suggesting the environment may play a less important role in
the evolution of bulgeless LSB galaxies. On the other hand, bulgeless LSB galaxies with
different morphologies have significant differences in the stellar populations. Different dynamic situations
make irregular LSB galaxies have more young stars in the center than regular LSB galaxies.
These results suggest that the
dynamics and mergers for at least the more massive ones may play a dominant role in the evolution of LSB galaxies
rather than environment.

\begin{acknowledgements}

We appreciate James Wicker for improving our English expression in the text from the native 
language. We appreciate Hector Flores, Guohu Zhong, Minnie Lam and Ming Yang for helpful discussions
about this paper. This study is supported by the Natural Science Foundation of China under grants 
No. 11273026. 

Funding for the SDSS and SDSS-II has been provided by the Alfred P. Sloan Foundation,
the Participating Institutions, the National Science Foundation, the U.S. Department of
Energy, the National Aeronautics and Space Administration, the Japanese Monbukagakusho,
the Max Planck Society, and the Higher Education Funding Council for England.
The SDSS Web Site is http://www.sdss.org/.

The SDSS is managed by the Astrophysical Research Consortium for the Participating Institutions.
The Participating Institutions are the American Museum of Natural History,
Astrophysical Institute Potsdam, University of Basel, University of Cambridge,
Case Western Reserve University, University of Chicago, Drexel University, Fermilab,
the Institute for Advanced Study, the Japan Participation Group, Johns Hopkins University,
the Joint Institute for Nuclear Astrophysics, the Kavli Institute for Particle Astrophysics and Cosmology,
the Korean Scientist Group, the Chinese Academy of Sciences (LAMOST),
Los Alamos National Laboratory, the Max-Planck-Institute for Astronomy (MPIA),
the Max-Planck-Institute for Astrophysics (MPA), New Mexico State University,
Ohio State University, University of Pittsburgh, University of Portsmouth, Princeton University,
the United States Naval Observatory, and the University of Washington.

\end{acknowledgements}

\end{document}